\documentclass[onecolumn,amsmath,amssymb,nofootinbib,superscriptaddress]{revtex4-2}

\usepackage[utf8]{inputenc}
\usepackage[margin=1in]{geometry}
\usepackage{amsmath, amsthm, amssymb,amscd, mathrsfs, amsfonts, mathtools,pgfplots}
\usepackage{appendix}
\usepackage{graphicx}
\usepackage{relsize}
\usepackage{mathtools}
\usepackage{dsfont}
\usepackage{float}
\usepackage{soul}
\usepackage{quantikz}
\usepackage{afterpage}
\usetikzlibrary{shapes, backgrounds}

\tikzset{
  qaoaBlock/.style={
    draw,
    rounded corners=2pt,
    minimum width=1.4cm,
    minimum height=0.8cm,
    align=center,
    font=\small,
  },
  problem/.style={qaoaBlock, fill=green!30},
  mixer/.style={qaoaBlock, fill=blue!30},
}
\usetikzlibrary{decorations.pathmorphing}

\usetikzlibrary{arrows.meta,positioning,calc}
\usetikzlibrary{patterns}
\usepackage{chngcntr}
\usepackage[bookmarks=true,
bookmarksnumbered=true,
breaklinks=true,
pdfstartview=FitH,
hyperfigures=false,
plainpages=false,
naturalnames=true,
colorlinks=true,
linkcolor=blue,       
citecolor=blue,       
urlcolor=blue,        
pagebackref=true,
pdfpagelabels]{hyperref}
\usepackage{ORCIDinREVTeX}
\theoremstyle{definition}
\newtheorem{thm}{Theorem}[section]
\newtheorem{prop}[thm]{Proposition}

\newtheorem{defn}[thm]{Definition}
\newtheorem{cor}[thm]{Corollary}
\newtheorem{rmk}[thm]{Remark}

\newtheorem{ex}[thm]{Example}

\allowdisplaybreaks

\usepackage[draft, commentmarkup=todo,
  todonotes={textsize=tiny, textwidth=0.83in}]{changes}

\definechangesauthor[name={Boris}, color=red]{new}
\definechangesauthor[name={Ilya Safro}, color=yellow]{IS}

\definechangesauthor[name={Boris}, color=blue]{BT}

\definechangesauthor[name={Bao Bach}, color=teal]{BB}

\usepackage{bbm}

\hypersetup{
	colorlinks = true,
	urlcolor = blue,
	linkcolor = blue,
	citecolor = red,
	pdfpagemode = UseNone
}

\usepackage{youngtab}
\usepackage{ytableau}

\definecolor{orbitBlue}{RGB}{230, 240, 255}
\definecolor{orbitBorder}{RGB}{50, 100, 180}
\definecolor{closureRed}{RGB}{200, 50, 50}
\definecolor{exactGreen}{RGB}{20, 130, 70}
\definecolor{flowBoxBg}{RGB}{250, 250, 252}
\definecolor{flowBoxBorder}{RGB}{180, 180, 200}

\begin{document}

\title{Circuit Synthesis for Variational Quantum Algorithms: The Depth of Algebra and Geometry}

\author{Boris Tsvelikhovskiy} 
\orcid{0000-0003-0798-7218}
\email{tsvelibor@gmail.com}
\affiliation{Department of Mathematics, University of California, Riverside, CA 92521, USA}

\begin{abstract}
The analysis of Hamiltonian Variational Ans\"atze centers on three fundamental questions: which states are exactly reachable from a given initial state, what native control depth suffices for local and global synthesis, and whether a bounded circuit extension can expose a direction of decreasing cost at a suboptimal output. We address these challenges by leveraging the dynamical Lie algebras (DLAs)  associated with the circuits and the geometry of the initial state's orbit  under the action of the corresponding dynamical Lie group $G$. We prove that an initial state has a closed projective orbit whenever it is a nondegenerate eigenstate of a control Hamiltonian, without requiring the dynamical Lie group itself to be closed. Combined with asymptotic QAOA convergence, this guarantees exact, finite depth ground state preparation for architectures employing diagonal problem Hamiltonians and irreducible stoquastic mixers. We also show that an orbit of quadratic dimension can contain exponentially many mutually orthogonal optimal states, demonstrating how a small reachable manifold can accommodate a large family of solutions. We bound the number of native control exponentials required for local unitary synthesis and state preparation using DLA commutator filtration growth alongside reachable orbit dimensions. Complementing these local results, we construct global frameworks on invariant sectors equipped with full unitary or special unitary dynamical Lie algebras. These constructions yield explicit applications to Pauli-generated systems and constrained maximum independent set multi-angle QAOA circuits; specifically, we show that their native control costs are directly determined by commutator filtration depth and independent-set toggle-group geometry, respectively. Finally, if the dynamical Lie group is compact, the observable $H_P$ generates an admissible evolution ($iH_P \in \mathfrak{g}$, where $\mathfrak{g}$ is the DLA), and the projective orbit of the initial state contains a ground state ray of $H_P$, then a bounded depth, identity-initialized circuit extension exhibits either first-order descent or negative curvature at every suboptimal circuit output.
\end{abstract}

\maketitle

\section{Introduction}\label{sec:introduction}

Variational circuits built from Hamiltonian evolutions employ a prescribed set of controls at every depth. For instance, the celebrated Quantum Approximate Optimization Algorithm (QAOA) employs a variational ansatz  produced by alternating unitary evolutions generated by the exponentials of the problem and mixer Hamiltonians~\cite{QAOA,HHR}; related constructions appear throughout variational quantum algorithms (VQA)~\cite{CABB}. Although asymptotic convergence to the optimal eigenspace is established for suitable QAOA mixers~\cite{BKZS}, it remains an open question whether the optimum can be attained exactly at a finite circuit depth. Likewise, establishing that a transformation is controllable does not quantify the number of unitary evolutions in the basic Hamiltonians required to implement it.  In this work, we establish general reachability and circuit synthesis theorems for Hamiltonian variational circuits. When applied to QAOA, these methods guarantee exact finite-depth ground state preparation for a broad class of mixers and yield explicit depth bounds across several standard architectures.

The dynamical Lie algebra $\mathfrak g$ associated with a family of control Hamiltonians (see \eqref{eq:intro-dynamical-algebra}) captures the infinitesimal  reachable directions of a quantum circuit. While Lie-algebraic techniques provide foundational controllability criteria in quantum control~\cite{dalessandro2009} and have recently proved instrumental in analyzing the expressivity and trainability of variational quantum algorithms~\cite{FHCKYHSP,RBSKMLC,larocca2025barren}, standard DLA analysis leaves a gap when applied to state-dependent reachability, as it typically abstracts away both the fixed initial state and the explicit circuit-depth bounds required to synthesize operations from a finite native gate set.
Specifically, the minimal \(\mathfrak{g}\)-invariant subspace containing the initial state $\ket{\xi}$ defines a subspace whose unit sphere contains all states generated by the action of parameterized VQA circuits on $\ket{\xi}$. However, because circuits implement unitary operations, the true reachable set is the dynamical Lie group  \(G\) (see \eqref{eq:intro-dynamical-group}) orbit of $\ket{\xi}$, which forms a submanifold of that unit sphere of oftentimes significantly lower dimension. Moreover, mere membership in the orbit's closure does not automatically guarantee attainability via a finite sequence of gate applications without explicit circuit-depth bounds. Finally, knowledge of the generated Lie algebra alone does not specify the required resource overhead, as different generator sets can yield identical algebras at drastically different synthesis costs. This paper's main goal is to bridge the algebraic, geometric, and circuit-level descriptions.

Let $H_1, \ldots, H_s$ be a collection of Hamiltonians on a finite-dimensional Hilbert space $W$ and $|\xi\rangle$ the initial state. We write
\begin{equation}
\label{eq:intro-dynamical-group}
G := \bigl\langle e^{-itH_j} \mid t \in \mathbb{R},\ 1 \le j \le s \bigr\rangle \subseteq U(W)
\end{equation}
for the group of finite native-control products. It is a connected subgroup with dynamical Lie algebra 
\begin{equation}
\label{eq:intro-dynamical-algebra}
\mathfrak{g} = \langle iH_1, \ldots, iH_k \rangle_{\mathrm{Lie}}
\end{equation}
and need not be closed in $U(W)$.

The elements of $G$ are the unitary transformations that can be implemented exactly by finite sequences of
native-control evolutions. In particular, the corresponding variational quantum ansatz generates unitaries of the form
\begin{equation}
\label{eq:general_VQA_circuit}
U_{L}(\boldsymbol{\theta})
=
e^{-i\theta_LH_{j_L}}
\cdots
e^{-i\theta_1H_{j_1}},
\quad j_\ell \in \{1, \ldots, k\}.
\end{equation}
Here, the sequence $\boldsymbol{j} = (j_1, \ldots, j_L)$ determines the Hamiltonians applied across successive layers, while $\boldsymbol{\theta} = (\theta_1, \ldots, \theta_L) \in \mathbb R^L$ provides the corresponding continuous variational parameters, with $L$ denoting the circuit depth.

Consider an initial state $|\xi\rangle$. Its projective orbit $G\cdot[\xi]$ consists of the states attainable (up to a global phase) from $\ket{\xi}$ by finite circuits as in~\eqref{eq:general_VQA_circuit}, whereas the closure of this orbit describes asymptotic reachability.

We first establish general criteria for converting asymptotic state preparation into exact, finite-depth realization using parameterized circuits of the form~\eqref{eq:general_VQA_circuit} (Section~\ref{sec:performance_guarantees}). We prove that this exactness property is guaranteed whenever the projective dynamical group, defined as the image of $G$ under the quotient map $U(W) \to PU(W)$, is closed, as well as when the initial state is a nondegenerate eigenstate of a generator, even when the group itself is not closed (Theorem~\ref{thm:closure_to_finite_word_reachability}). A related result shows that if an observable $H_P$ satisfies $iH_P \in \mathfrak{g}$, its energy infimum over a reachable state orbit is attained by a finite circuit (Theorem~\ref{thm:infimum_attainability}). These statements apply to finite-dimensional Hamiltonian variational ans\"atze generally; neither full unitary controllability nor a specific layer structure is required.

For QAOA, the exact-reachability criterion yields a broad consequence. If the problem Hamiltonian is diagonal and the mixer is irreducible and stoquastic in the computational basis, the mixer possesses a unique Perron--Frobenius ground state. Prior work demonstrates that QAOA initialized in this state approaches the optimal problem Hamiltonian eigenspace (ground space) asymptotically as the depth increases~\cite{BKZS}. Our criterion upgrades this conclusion: some \emph{finite} alternating circuit, utilizing only the original mixer and problem Hamiltonian, prepares an exact ground state (Theorem~\ref{thm:PF_orbit_closure}). This is a statement regarding attainability; it does not bound the minimum circuit depth nor provide an efficient method for computing the optimal angles.


Exact reachability leaves open the quantitative question of the circuit depth required to attain it. Building on conjugation-based iterative DLA basis construction proposed in~\cite{dalessandro2009}, we derive exact local synthesis depth bounds from the commutator order filtration of the DLA (Theorem~\ref{thm:local_synthesis_depth_bound}). Importantly, the filtration is constructed directly from the set of generators $\mathbf{a}$ rather than the DLA $\mathfrak{g}$ itself. If $c_q$ new linearly independent Lie algebra directions emerge at filtration level $q$, and $\mathcal{D}_\mathbf{a}$ is the stabilization depth of the filtration, defined as the minimal index at which the filtration stabilizes (see Definition~\ref{def:stabilization_depth}), any element within a sufficiently small open neighborhood of the identity in the dynamical Lie group can be synthesized using at most
\begin{equation}
\label{eq:intro-local-depth}
\mathcal{T}_{\mathrm{loc}} \le \sum_{j=1}^{\mathcal{D}_\mathbf{a}} (2j-1)c_j \le (2\mathcal{D}_\mathbf{a} - 1)\dim(\mathfrak{g})
\end{equation}
parametrized exponentials of the basic control Hamiltonians. The corresponding infinitesimal state orbit preparation uses the directions tangent to the orbit, which can be far fewer than the dimension of the group because the state may possess a large \emph{stabilizer} (i.e., the subgroup of unitaries leaving the state invariant up to a global phase). Thus, the filtration measures how rapidly the physical controls acquire new directions, whereas the orbit determines how many of those directions matter for a particular initial state. The resulting formulas translate abstract Lie-algebraic structure into quantitative \emph{local} circuit-depth guarantees.

In Section~\ref{sec:stab_depth_QAOA} we evaluate these bounds across several concrete architectures. Leveraging the general classification framework of~\cite{WKKB1} and the explicit characterization of free Maximum Cut (MaxCut) dynamical Lie algebras~\cite[Theorem 1]{KLFCCZ}, we establish system-size-linear stabilization depth bounds for multi-angle QAOA applied to MaxCut on paths, cycles, and complete graphs, accompanied by explicit analytical bounds on  circuit depth for local element synthesis. We proceed by establishing the corresponding bounds for reduced MaxCut, Maximum Independent Set (MIS), and Grover-mixer QAOA circuits, whose dynamical Lie algebras have been classified in~\cite{TBFS, TBS, TNB}.

As local charts fail to yield uniform depth bounds across an entire group manifold or orbit, we develop a complementary \emph{global} synthesis framework for invariant subspaces $V \subseteq W$ of the ambient Hilbert space on which the DLA restricts to the full or special unitary Lie algebra, $\mathfrak{u}(V)$ or $\mathfrak{su}(V)$ (Theorem~\ref{thm:uniform_orbit_generation}). While the implications of these algebras for controllability are standard~\cite{dalessandro2021quantum}, the quantitative challenge lies in implementing the required elementary operations using the native control evolutions. To address this, we first establish depth bounds for synthesizing diagonal operations and elementary two-level rotations, and subsequently assemble them via the unitary decomposition techniques of~\cite{spengler2010composite}. This yields explicit analytic depth bounds for both arbitrary group elements and state-to-state transitions within the sector and its projectivization. Notably, preparing an arbitrary target quantum state can require substantially fewer control steps than implementing an arbitrary unitary operator over the same subspace.

Alternatively, a second global synthesis strategy applies when a fully controllable sector of dimension $D = 2^q$ encodes $q$ logical qubits and the restricted DLA is isomorphic to $\mathfrak{u}(D)$ or $\mathfrak{su}(D)$. By combining the quantum Shannon decomposition with the state-preparation framework of~\cite{shende2006synthesis}, we express the circuit depth bounds directly in terms of the primitive native implementation costs for single-qubit gates ($K_{\mathrm{1q}}$) and CNOT gates ($K_{\mathrm{CX}}$). This yields asymptotic projective unitary-synthesis and state-to-state transfer complexities of $\mathcal{O}\left(D^2 (K_{\mathrm{1q}} + K_{\mathrm{CX}})\right)$ and $\mathcal{O}\left(D (K_{\mathrm{1q}} + K_{\mathrm{CX}})\right)$, respectively (see Theorem~\ref{thm:uniform_orbit_generation2}).

Specifically, our linear stabilization depth estimate for reduced free MaxCut on complete graphs (Theorem~\ref{thm:reduced_free_DLA_filtration_classical}) yields explicit global bounds over its $2^{n-1}$-dimensional reduced Hilbert space (Corollary~\ref{cor:reduced_free_maxcut_K_n}). More broadly, as highlighted in Remark~\ref{rmk:unreduced_vs_reduced_DLAs}, symmetry-reduced ma-QAOA MaxCut instances on graphs, whose reduction is connected and neither bipartite nor cyclic, generate a simple DLA isomorphic to $\mathfrak{su}(2^{n-1})$ on the $(n-1)$-qubit reduced space \cite[Theorem~IV.14]{TBFS}. Consequently, Theorem~\ref{thm:Pauli_global_synthesis} provides a direct route to global synthesis for these reduced architectures once the generator stabilization depth $\mathcal{D}_{\mathbf{a}}$ is determined.

For Maximum Independent Set circuits, whose generators comprise the local \emph{Flip-or-Stay} mixers introduced in~\cite{TBS} alongside individual vertex controls $Z_v$, the DLA realizes the full unitary algebra on the feasible sector, as established in~\cite{TBS}. Letting $D$ denote the number of feasible configurations and $\Delta_\Gamma$ the diameter of the associated toggle-group Cayley graph, our construction yields explicit circuit-depth bounds of $\mathcal{O}(D^3 \Delta_\Gamma)$ for synthesizing arbitrary unitaries and $\mathcal{O}(D^2 \Delta_\Gamma)$ for arbitrary state-to-state transfers (Theorem~\ref{thm:MIS_connected_global_bound}). This concrete application exemplifies our broader methodological transition: moving past qualitative reachability criteria to rigorously bound the physical resources required to execute quantum operations across the reachable manifold.

Local synthesis bridges reachable-state geometry and variational optimization. Specifically, when the dynamical Lie group \(G\) is closed, the problem generator satisfies \(iH_P \in \mathfrak{g}\), and an exact ground state of \(H_P\) is reachable within the orbit, the resulting energy landscape is free of suboptimal local minima. More concretely, from any suboptimal circuit output, one can append an identity-initialized block of bounded native-control depth that exposes either a nonzero energy gradient or a negative-curvature direction (Theorem~\ref{thm:finite_depth_exposure_descent}). This statement complements QAOA layer-extension results~\cite{sack2023recursive} and applies to the broader Hamiltonian-control setting under its hypotheses, guaranteeing the existence of a descent direction.

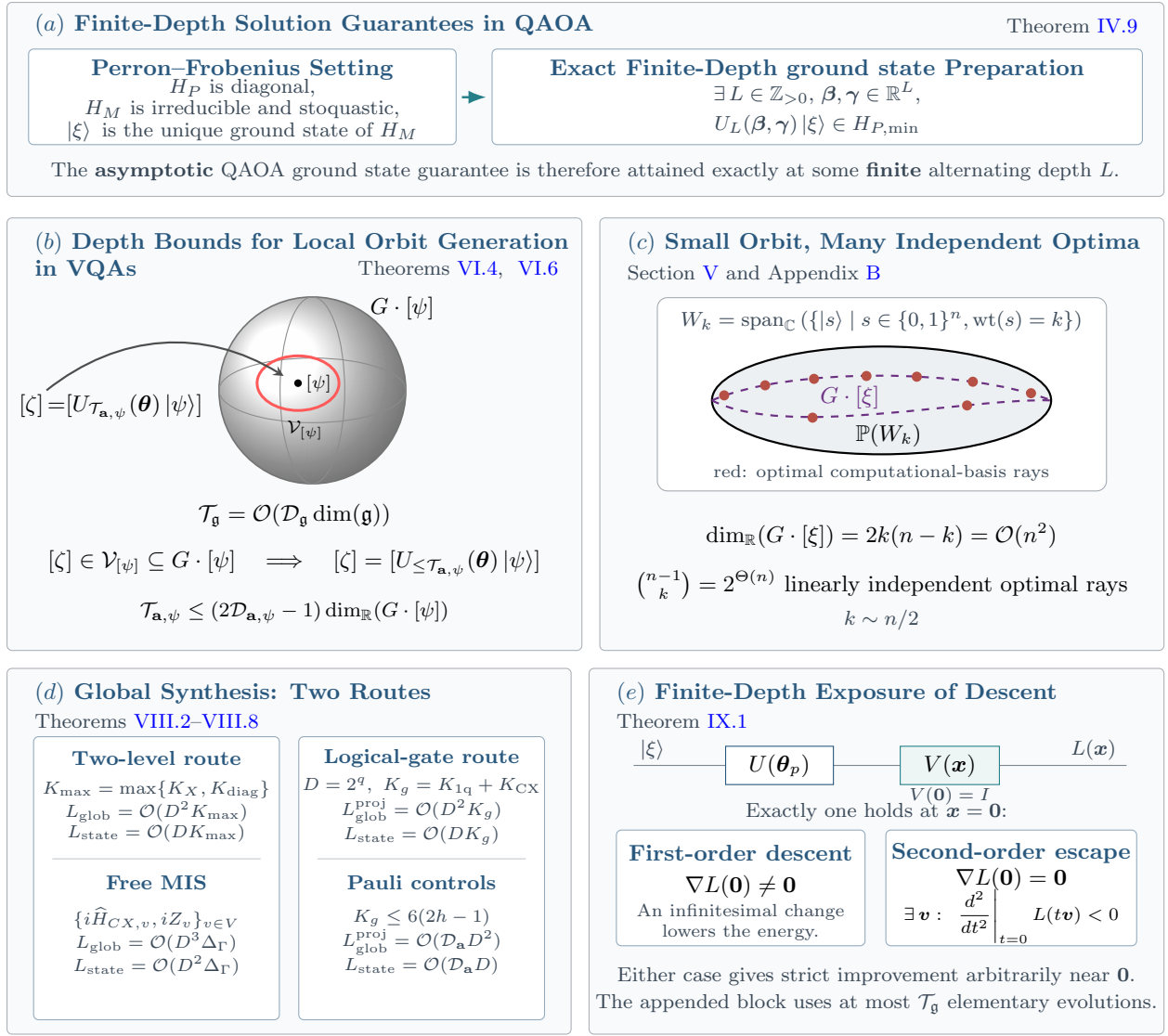
\begin{figure*}[t]
\centering
\begingroup
\definecolor{ovblue}{HTML}{28577D}
\definecolor{ovteal}{HTML}{247F88}
\definecolor{ovrust}{HTML}{B65045}
\definecolor{ovink}{HTML}{364554}
\definecolor{sage}{HTML}{6C3082}
\resizebox{\textwidth}{!}{%
\begin{tikzpicture}[x=1cm,y=1cm,>=Latex,
  frame/.style={draw=ovink!42,rounded corners=2pt,fill=ovblue!3},
  ribbon/.style={draw=ovblue!50,rounded corners=2pt,fill=white},
  title/.style={anchor=west,font=\bfseries\small,text=ovblue},
  detail/.style={font=\footnotesize,text=ovink,align=center}]

\draw[frame] (0,0) rectangle (16.1,-2.70);

\node[title] at (.27,-.32)
  {$(a)$ Finite-Depth Solution Guarantees in QAOA};

\node[
  font=\footnotesize,
  text=ovink,
  anchor=east
] at (15.85,-.32)
  {Theorem~\ref{thm:PF_orbit_closure}};

\draw[ribbon] (.30,-.65) rectangle (6.25,-1.98);

\node[
  font=\small\bfseries,
  text=ovblue
] at (3.275,-.94)
  {Perron--Frobenius Setting};

\node[
  font=\footnotesize,
  text=ovink,
  align=center,
  text width=5.55cm,
  inner sep=1pt
] at (3.275,-1.19)
  {$\displaystyle
      H_P \text{ is diagonal},
  $};

\node[
  font=\footnotesize,
  text=ovink,
  align=center,
  text width=5.55cm,
  inner sep=1pt
] at (3.275,-1.49)
  {$\displaystyle
      H_M \text{ is irreducible and stoquastic,}
  $};

\node[
  font=\footnotesize,
  text=ovink,
  align=center,
  text width=5.55cm,
  inner sep=1pt
] at (3.275,-1.79)
  {$\displaystyle
      \ket{\xi}\text{ is the unique ground state of }H_M
  $};

\draw[
  ->,
  ovteal,
  line width=.9pt
] (6.33,-1.32)--(6.67,-1.32);

\draw[ribbon] (6.75,-.65) rectangle (15.80,-1.98);

\node[
  font=\small\bfseries,
  text=ovblue
] at (11.275,-.94)
  {Exact Finite-Depth ground state Preparation};

\node[
  font=\footnotesize,
  text=ovink,
  align=center,
  text width=8.55cm,
  inner sep=1pt
] at (11.275,-1.49)
  {$\displaystyle
    \begin{gathered}
      \exists\,L\in\mathbb Z_{>0},\,
      \boldsymbol{\beta},\boldsymbol{\gamma}\in\mathbb R^L,\\[-1pt]
      U_L(\boldsymbol{\beta},\boldsymbol{\gamma})\ket{\xi}
      \in H_{P,\min}
    \end{gathered}
  $};

\node[
  detail,
  text width=15.35cm
] at (8.05,-2.35)
  {The \textbf{asymptotic} QAOA ground state guarantee is therefore attained
   exactly at some \textbf{finite} alternating depth \(L\).};

\draw[frame] (0,-3.00) rectangle (8,-9.02);
\node[title] at (.27,-3.35)
 {$(b)$ Depth Bounds for Local Orbit Generation};
 \node[title] at (.27,-3.73)
 {in VQAs};
\node[font=\footnotesize,text=ovink,anchor=west]
 at (4.77,-3.72)
 {Theorems~\ref{thm:local_synthesis_depth_bound},
  ~\ref{thm:local_state_reachability}};

\coordinate (depth) at (3.25,-1.90);
\begin{scope}[shift={([xshift=8mm,yshift=-35mm]depth.center)}]

    \shade[
        ball color=white!40,
        opacity=0.90
    ] (0.2,0) circle (13mm);

    \draw[black!55, line width=0.6pt]
        (0.2,0) circle (13mm);

    \draw[black!45]
        (0.2,0) ellipse [x radius=13mm, y radius=4.5mm];

    \draw[black!45]
        (0.2,0) ellipse [x radius=4.5mm, y radius=13mm];

    \node[
        font=\small,
        text=black,
        anchor=south west
    ] at (8.9mm,8.7mm) {$G\cdot[\psi]$};

    \node[
        font=\small,
        text=black,
        anchor=south west
    ] at (-40mm,-5.2mm)
      {$[\zeta]\operatorname{=}[U_{\mathcal T_{\mathbf a,\psi}}(\boldsymbol\theta)\ket{\psi}]$};

    \coordinate (identity) at (0mm,1mm);

    \fill[black]
        (identity) circle (1.5pt);

    \node[
        font=\scriptsize,
        anchor=south west
    ] at ([xshift=-0.05mm,yshift=-2.5mm]identity) {$[\psi]$};

    \draw[
        red!65,
        line width=1.1pt
    ]
        (identity)
        ellipse [x radius=5.7mm, y radius=3.8mm];

    \node[
        font=\scriptsize,
        text=black,
        anchor=north
    ] at (1mm,-3.1mm) {$\mathcal V_{[\psi]}$};

    \draw[->, >=stealth, line width=0.8pt,
          black!70, bend left=35]
        (-35mm, 0mm) to (-2mm,2mm);
\end{scope}

\node[font=\small,align=center] at (4,-7.14)
 {$\mathcal T_{\mathfrak g}
     =\mathcal O(\mathcal D_{\mathfrak g}\dim(\mathfrak g))$};
\node[font=\small,align=center] at (4,-7.80)
 {$[\zeta]\in\mathcal V_{[\psi]}\subseteq G\cdot[\psi]
   \quad\Longrightarrow\quad
   [\zeta]=[U_{\le\mathcal T_{\mathbf a,\psi}}(\boldsymbol\theta)\ket{\psi}]$};
\node[font=\footnotesize,align=center] at (4,-8.49)
 {$\mathcal T_{\mathbf a,\psi}\le
   (2\mathcal D_{\mathbf a,\psi}-1)
   \dim_{\mathbb R}(G\cdot[\psi])$};

\draw[frame] (8.25,-3.00) rectangle (16.1,-9.02);
\node[title] at (8.52,-3.35)
 {$(c)$ Small Orbit, Many Independent Optima};
\node[font=\footnotesize,text=ovink,anchor=west]
 at (8.52,-3.79)
 {Section \ref{subsec:fixed_particle_XY_example} and Appendix~\ref{app:fixed_particle_XY}};
\draw[rounded corners=4pt,draw=ovink!45,fill=white]
 (9.05,-4.11) rectangle (15.30,-6.80);
\node[detail] at (12.17,-4.43)
 {$
   W_k=\text{span}_{\mathbb {C}}\left( \left\{\ket{s}\mid s \in \{0,1\}^n, \text{wt}(s) = k\right\}\right)$};
\draw[fill=ovblue!9,thick]
 (12.17,-5.54) ellipse (2.36 and .75);
\draw[sage,dashed,thick] (9.81,-5.54) .. controls (10.86,-5.08) and (13.36,-5.08) .. (14.53,-5.54);
\draw[sage,dashed,thick] (9.81,-5.54) .. controls (10.86,-6.08) and (13.36,-5.5) .. (14.53,-5.54);
\foreach \x/\y in {
 9.98/-5.47,10.55/-5.33,11.23/-5.24,
 11.96/-5.20,12.66/-5.21,13.44/-5.28,
 14.25/-5.44,13.36/-5.61,11.21/-5.78}
 \fill[ovrust] (\x,\y) circle (2pt);
\node[font=\small,align=center] at (12.27,-6.03)
 {$ \mathbb P(W_k)$};
\node[font=\small,align=center] at (11.73,-5.50)
 {$\textcolor{sage}{G\cdot[\xi]}$};
\node[font=\scriptsize,text=ovink] at (12.17,-6.55)
 {red: optimal computational-basis rays};
\node[font=\small,align=center] at (12.17,-7.42)
 {$ \dim_{\mathbb R}(G\cdot[\xi])
     =2k(n-k)=\mathcal O(n^2)$};
\node[font=\small,align=center] at (12.17,-8.10)
 {$\binom{n-1}{k}=2^{\Theta(n)}$
   linearly independent optimal rays};
\node[detail] at (12.17,-8.59) {$k\sim n/2$};

\begin{scope}[xshift=-8.25cm]
\draw[frame] (8.25,-9.27) rectangle (16.1,-14.36);
\node[title] at (8.52,-9.62)
 {$(d)$ Global Synthesis: Two Routes};
\node[font=\footnotesize,text=ovink,anchor=west] at (8.52,-10.00)
 {Theorems~\ref{thm:uniform_orbit_generation}--\ref{thm:MIS_connected_global_bound}};

\draw[ribbon] (8.62,-10.22) rectangle (12.04,-14.16);
\draw[ribbon] (12.31,-10.22) rectangle (15.73,-14.16);

\node[font=\footnotesize\bfseries,text=ovblue]
 at (10.33,-10.52) {Two-level route};
\node[font=\scriptsize,text=ovink,align=center]
 at (10.33,-11.25)
 {$K_{\max}=\max\{K_X,K_{\mathrm{diag}}\}$\\
  $L_{\mathrm{glob}}=\mathcal O(D^2K_{\max})$\\
  $L_{\mathrm{state}}=\mathcal O(DK_{\max})$};
\draw[ovink!30] (8.88,-11.92)--(11.78,-11.92);
\node[font=\footnotesize\bfseries,text=ovblue]
 at (10.33,-12.25) {Free MIS};
\node[font=\scriptsize,text=ovink,align=center]
 at (10.33,-13.07)
 {$\{i\widehat H_{CX,v},iZ_v\}_{v\in V}$\\
  $L_{\mathrm{glob}}=\mathcal O(D^3\Delta_\Gamma)$\\
  $L_{\mathrm{state}}=\mathcal O(D^2\Delta_\Gamma)$};

\node[font=\footnotesize\bfseries,text=ovblue]
 at (14.02,-10.52) {Logical-gate route};
\node[font=\scriptsize,text=ovink,align=center]
 at (14.02,-11.25)
 {$D=2^q,\ K_g=K_{\mathrm{1q}}+K_{\mathrm{CX}}$\\
  $L_{\mathrm{glob}}^{\mathrm{proj}}=\mathcal O(D^2K_g)$\\
  $L_{\mathrm{state}}=\mathcal O(DK_g)$};
\draw[ovink!30] (12.57,-11.92)--(15.47,-11.92);
\node[font=\footnotesize\bfseries,text=ovblue]
 at (14.02,-12.25) {Pauli controls};
\node[font=\scriptsize,text=ovink,align=center]
 at (14.02,-13.07)
 {$ K_g\le6(2h-1)$\\
  $L_{\mathrm{glob}}^{\mathrm{proj}}=\mathcal O(\mathcal D_{\mathbf a}D^2)$\\
  $L_{\mathrm{state}}=\mathcal O(\mathcal D_{\mathbf a}D)$};
\end{scope}

\begin{scope}[xshift=8.1cm]
\draw[frame] (0,-9.27) rectangle (8,-14.36);
\node[title] at (.27,-9.62)
 {$(e)$ Finite-Depth Exposure of Descent};
\node[font=\footnotesize,text=ovink,anchor=west]
 at (.27,-10.00)
 {Theorem~\ref{thm:finite_depth_exposure_descent}};

\draw[ovink!65] (.62,-10.62)--(7.48,-10.62);
\node[font=\footnotesize,text=ovink]
 at (.88,-10.40) {$|\xi\rangle$};
\node[draw=ovblue,fill=white,
      minimum width=1.50cm,minimum height=.45cm]
 at (2.65,-10.62) {$U(\boldsymbol\theta_p)$};
\node[draw=ovteal,fill=ovteal!8,
      minimum width=1.38cm,minimum height=.45cm]
 at (5.02,-10.62) {$V(\boldsymbol x)$};
\node[font=\scriptsize,text=ovink]
 at (5.02,-11.04) {$V(\boldsymbol0)=I$};
\node[font=\footnotesize,text=ovink]
 at (7.02,-10.40) {$L(\boldsymbol x)$};

\node[font=\footnotesize,text=ovink]
 at (4,-11.27)
 {Exactly one holds at $\boldsymbol x=\boldsymbol0$:};

\draw[ribbon] (.38,-11.52) rectangle (3.84,-13.13);
\node[font=\small\bfseries,text=ovblue]
 at (2.11,-11.84) {First-order descent};
\node[font=\small] at (2.11,-12.31)
 {$\nabla L(\boldsymbol0)\neq\boldsymbol0$};
\node[font=\scriptsize,text=ovink,align=center]
 at (2.11,-12.80)
 {An infinitesimal change\\ lowers the energy.};

\draw[ribbon] (4.13,-11.52) rectangle (7.62,-13.13);
\node[font=\small\bfseries,text=ovblue]
 at (5.875,-11.84) {Second-order escape};
\node[font=\small] at (5.875,-12.2)
 {$\nabla L(\boldsymbol0)=\boldsymbol0$};
\node[font=\scriptsize,align=center] at (5.875,-12.72)
 {$\displaystyle
   \exists\,\boldsymbol v:\ 
   \left.\frac{d^2}{dt^2}\right|_{t=0}
   L(t\boldsymbol v)<0$};

\node[font=\footnotesize,text=ovink,align=center]
 at (4,-13.55)
 {Either case gives strict improvement arbitrarily near
  $\boldsymbol0$.};
\node[font=\footnotesize,text=ovink,align=center]
 at (4,-13.94)
 {The appended block uses at most
  $\mathcal T_{\mathfrak g}$ elementary evolutions.};
\end{scope}

\end{tikzpicture}%
}
\endgroup

\caption{\textbf{Overview of main theoretical results.} 
\textbf{$(a)$~QAOA: Finite-Depth Solution Guarantees.} If the problem Hamiltonian $H_P$ is diagonal and the mixer $H_M$ is irreducible, stoquastic in the computational basis, QAOA asymptotic solution reachability promotes to exact finite-depth realization.
\textbf{$(b)$ Local Orbit Generation.} We establish parametrized exact synthesis bounds for all states in a local open neighborhood $\mathcal{V}_{[\psi]} \subseteq G \cdot [\psi]$ around the projectivized initial state $[\psi]$ within its dynamical Lie group orbit. The provided depth upper bounds scale proportionally with the product of the DLA dimension and dimension of the projective orbit of $[\psi]$.
\textbf{$(c)$~Orbit vs.\ Optima Bottleneck.} We present a family of instances exhibiting an $\mathcal{O}(n^2)$-dimensional orbit that encompasses $2^{\Theta(n)}$ ground state rays: a low-dimensional dynamical Lie group orbit can contain exponentially many distinct optimal computational basis states. \textbf{$(d)$~Global Synthesis Depth Bounds.}
For invariant sectors with $\mathfrak{g}\big|_V \simeq \mathfrak{u}(V)$ or $\mathfrak{su}(V)$, we derive complementary uniform depth bounds using diagonal/two-level synthesis and on qubit subsectors, one-qubit + CNOT gate compilation. This yields explicit depth bounds for ma-QAOA MIS and Pauli-generated DLAs across both unitary synthesis and state-to-state transfer.
\textbf{$(e)$~Finite-Depth Exposure.} Appending a bounded-depth, identity-initialized circuit block $V(\boldsymbol{x})$ at any suboptimal output exposes either a first-order descent direction or a strictly negative second-order direction, allowing escape from local non-optimal stationary points. 
}
\label{fig:unified_overview}
\end{figure*}

A summary of our main results is provided in Figure~\ref{fig:unified_overview}.

\textbf{Acknowledgements.}
I would like to thank Bojko Bakalov and Dimitar Grantcharov for their interest in this work and many insightful discussions. I am also grateful to Domenico D'Alessandro for introducing me to Ref.~\cite{dalessandro2009} and providing an overview of its central results, which played a fundamental role in establishing the circuit-depth estimates for local element generation presented in Section~\ref{sec:infinitesimal}. Finally, I thank Bao Bach for his help in generating the numerical results reported in Tables~\ref{table:SplittingStatisticsTable}, \ref{AsymGraphsSixNodesFree}, and~\ref{AsymGraphsSixNodesStandard}.

\textbf{AI Disclosure.}
During the preparation of this manuscript, the author used OpenAI ChatGPT (GPT-5.6 Sol and GPT-6 Astra) to assist in reviewing relevant literature, verifying the correctness, and refining the exposition. All AI-assisted material, including citations and bibliographic information, was reviewed and verified by the author, who assumes full responsibility for the manuscript.

\section{Reachability Constraints, Level I (Algebra): Cyclic Representation of the DLA Generated by the Initial State}

To analyze state space reachability in variational quantum ans\"atze,  we first identify the portion of Hilbert space accessible from a given initial state.  Such an ansatz generates unitaries of the form
\begin{equation}
    \label{eq:general_vqa_circuit}
    U_L(\boldsymbol{\theta}) = \prod_{j=1}^L e^{-i\theta_j A_j}, \quad \text{with } A_j \in \{H_1, \dots, H_k\},
\end{equation}
where $L \in \mathbb{Z}_{>0}$ denotes the total number of gate applications, $\boldsymbol{\theta} = (\theta_1, \dots, \theta_L) \in \mathbb{R}^L$ is the vector of real variational parameters, and $\{H_1, \dots, H_k\}$ is the set of Hermitian operators on the $n$-qubit Hilbert space $W \cong \mathbb{C}^{2^n}$ corresponding to the circuit's gate generators (or driving Hamiltonians). 

We begin by defining the \emph{dynamical Lie algebra} associated with the above ansatz as the real Lie algebra generated by the nested commutators of the skew-Hermitian operators $\{iH_1, \dots, iH_k\}$:
\begin{equation}
    \label{eq:stdDLAdefn}
    \mathfrak{g} := \langle iH_1, \dots, iH_k \rangle_{\mathrm{Lie}} \subseteq \mathfrak{u}(W).
\end{equation}
The algebra $\mathfrak{g}$ governs the complete set of unitary operations accessible to the ansatz and plays a central role in characterizing its control-theoretic properties.

\begin{rmk}
    It is important to emphasize that the correspondence between a parameterized quantum circuit architecture and its underlying dynamical Lie algebra is far from unique. Entirely distinct physical generator pools can give rise to the exact same DLA $\mathfrak{g}$. This occurs because the DLA is determined solely by the full algebraic closure under the Lie bracket; consequently, different sets of generating operators can share the same DLA.
\end{rmk}

The following representation-theoretic framework is essential for characterizing the state space limitations of the ansatz.

\begin{defn}
\label{defn:Cyclic_Subrepresentation}
Let $V$ be a representation space of a Lie algebra $\mathfrak{g}$ and let $\ket{\xi} \in V$ be a vector. The \emph{cyclic subrepresentation} $V_{\mathfrak{g},\ket{\xi}}$ generated by $\ket{\xi}$ is the minimal complex $\mathfrak{g}$-invariant subspace of $V$ containing $\ket{\xi}$.
\end{defn}

\begin{rmk}
\label{rmk:krylov-comparison}
The cyclic subrepresentations considered above share conceptual roots with \emph{Krylov subspaces}, which are ubiquitous in control theory and numerical linear algebra (see, e.g.,~\cite{boley1994krylov}). While a standard Krylov subspace is generated by the successive powers of a single operator $A$ acting on a reference state $\ket{\xi}$ (i.e., $\operatorname{span}\{A^k \ket{\xi}\}_{k \ge 0}$, corresponding to the polynomial algebra $\mathbb{C}[A]$), our cyclic subrepresentation arises from the joint action of a multi-generator Lie algebra $\mathfrak{g}$ (or equivalently, its universal enveloping algebra) on the state space.
\end{rmk}

\begin{rmk}
\label{rmk:cyclic_space_containment}
The physical motivation for restricting our analysis to these invariant subspaces stems from the structural constraints of parameterized quantum circuits. As the algorithm is initialized in a reference state $\ket{\xi}$, the DLA generators preserve the corresponding cyclic subrepresentation of $\mathfrak{g}$ by definition. Furthermore, since the circuit unitaries are analytic expressions in the generators of $\mathfrak{g}$ via the group exponential map, the entire state trajectory remains strictly confined within this invariant subspace throughout the unitary evolution prior to final measurement.
\end{rmk}

In other words, this subspace delineates the dynamically accessible sector of the full Hilbert space. Computing its dimension establishes a fundamental upper bound on the effective variational degrees of freedom that govern the algorithm's expressiveness—a quantity that is frequently substantially (and in some cases, exponentially) smaller than the ambient dimension $2^n$.

\section{Reachability Constraints, Level II (Geometry): Dynamical Group Orbit of the Initial State}
\label{sec:reachability_2}

A common paradigm in assessing the expressibility of a parameterized quantum circuit ansatz relies on analyzing its associated dynamical Lie algebra $\mathfrak{g}$ (see \eqref{eq:stdDLAdefn})
and examining the corresponding cyclic subspace $W_{\mathfrak{g},\xi}$ (see Definition~\ref{defn:Cyclic_Subrepresentation}). 
However, the true set of reachable states is in general confined to a lower-dimensional manifold.

Specifically, let $G$ denote the connected analytic Lie subgroup of $\mathrm{U}(W)$ generated by the control Hamiltonians,
\begin{equation}
\label{eq:generated_analytic_subgroup}
    G\coloneqq\langle e^{tA}\mid A\in\{iH_1,\ldots,iH_k\},\,t\in\mathbb R\rangle \subseteq \mathrm{U}(W).
\end{equation}
By construction, every parameterized VQA unitary $U_L(\boldsymbol{\theta})$ (see~\eqref{eq:general_vqa_circuit}) is an element of $G$. Thus, the true reachable set of the parameterized quantum circuits initialized in a unit norm state $\ket{\xi}$ is given precisely by the dynamical Lie group orbit of the initial state:
\begin{equation}
\label{eq:G-orbit}
    \mathcal{O}_\xi \coloneqq G\cdot\ket{\xi} = \{ U\ket{\xi} \mid U \in G \} \subseteq \mathbb{S}(W_{\mathfrak{g},\xi}),
\end{equation}
where $\mathbb{S}(W_{\mathfrak{g},\xi})\coloneqq \{ \ket{\psi} \in W_{\mathfrak{g},\xi} \mid \|\ket{\psi}\| = 1 \}$ denotes the unit sphere of the cyclic subspace $W_{\mathfrak{g},\xi}$. The  containment $\mathcal{O}_\xi \subseteq \mathbb{S}(W_{\mathfrak{g},\xi})$ is detailed in Remark~\ref{rmk:cyclic_space_containment}.

Next, we establish precise criteria for when the group orbit achieves \emph{maximal reachability} within the cyclic subspace (modulo global phase)—that is, when the projective orbit of the ray $[\xi]$ generated by $\ket{\xi}$ completely saturates the projective space:
\begin{equation}
\label{eq:maximal_orbit}
    \mathcal{O}_{[\xi]} \coloneqq G\cdot[\xi] = \mathbb{P}(W_{\mathfrak{g},\xi}),
\end{equation}
where $\mathbb{P}(W_{\mathfrak{g},\xi}) \coloneqq \mathbb{S}(W_{\mathfrak{g},\xi})/\mathrm{U}(1)$ denotes the space of physical quantum states modulo global phase.

\begin{rmk}
Although the classification of Lie subalgebras yielding transitive actions on complex projective spaces is classical, we invoke it here to isolate the precise algebraic criteria ensuring projective transitivity within our framework. The detailed proof is deferred to Appendix~\ref{sec:proofs_reachability_2_and_performance_guarantees}.
\end{rmk}

\begin{thm}
\label{thm:projective_transitivity_classification}
Let $V$ be a complex Hilbert space of dimension $d \ge 2$, and let $\mathfrak{g} \subseteq \mathfrak{u}(V)$ be a Lie subalgebra of skew-Hermitian operators. Let $G = e^{\mathfrak{g}} \subseteq \mathrm{U}(V)$ be the connected Lie subgroup generated by $\mathfrak{g}$, and fix a normalized state $\ket{\xi} \in V$.

The projective orbit $\mathcal{O}_{[\xi]}$ equals the full projective space $\mathbb{P}(V)$ if and only if, up to a unitary change of basis on $V$, $\mathfrak{g}$ is one of the following Lie subalgebras of $\mathfrak{u}(d)$:
\begin{enumerate}
    \item \emph{full unitary or special unitary algebra:} $\mathfrak{su}(d)$ or $\mathfrak{u}(d)$ for any $d \ge 2$.
    \item \emph{symplectic algebra (or its $\mathfrak{u}(1)$-central extension):} $\mathfrak{sp}(m)$ or $\mathfrak{sp}(m) \oplus \mathfrak{u}(1)$ when $d = 2m \ge 4$.
\end{enumerate}
For $d = 2$, $\mathfrak{sp}(1) \cong \mathfrak{su}(2)$, which coincides with case~(1).
\end{thm}





\section{Performance Guarantees and  Limitations}
\label{sec:performance_guarantees}

A fundamental geometric property of an orbit $\mathcal{O}_\xi$ (resp. $\mathcal{O}_{[\xi]}$) is its topological closedness—namely, whether the orbit coincides with its closure $\overline{\mathcal{O}_\xi}$ in the ambient unit sphere $\mathbb{S}(W_{\mathfrak{g},\xi})$ (resp. $\mathcal{O}_{[\xi]}$ coincides with its closure in the ambient projective space $ \mathbb{P}(W_{\mathfrak{g},\xi})$). Closedness is the exact orbit property that bridges the gap between asymptotic expressibility and exact reachability: it allows us to elevate convergence guarantees in the infinite depth limit ($L \to \infty$) to rigorous existence guarantees for finite depth parametrized circuits. 

Below, we outline several practical and efficiently verifiable criteria ensuring the closedness of these dynamical group orbits. Complete proofs are deferred to Appendix~\ref{sec:proofs_reachability_2_and_performance_guarantees}.

\begin{thm}
\label{thm:closed_dynamical_lie_group}
Let $H_1, \ldots, H_k$ be Hermitian operators on a finite-dimensional Hilbert space $W$. Let $\mathfrak{g} \subseteq \mathfrak{u}(W)$ denote the dynamical Lie algebra as in \eqref{eq:stdDLAdefn} and let $G \subseteq \mathrm{U}(W)$ be the corresponding connected analytic subgroup as in \eqref{eq:generated_analytic_subgroup}.

If each generator is periodic—that is, for every $j \in \{1, \ldots, k\}$ there exists $\tau_j \in \mathbb{R}_{>0}$ such that
\begin{equation}
\label{eq:generator_periodicity}
    e^{i \tau_j H_j} = I,
\end{equation}
then $G$ is a closed subgroup of $\mathrm{U}(W)$. 
Consequently, for any normalized state $\ket{\xi} \in W$, the orbit $\mathcal{O}_\xi$ is  closed in $\mathbb{S}(W)$.
\end{thm}

This  periodicity framework applies directly to standard quantum approximate optimization algorithm and VQA ans\"atze featuring integer or rationally commensurate Hamiltonian spectra.

\begin{ex}
\label{ex:X_Ising_closed_group}
Let $H_M$ and $H_P$ be the transverse-field driver and Ising target Hamiltonians on $n$ qubits, defined respectively by
\begin{equation*}
    H_M = \sum_{v=1}^n X_v  \text{ and }  H_P = \sum_{v=1}^n h_v Z_v + \sum_{\{u,v\}\in E} J_{uv} Z_u Z_v.
\end{equation*}

Suppose that the parameters are rationally commensurate, i.e. there exists $q \in \mathbb{R}_{>0}$ such that
\begin{equation*}
    h_v \in q\mathbb{Z} \quad \text{and} \quad J_{uv} \in q\mathbb{Z} \qquad \text{for all } v \in \{1,\ldots,n\} \text{ and } \{u,v\}\in E.
\end{equation*}

Then $H_M$ and $H_P$ are strictly periodic: $H_M$ has integer spectrum, yielding $e^{i 2\pi H_M} = I$ with period $\tau_M = 2\pi$, while the commuting diagonal terms of $H_P$ restrict its spectrum to $q\mathbb{Z}$, yielding $e^{i (2\pi/q) H_P} = I$ with period $\tau_P = 2\pi/q$.

By Theorem~\ref{thm:closed_dynamical_lie_group}, the connected analytic subgroup of $\mathrm{U}((\mathbb{C}^2)^{\otimes n})$ generated by $\{e^{it H_M}, e^{it H_P} \mid t \in \mathbb{R}\}$ is closed, and so is the orbit $\mathcal{O}_{\xi}\subseteq \mathbb{S}((\mathbb{C}^2)^{\otimes n})$ of any initial state $\xi$.
\end{ex}

\begin{rmk}
\label{rmk:projective_periodic_generators}
For state-preparation questions, the assumption in
\eqref{eq:generator_periodicity} can be weakened to projective
periodicity. Namely, suppose that, for every $j$, there exist
$\tau_j \in \mathbb{R}_{>0}$ and $\theta_j\in\mathbb R$ such that
\begin{equation}
\label{eq:projective_generator_periodicity}
    e^{i\tau_jH_j}=e^{i\theta_j}I.
\end{equation}
Applying Theorem~\ref{thm:closed_dynamical_lie_group} to the
images of the generators in the projective unitary group
$\mathrm{PU}(W) \cong \mathrm{U}(W)/\mathrm{U}(1)$ shows that the projective image of $G$ is
closed. 
Consequently, every projective $G$-orbit in
$\mathbb P(W)$ is closed.
\end{rmk}

\subsection{Orbit Closedness and Finite Depth Guarantees}

While expressiveness analyses often characterize the states reachable in the asymptotic limit of infinite depth, physical VQA implementations require exact state preparation using finite-depth architectures. The topological closedness of the dynamical Lie group orbit plays a pivotal role here, establishing a direct equivalence between infinite-depth asymptotic limits and exact finite-depth expressiveness.

\begin{figure}[htbp]
\centering
\begin{tikzpicture}[
    >=Stealth,
    font=\small,
    nodeStyle/.style={
        rectangle, 
        rounded corners=4pt, 
        draw=flowBoxBorder, 
        fill=flowBoxBg, 
        thick, 
        inner sep=6pt, 
        align=center, 
        minimum height=1.8cm, 
    },
    condStyle/.style={
        thick, 
        inner sep=6pt, 
        align=center
    }
]

    \node[nodeStyle] (step1) {
        \textbf{Asymptotic Reachability}\\[0.15cm]
        \vspace{0.10in}
        $[t]\in\overline{
   \bigcup\limits_{\substack{L \ge 0 \\ \mathbf{\theta} \in \mathbb{R}^L}}
   [U_L(\mathbf \theta)\xi]
}=\overline{G\cdot[\xi]}.$
    };

    \node[nodeStyle, right=1.45cm of step1] (step2) {
        \textbf{Exact Group Action}\\[0.15cm]
        $\exists g \in G: \, [t] = g\cdot[\xi]$
    };

    \node[nodeStyle, fill=exactGreen!10, draw=exactGreen, right=1.0cm of step2] (step3) {
        \textbf{Exact Finite-Depth Reachability}\\[0.15cm]
        $\ket{t} = e^{i\phi}\left(\prod\limits_{l=1}^{L_0} e^{-i \theta_l H_{j_l}}\right) \ket{\xi}, \quad L_0 < \infty$
    };

    \draw[->, thick] (step1.north) to[out=35, in=145] 
        node[midway, above=4pt, condStyle] (bridge) {
            \textbf{Theorem~\ref{thm:closed_dynamical_lie_group} / Remark~\ref{rmk:projective_periodic_generators}}\\
            Orbit Closedness ($\mathcal{O}_{[\xi]} = \overline{\mathcal{O}_{[\xi]}}$)
        } 
        (step2.north);

    \draw[->, ultra thick, exactGreen] (step2.east) -- (step3.west);

\end{tikzpicture}
\caption{Logical pipeline establishing how projective orbit closedness ($\mathcal{O}_{[\xi]} = \overline{\mathcal{O}_{[\xi]}}$) bridges asymptotic reachability and exact finite-depth circuit synthesis. When $G$ is closed, any projectivized target state $[t]$ in the topological closure $\overline{\mathcal{O}_{[\xi]}}$ lies within the orbit itself, enabling an exact finite-depth state preparation (up to a global phase).}
\label{fig:orbit_closure_summary}
\end{figure}
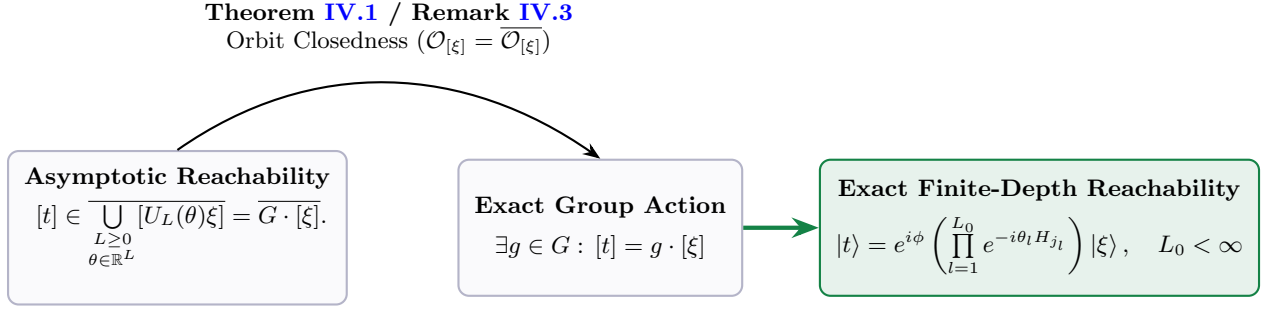

The primary practical significance of orbit closedness is illustrated in Figure~\ref{fig:orbit_closure_summary}. Specifically, any projectivized target state $[t]$ obtained as a limit of state sequences produced by parameterized quantum circuits of the form \eqref{eq:general_vqa_circuit} as the circuit depth $L \to \infty$ lies by definition in the topological closure $\overline{\mathcal{O}_{[\xi]}}$ of the dynamical Lie group orbit $\mathcal{O}_{[\xi]} = G\cdot [\xi]$. 

When the orbit $\mathcal{O}_{[\xi]}$ is closed—as guaranteed, for instance, when $G$ is compact under the periodic conditions of Theorem~\ref{thm:closed_dynamical_lie_group} (or Remark~\ref{rmk:projective_periodic_generators}), the orbit coincides with its closure ($\mathcal{O}_{[\xi]} = \overline{\mathcal{O}_{[\xi]}}$). Consequently, any state $\ket{t}$ reachable in the infinite-depth limit can in fact be represented \emph{exactly} as $\ket{t} = e^{i\phi}g\ket{\xi}$ for some $g \in G$ and $\phi \in [0,2\pi]$. In turn, because $G$ is the connected Lie group generated by the skew-Hermitian operators $\{iH_j\}_{j=1}^k$, every $g \in G$ admits an exact synthesis as a finite-depth parameterized circuit of the form \eqref{eq:general_vqa_circuit}. Thus, orbit closedness closes the gap between asymptotic and exact expressiveness, ensuring that no target state requires infinite depth to be realized.

We establish the following general reachability criteria.

\begin{thm}
\label{thm:closure_to_finite_word_reachability}
Let $H_1, \ldots, H_k$ be a collection of Hamiltonians on a finite-dimensional Hilbert space $W$, let $\mathfrak{g} \subseteq \mathfrak{u}(W)$ denote the corresponding dynamical Lie algebra (see \eqref{eq:stdDLAdefn}), and let $G \subseteq \mathrm{U}(W)$ be the corresponding connected analytic subgroup. 

Suppose that $\ket{s}, \ket{t} \in W$ are nonzero vectors satisfying
\begin{equation}
\label{eq:target_in_projective_orbit_closure}
    [t] \in \overline{\mathcal O_{[s]}}.
\end{equation}
Let
\[
    \pi \colon \mathrm{U}(W) \longrightarrow \mathrm{PU}(W)
\]
denote the canonical projective quotient map.

Assume, in addition, that at least one of the following conditions holds:
\begin{enumerate}
    \item  The projective subgroup \(\pi(G)\subseteq\mathrm{PU}(W)\) is closed. This holds, in particular, under the periodicity conditions of \eqref{eq:projective_generator_periodicity}.

    \item The vector $\ket{s}$ is an eigenvector of some $H_r$, $1 \le r \le k$, and its corresponding eigenspace is one-dimensional.
\end{enumerate}
Then there exist an integer $L \ge 0$, indices
\[
    j_1, \ldots, j_L \in \{1, \ldots, k\},
\]
and collection of parameters $\boldsymbol{\theta}=(\theta_1, \ldots, \theta_L) \in \mathbb{R}^L$ such that

\begin{equation}
\label{eq:exact_projective_finite_word}
    U_{L}(\boldsymbol{\theta}) \cdot [s] = [t]
\end{equation}
with $  U_{L}(\boldsymbol{\theta})$ as in \eqref{eq:general_vqa_circuit}.
\end{thm}

\begin{rmk}
\label{rmk:finite_word_to_layered_ansatz}
Every product ansatz of the form \eqref{eq:general_vqa_circuit} in Theorem~\ref{thm:closure_to_finite_word_reachability} can also be expressed using the fixed-order layered ansatz
\begin{equation}
\label{eq:k_generator_layered_ansatz}
    U_L(\boldsymbol{\theta}) = \prod_{\ell=1}^{L} \left( e^{-i\theta_{\ell,k} H_k} \cdots e^{-i\theta_{\ell,1} H_1} \right),
\end{equation}
where the product is time-ordered from right to left. Indeed, one may use a single nonzero parameter in each layer $\ell$ corresponding to generator $H_{j_\ell}$, setting all remaining parameters within that layer equal to zero.

For $k = 2$, with $H_1 = H_P$ and $H_2 = H_M$, equation \eqref{eq:k_generator_layered_ansatz} reduces to the standard QAOA circuit 
\begin{equation}
\label{eq:qaoa_chain}
U_p(\boldsymbol{\beta},\boldsymbol{\gamma}) = \prod_{k=1}^p e^{-i\beta_k H_M} e^{-i\gamma_k H_P} = e^{-i\beta_p H_M} e^{-i\gamma_p H_P} \cdots e^{-i\beta_1 H_M} e^{-i\gamma_1 H_P},
\end{equation}
where $\boldsymbol{\beta} = (\beta_1, \dots, \beta_p) \in \mathbb{R}^p$ and $\boldsymbol{\gamma} = (\gamma_1, \dots, \gamma_p) \in \mathbb{R}^p$ are variational parameters.
\end{rmk}

\begin{cor}
\label{cor:projectively_semisimple_exact_reachability}
Suppose that the DLA $\mathfrak{g}$ admits a  decomposition
\begin{equation}
\label{eq:projectively_semisimple_DLA}
    \mathfrak{g} = [\mathfrak{g}, \mathfrak{g}] \oplus \mathfrak{c}, 
    \qquad 
    \mathfrak{c} \subseteq i \mathbb{R} I_W.
\end{equation}
Then every projective $G$-orbit is closed. Consequently, for any non-zero states $\ket{s}, \ket{t} \in W$,
\[
    [t] \in \overline{G \cdot [s]}
\]
implies that $\ket{t}$ is exactly (up to a global phase) reachable from $\ket{s}$ by a finite parameterized circuit.

In particular, the conclusion holds whenever $\mathfrak{g}$ is semisimple.
\end{cor}

\begin{ex}
\label{ex:projectively_semisimple_MaxCut}
The preceding corollary applies to several classes of dynamical Lie algebras arising in quantum optimization.

First, consider the free (multi-angle) MaxCut QAOA ansatz defined on a connected graph $\Gamma = (V, E)$ with $n := |V| \ge 2$. Its dynamical Lie algebra is given by
\begin{equation}
\label{eq:free_MaxCut_DLA}
    \mathfrak{g}_{\Gamma,\mathrm{free}} 
    = 
    \left\langle 
        iX_v,\, iZ_u Z_v 
        \;\middle|\; 
        v \in V,\, (u,v) \in E 
    \right\rangle_{\mathrm{Lie}}.
\end{equation}

By the structural classification in \cite[Theorem~1]{KLFCCZ}, $\mathfrak{g}_{\Gamma,\mathrm{free}}$ belongs to one of six families, each of which is either a compact simple Lie algebra or a direct sum of compact simple Lie algebras. Consequently, $\mathfrak{g}_{\Gamma,\mathrm{free}}$ is semisimple, which implies that the corresponding dynamical Lie group is closed~\cite[Proposition~7.9]{Knapp}. Thus, Theorem~\ref{thm:closure_to_finite_word_reachability} implies that whenever a target ray belongs to the closure of the orbit of the initial one, it is exactly reachable by a finite multi-angle circuit.

This conclusion extends directly to general (not necessarily connected) graphs. If a graph has no isolated vertices, its free DLA decomposes into a direct sum of the DLAs associated with its connected components, thereby satisfying the hypotheses of Corollary~\ref{cor:projectively_semisimple_exact_reachability} with $\mathfrak{c}=0$.

More generally, let $\mathcal{A} = \{iP_1, \ldots, iP_m\}$ be a Lie-algebraically independent set of skew-Hermitian Pauli strings, and let $\mathcal{C}(\mathcal{A})$ be its anticommutation graph, whose vertices represent the generators $iP_j$ and whose edges join pairs $(iP_j, iP_\ell)$ that anticommute ($P_j P_\ell = -P_\ell P_j$). Suppose that $\mathcal{C}(\mathcal{A})$ is connected and contains at least two vertices.

According to the classification of Pauli-generated Lie algebras in \cite[Theorem~2]{AguilarEtAl2024}, the generated Lie algebra
\begin{equation}
\label{eq:pauli_generated_DLA}
    \mathfrak{g}_{\mathcal{A}} 
    := 
    \langle iP_1, \ldots, iP_m \rangle_{\mathrm{Lie}}
\end{equation}
is isomorphic to a direct sum of classical compact simple Lie algebras of type $\mathfrak{so}(d)$, $\mathfrak{sp}(d)$, or $\mathfrak{su}(d)$, where the specific algebra and rank are uniquely determined by the topology of $\mathcal{C}(\mathcal{A})$ and the algebraic relations among its generators. Thus, $\mathfrak{g}_{\mathcal{A}}$ is semisimple, every projective orbit generated by the corresponding circuit architecture is closed, and exact finite reachability holds due to Theorem~\ref{thm:closure_to_finite_word_reachability} .
\end{ex}

\begin{thm}
\label{thm:infimum_attainability}
Let $H_1, \ldots, H_k$ be a collection of Hamiltonians on a finite-dimensional Hilbert space $W$, let $\mathfrak{g} \subseteq \mathfrak{u}(W)$ denote the corresponding dynamical Lie algebra (see~\eqref{eq:stdDLAdefn}), and let $G \subseteq \mathrm{U}(W)$ be the corresponding connected Lie subgroup. Suppose that the target observable $H_P$ satisfies 
\[
i H_P \in \mathfrak{g}.
\]
Then, for any initial state $\ket{s}$ with density matrix $\rho_s := \ket{s}\!\bra{s}$, the infimum of the expectation value of $H_P$ over the reachable state manifold,
\[
\inf_{g \in G} \operatorname{Tr}\left( H_P g \rho_s g^\dagger \right)
\]
is attained by a finite parametrized circuit of the form given in~\eqref{eq:general_vqa_circuit} in the original controls. In particular, if the infimum coincides with the global ground state energy of $H_P$, an exact ground state can be reached by a finite-depth circuit within the ansatz.
\end{thm}

\subsection{Finite-Depth Ground Space Reachability Guarantees in QAOA}

While general state controllability concerns the reachability of arbitrary pure states, in many practical applications, particularly classical combinatorial optimization, our primary interest lies in the reachability of the optimal solution subspace rather than a specific target state.

Let $W = (\mathbb{C}^2)^{\otimes n}$ denote the $n$-qubit Hilbert space equipped with the computational basis $\{\ket{s}\}_{s \in \{0,1\}^n}$. Let $H_P$ be the problem Hamiltonian acting on $W$, which encodes a classical objective function 
\[
F \colon \{0,1\}^n \to \mathbb{R}
\]
via $H_P \ket{s} = F(s)\ket{s}$. Let $\lambda_{\min} := \min\limits_{s \in \{0,1\}^n} F(s)$ denote the ground state energy of $H_P$, and let $H_{P,\min} \subseteq W$ denote its corresponding ground space:
\begin{equation}
    \label{eq:HP_min_definition}
    H_{P,\min} := \operatorname{span}\left\{ \ket{s}\mid s\in \{0,1\}^n, F(s) = \lambda_{\min}\right\}.
\end{equation}

Suppose that the mixer Hamiltonian $H_M$ in QAOA is \emph{stoquastic} with respect to the computational basis, meaning its off-diagonal elements satisfy $\bra{s} H_M \ket{t} \le 0$ for all distinct $s, t \in \{0,1\}^n$, and \textit{irreducible}, in the sense that it has no stable proper coordinate subspaces.

By the Perron-Frobenius theorem, the ground state space of $H_M$ is one-dimensional. Let $\ket{\xi}$ denote a unit vector spanning this space. 


Given $H_M$ as above and $H_P$ diagonal in the computational basis, it was shown in~\cite{BKZS} that the QAOA ansatz, initialized at $\ket{\xi}$, asymptotically prepares a state within the minimum-energy sector $H_{P,\min}$ as the circuit depth $p \to \infty$.

Since every parametrized quantum circuit representing the QAOA ansatz (see~\eqref{eq:qaoa_chain}) is structurally confined to the dynamical Lie group $G$, satisfying the Perron--Frobenius conditions imposes the non-trivial overlap constraint on the initial state orbit closure:
\begin{equation}
\label{eq:nontriv_overlap}
    \overline{G \cdot [\xi]} \cap \mathbb{P}(H_{P,\min}) \neq \varnothing.
\end{equation}

The following result is a particular instance of Theorem~\ref{thm:closure_to_finite_word_reachability} (by setting $H_r=H_M$). It demonstrates that under mild conditions on the mixer, the orbit of the initial state is closed, thereby upgrading the standard infinite-depth asymptotic convergence of QAOA to an exact, finite-depth reachability guarantee.

\begin{thm}
\label{thm:PF_orbit_closure}
Let $W$ be a finite-dimensional Hilbert space equipped with a computational basis. Let $H_P, H_M \in \operatorname{Herm}(W)$ be Hermitian operators such that $H_P$ is diagonal in the computational basis, and $H_M$ is irreducible and stoquastic in the same basis. Let $\ket{\xi}$ be the unique normalized ground state of $H_M$ .

Then, there exist a finite circuit depth $L \in \mathbb{Z}_{>0}$, parameter vectors $\boldsymbol{\beta}, \boldsymbol{\gamma} \in \mathbb{R}^L$, a normalized ground state $\ket{\zeta} \in H_{P,\min}$, and a global phase $\phi \in \mathbb{R}$ such that
\begin{equation}
\label{eq:exact_finite_depth_prep}
U_L(\boldsymbol{\beta},\boldsymbol{\gamma})\ket{\xi} = \ket{\zeta},
\end{equation}
where $U_L(\boldsymbol{\beta},\boldsymbol{\gamma})$ is defined as in \eqref{eq:qaoa_chain}.

Consequently, the infinite-depth asymptotic convergence guarantee of QAOA is promoted to an exact finite-depth reachability guarantee.
\end{thm}

\begin{rmk}
\label{rmk:qualitative_reachability_guarantee}
While Theorem~\ref{thm:PF_orbit_closure} establishes a qualitative existence guarantee, it is non-constructive in nature, as it neither provides explicit target-specific depth bounds nor yields concrete parameter optimization procedures.
\end{rmk}

\section{Geometry Vs Algebra: Small Orbits with Exponentially Many Linearly Independent Optima}
\label{subsec:fixed_particle_XY_example}

A fundamental geometric feature of the dynamical Lie group orbit $\mathcal{O}_{[\xi]}\subseteq \mathbb{P}(W)$ generated by the ray of an initial state $\ket{\xi}$ is that it forms a smooth, connected manifold rather than a linear subspace. Consequently, a low-dimensional orbit can \textit{curve} through projective Hilbert space in such a way that the states inside it span a high-dimensional vector subspace.

In this section, we present an explicit example of this phenomenon, arising from a fixed-particle $XY$ ansatz, where the orbit of quadratic dimension $\mathcal{O}(n^2)$ in the number of qubits $n$ contains exponentially many linearly independent ground states of the problem Hamiltonian $H_P$. Although the underlying classical optimization task in this construction is intentionally simple, its primary purpose is to vividly demonstrate that low orbit dimensionality poses no intrinsic barrier to capturing exponentially many solutions or achieving exact ground space reachability in variational quantum algorithms.

Let $n \ge 2$, $1 \le k \le n-1$, and consider the fixed-excitation (fixed-Hamming-weight) subspace
\begin{equation}
\label{eq:fixed_weight_subspace}
    W_k \coloneqq \operatorname{span}\left\{ \ket{S} \;\middle|\; S \subseteq \{1,\ldots,n\},\, |S| = k \right\}.
\end{equation}
Here, $S$ records the occupied qubit sites, and $N_j \coloneqq \frac{I - Z_j}{2} = \ket{1}\!\bra{1}_j$ is the occupation-number operator at site $j$. We consider the nearest-neighbor $XY$ path mixer $H_M$ paired with an endpoint problem Hamiltonian $H_P$:
\begin{equation}
\label{eq:fixed_weight_XY_pair}
    H_M \coloneqq -\frac{1}{2} \sum_{j=1}^{n-1} \left( X_j X_{j+1} + Y_j Y_{j+1} \right), \qquad H_P \coloneqq N_1.
\end{equation}

\begin{rmk}
    \label{rmk:physical_interpretation}
Physically, $H_M$ corresponds to the isotropic limit ($\gamma = 0$) of the 1D $XY$ spin chain with free boundary conditions introduced by Lieb, Schultz, and Mattis \cite{LiebSchultzMattis1961}.  The problem Hamiltonian $H_P = N_1$ acts as a localized boundary potential (or single-site occupation pin). Particle-number-preserving $XY$ mixers of this form serve as a standard mechanism for hard-constrained optimization in the Quantum Alternating Operator Ansatz \cite{HWORVB}.
\end{rmk}

The mixer $H_M$ in \eqref{eq:fixed_weight_XY_pair} generates nearest-neighbor excitation-hopping transitions, thereby commuting with the total particle-number operator $\sum\limits_{j=1}^n N_j$ and preserving each fixed-particle subspace $W_k$ defined in \eqref{eq:fixed_weight_subspace}. When restricted to $W_k$, the operator $H_M\big|_{W_k}$ is irreducible with entrywise non-positive off-diagonal matrix elements in the computational basis. Consequently, by the Perron--Frobenius theorem, $H_M|_{W_k}$ possesses a unique ground state $\ket{\xi_k}$, whose computational basis expansion coefficients are strictly positive. Furthermore, the optimal configurations for the problem Hamiltonian $H_P = N_1$ within $W_k$ are precisely those computational basis states $\ket{S}$ for which site $1$ is unoccupied ($1 \notin S$):

\begin{equation}
\label{eq:fixed_weight_problem_ground_space}
    W_{P,\min}
    =
    \ker\!\left(N_1\big|_{W_k}\right)
    =
    \operatorname{span}
    \left\{
        \ket{S}
        \;\middle|\;
        S \subseteq \{1,\ldots,n\},\, |S|=k,\, 1\notin S
    \right\}
\end{equation}

and has real dimension

\begin{equation}
\label{eq:fixed_weight_problem_ground_dimension}
    \dim_{\mathbb R}(W_{P,\min})
    =2\binom{n-1}{k}.
\end{equation}

The following proposition establishes that the initial state $\ket{\xi_k}$ generates a low-dimensional orbit that nonetheless spans the entire subspace $W_k$ and contains exponentially many exact optima.

\begin{prop}
\label{prop:fixed_weight_reachable_optima}
Let $W_k \subset (\mathbb{C}^2)^{\otimes n}$ be the $k$-Hamming weight subspace defined in \eqref{eq:fixed_weight_subspace}, let
\[
    \mathfrak{g}_{n,k} = \left\langle i H_M\big|_{W_k},\, i H_P\big|_{W_k} \right\rangle_{\mathrm{Lie}}
\]
be the associated dynamical Lie algebra, and let $G_{n,k} \subseteq \mathrm{U}(W_k)$ be its connected dynamical Lie group. Then,  for an initial reference state $\ket{\xi_k}$ corresponding to a single Slater determinant formed by $k$ occupied single-particle orbitals, the DLA $\mathfrak{g}_{n,k}$, the projective orbit $\mathcal{O}_{[\xi_k]} = G_{n,k} \cdot [\xi_k]$, and the cyclic subrepresentation $W_{\mathfrak{g}_{n,k},\ket{\xi_k}}$ satisfy:
\begin{align}
\mathfrak{g}_{n,k} &\cong \mathfrak{u}(n), \qquad \dim_{\mathbb{R}}(\mathfrak{g}_{n,k}) = n^2,
\label{eq:fixed_weight_orbit_dim} \\
\dim_{\mathbb{R}}\mathcal{O}_{[\xi_k]} &= 2k(n-k),
\label{eq:fixed_Hamming_weight_subspace_dim} \\
W_{\mathfrak{g}_{n,k},\ket{\xi_k}} &= W_k, \qquad \dim_{\mathbb{R}}(W_{\mathfrak{g}_{n,k},\ket{\xi_k}}) = 2\binom{n}{k}.
\label{eq:fixed_weight_cyclic_space}
\end{align}
All $\binom{n-1}{k}$ optimal computational basis states are exactly reachable, up to a global phase, by finite alternating circuits as in \eqref{eq:qaoa_chain}.
\end{prop}

\begin{rmk}
    \label{rmk:half_filling_asymptotics}
    At half-filling $k = \lfloor (n-1)/2 \rfloor$, both the cyclic space dimension and the number of mutually orthogonal reachable optimal states grow exponentially as $n \to \infty$:
    \begin{equation}
    \label{eq:fixed_weight_exponential_span}
        \binom{n}{k} = \Theta\!\left(\frac{2^n}{\sqrt{n}}\right)
        \qquad \text{and} \qquad
        \binom{n-1}{k} = \Theta\!\left(\frac{2^n}{\sqrt{n}}\right),
    \end{equation}
    whereas the dynamical Lie algebra orbit dimension scales only quadratically as $\mathcal{O}(n^2)$.
\end{rmk}

\begin{figure}[htbp]
\centering
\resizebox{0.98\linewidth}{!}{%
\begin{tikzpicture}[
    font=\small,
    >={Stealth[length=2mm]},
    optimum/.style={
        circle,fill=orange!85!black,draw=white,
        line width=0.5pt,inner sep=2.3pt
    }
]
\filldraw[
    fill=black!2,draw=black!22,rounded corners=5pt
] (0,-0.45) rectangle (10,5.75);

\node[align=center] at (5,5.05) {
  \footnotesize  $\text{Projective cyclic space: }\mathbb P(W_k)$\\
    \footnotesize
    real dimension: $2\displaystyle\binom nk-2$
};

\path[fill=teal!13,draw=teal!70!black,line width=0.9pt]
    plot[domain=0:1,samples=65]
        ({0.7+8*\x},{1.2-0.6*sin(360*\x)})
    -- (9.3,3.1)
    -- plot[domain=1:0,samples=65]
        ({1.3+8*\x},{3.1+0.6*sin(360*\x)})
    -- cycle;

\foreach \v in {0.2,0.4,0.6,0.8}{
    \draw[teal!40,line width=0.35pt]
    plot[domain=0:1,samples=65]
      ({0.7+8*\x+0.6*\v},
       {1.2+1.9*\v+0.6*sin(360*\x)*(2*\v-1)});
}
\foreach \u in {0.1,0.2,...,0.9}{
    \draw[teal!40,line width=0.35pt]
    plot[domain=0:1,samples=2]
      ({0.7+8*\u+0.6*\x},
       {1.2+1.9*\x+0.6*sin(360*\u)*(2*\x-1)});
}

\foreach \u/\v in {
    0.08/0.30,0.13/0.78,0.22/0.18,0.29/0.86,
    0.35/0.40,0.42/0.70,0.49/0.22,0.56/0.82,
    0.63/0.42,0.70/0.72,0.77/0.20,0.83/0.60,
    0.91/0.34,0.94/0.82
}{
    \node[optimum] at
      ({0.7+8*\u+0.6*\v},
       {1.2+1.9*\v+0.6*sin(360*\u)*(2*\v-1)}) {};
}

\coordinate (initial) at
    ({0.7+8*0.24+0.6*0.57},
     {1.2+1.9*0.57+0.6*sin(360*0.24)*(2*0.57-1)});

\node[
    circle,fill=blue!70!black,draw=white,
    line width=0.6pt,inner sep=2.7pt
] at (initial) {};

\node[above left=2pt,inner sep=1.5pt]
    at (initial) {${\scriptstyle[\xi_k]}$};

\node[below=4pt] at
    ({0.7+8*0.08+0.6*0.30},
     {1.2+1.9*0.30+0.6*sin(360*0.08)*(2*0.30-1)})
    {${\scriptstyle[S_1]}$};

\node[above=4pt] at
    ({0.7+8*0.29+0.7*0.86-0.5},
     {1.2+1.9*0.86+0.5*sin(360*0.29)*(2*0.86-1)-0.25})
    {${\scriptstyle[S_2]}$};

\node[below=4pt] at
    ({0.7+8*0.91+0.6*0.34+0.4},
     {1.2+1.9*0.34+0.6*sin(360*0.91)*(2*0.34-1)+0.7})
    {${\scriptstyle [S_M]}$};

\node[align=center,text=teal!55!black]
    at (7.65,3.70) {
  \footnotesize     $\text{Orbit }\mathcal O_{[\xi_k]}$\\
       \footnotesize real dimension $2k(n-k)$
    };

\draw[->,teal!60!black]
    (7.7,3.32) -- (7.7,2.85);

\node[align=center] at (5,0.05) {
   \footnotesize \textcolor{orange!85!black}{Optimal space:}
    $\dim_{\mathbb R}(\mathbb P(W_{P,\min}))
    =2\binom{n-1}{k}-2$
};

\node[align=center] at (1.7,4.2) {
   \footnotesize \textcolor{blue!85!black}{Initial state:}
    $\xi_k$
};
\end{tikzpicture}%
}
\caption{
  Schematic illustration of the setup in (Subsection~\ref{subsec:fixed_particle_XY_example}). A low-dimensional orbit can host exponentially many linearly independent optimal states. The shaded surface represents the dynamical Lie group orbit $\mathcal{O}_{[\xi_k]}$ of the initial state $[\xi_k]$ inside the projectivized cyclic subspace, generated by the ray of $\ket{\xi_k}$ under the action of the dynamical Lie algebra $\mathfrak{g}$. Orange points represent optimal computational basis rays. While the orbit dimension scales quadratically in $n$, both the cyclic space dimension and the number of optimal states grow exponentially. \textit{Note:} the drawing is purely schematic; planar distances do not represent Hilbert space inner products or linear relations.%
}
\label{fig:small_orbit_many_optima}
\end{figure}
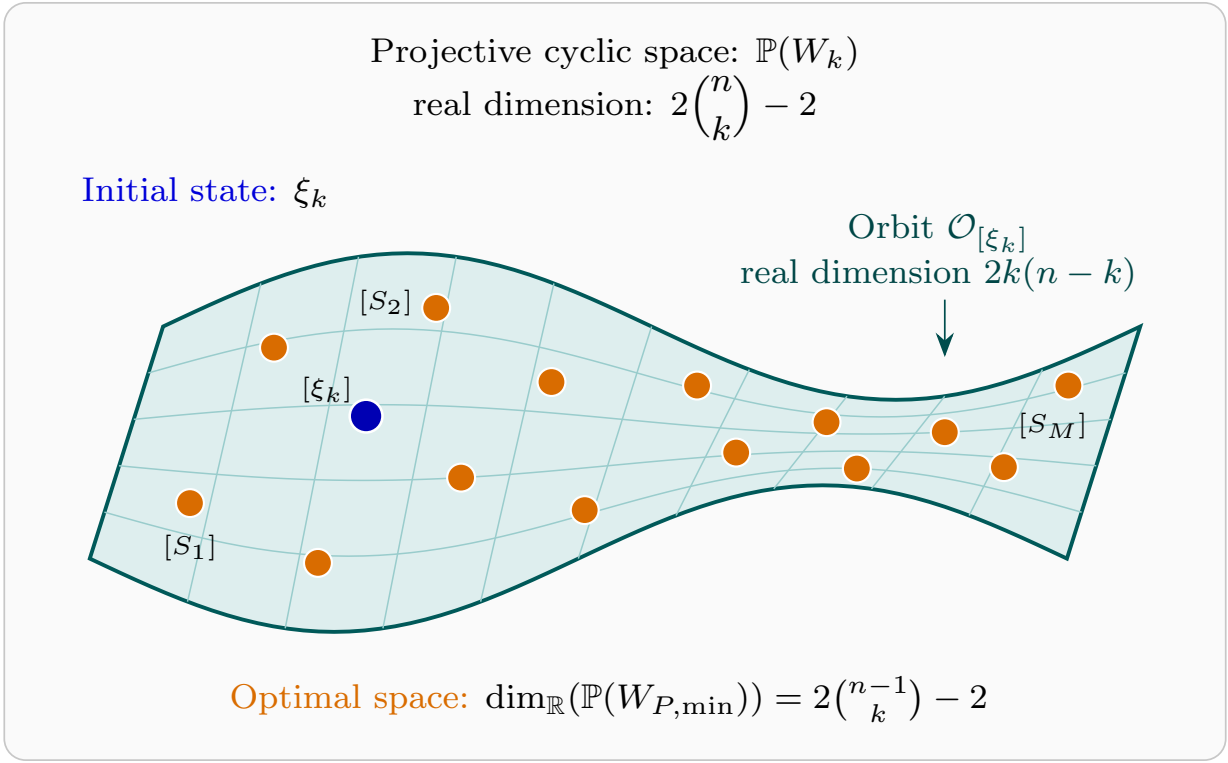

A complete, rigorous proof of Proposition~\ref{prop:fixed_weight_reachable_optima}, together with its representation-theoretic details and geometric interpretations, is provided in Appendix~\ref{app:fixed_particle_XY}; a schematic overview of this geometric phenomenon is illustrated in Figure~\ref{fig:small_orbit_many_optima}.

\section{Circuit Depth Bounds for Near-Identity Group Elements and Local Flows along the State Orbit Manifold}
\label{sec:infinitesimal}

In this section, we analyze the expressibility of parameterized variational quantum algorithm circuits of the form \eqref{eq:general_vqa_circuit} when synthesizing elements within a sufficiently small open neighborhood $U_I \subseteq G$ of the identity element $I$ in the dynamical Lie group $G$. Our primary objective is to derive an upper bound on the circuit depth $L$, defined as the minimum number of exponential factors in the DLA generators required to generate any element $g \in U_I$. We first establish a uniform bound for local element synthesis expressed in terms of the \textit{stabilization depth} of a natural filtration associated with the generators of a dynamical Lie algebra $\mathfrak{g}$ (see \eqref{eq:stdDLAdefn}). We then sharpen this bound for local state evolution along the orbit manifold $\mathcal{O}_{[\psi]}$ of a reference state $\ket{\psi}$. While points states in an open neighborhood of $[\psi]$ within the orbit $\mathcal{O}_{[\psi]}$ can be expressed as $g \cdot [\psi]$ for $g \in U_I$, elements of the isotropy subgroup 
\[
\operatorname{Stab}_G([\psi]) \coloneqq \{g \in G \mid g \cdot \ket{\psi} = e^{i\phi} \ket{\psi}\}
\]
leave the projective state invariant up to a global phase. By factoring out these redundant stabilizer directions, we obtain a tighter, state-dependent bound on the circuit depth required to generate local flows along the state orbit. We note that the bounds derived in this section are structural existence guarantees for short-depth local control; they do not explicitly yield the optimal parameter values needed to synthesize a target gate or state.

\subsection{Local Canonical Coordinates and Stabilization Depth}

A central tool for establishing the existence of the aforementioned local parametrization, we employ the product-of-exponentials map, defining \textit{canonical coordinates of the second kind} on the dynamical Lie group, a foundational tool in geometric control theory and variational quantum circuit design.

 Given an ordered basis $\{b_1, \dots, b_n\}$ of $\mathfrak{g}$ (where $n = \dim \mathfrak{g}$), this map is defined as
\begin{equation}
\label{eq:phi_map}
\begin{aligned}
\varphi_{\mathfrak{g}}: \mathfrak{g} &\rightarrow G, \\
\varphi_{\mathfrak{g}}(x) &:= e^{x_1 b_1} \cdots e^{x_n b_n},
\end{aligned}
\end{equation}
where $x = \sum\limits_{i=1}^n x_i b_i \in \mathfrak{g}$.

\begin{rmk}
\label{rmk:inverse_fun}
The map $\varphi_{\mathfrak{g}}$ provides a factorized local parametrization around the identity in terms of a product of one-parameter subgroups, in contrast to the standard exponential map $x \mapsto e^x$. By construction, $\varphi_{\mathfrak{g}}$ maps the origin of the Lie algebra to the group identity, i.e., $\varphi_{\mathfrak{g}}(0) = I$. 

As a finite product of smooth exponential maps, $\varphi_{\mathfrak{g}}$ is smooth. Furthermore, its differential at the origin is the identity transformation, $(d\varphi_{\mathfrak{g}})_0 = \mathrm{id}_{\mathfrak{g}}$. By the Inverse Function Theorem, there exist open neighborhoods $V_0 \subseteq \mathfrak{g}$ containing $0$ and $U_{I, \varphi_{\mathfrak{g}}} \subseteq G$ containing $I$ such that 
\begin{equation}
\label{eq:local_diffeo}
\varphi_{\mathfrak{g}} : V_0 \xrightarrow{\ \cong\ } U_{I, \varphi_{\mathfrak{g}}}
\end{equation}
is a smooth local diffeomorphism. Consequently, any element $g \in U_{I, \varphi_{\mathfrak{g}}}$ can be uniquely factorized into the product form \eqref{eq:phi_map}.
\end{rmk}

Following the framework of~\cite[Section~2]{dalessandro2009}, we represent elements of the open neighborhood $U_{I, \varphi_{\mathfrak{g}}} \subseteq G$ (see \eqref{eq:local_diffeo}) in the parameterized circuit form as in~\eqref{eq:general_vqa_circuit}. To analyze and bound the resulting circuit depth~$L$, we leverage an ascending vector space filtration  that stratifies $\mathfrak{g}$ according to the minimal bracket depth required to generate its elements from the set of original generators
\(\mathbf a=\{a_1,\ldots,a_k\}\) of $\mathfrak g$ (here we assume that the generators are linearly independent over $\mathbb{R}$).

More precisely, let $\mathfrak{g}$ be a finite-dimensional (real) Lie algebra presented by a set of generators. It is natural to organize the elements of $\mathfrak{g}$ according to the minimum bracket depth required to generate them. This induces an ascending vector space filtration on $\mathfrak{g}$:
\begin{equation}
\label{eq:LieAlgFiltration}
\{0\} = \mathcal{F}^{\mathbf{a}}_0 \subseteq \mathcal{F}^{\mathbf{a}}_1 \subseteq \mathcal{F}^{\mathbf{a}}_2 \subseteq \dots \subseteq \mathcal{F}^{\mathbf{a}}_j \subseteq \dots \subseteq \mathfrak{g},
\end{equation}
defined recursively as follows.
\begin{itemize}
    \item \textbf{Base layer:} the subspace spanned by the generators, $\mathcal{F}^{\mathbf{a}}_1 := \mathrm{span}_\mathbb{R}\{a_1, \dots, a_k\}$.
    \item \textbf{Higher layers:} $\mathcal{F}^{\mathbf{a}}_j := \mathcal{F}^{\mathbf{a}}_{j-1} + [\mathcal{F}^{\mathbf{a}}_1, \mathcal{F}^{\mathbf{a}}_{j-1}]$ for $j \ge 2$, which ensures the bracket compatibility condition
    \begin{equation}
    \label{eq:bracket_compatibility}   
    [\mathcal{F}^{\mathbf{a}}_{d_1}, \mathcal{F}^{\mathbf{a}}_{d_2}] \subseteq \mathcal{F}^{\mathbf{a}}_{d_1+d_2}.
    \end{equation}
\end{itemize}

As $\mathfrak{g}$ is finite-dimensional, this ascending filtration of subspaces must eventually stabilize at $\mathfrak{g}$. This allows us to formalize the notion of \textit{stabilization depth}.

\begin{defn}
    \label{def:stabilization_depth}
 Let $\mathbf{a} = \{a_1, \dots, a_k\}$ be a set of generators. The \emph{stabilization depth} of $\mathbf{a}$ is the minimal index at which the filtration sequence \eqref{eq:LieAlgFiltration} stabilizes:
    \begin{equation}
        \label{eq:stabilization_num}
        \mathcal{D}_\mathbf{a} := \min \{ j \in \mathbb{N} \mid \mathcal{F}^{\mathbf{a}}_j = \mathcal{F}^{\mathbf{a}}_{j+1} \}.
    \end{equation}
    Equivalently, \(\mathcal{D}_\mathbf{a}\) is the minimal integer \(j\) such that any Lie word of length greater than \(j\) lies within the linear span of Lie words of length at most \(j\).
\end{defn}

To express every element
\(g\in U_{I,\varphi_{\mathfrak g}}\subseteq G\)
(see~\eqref{eq:local_diffeo}) in the circuit
form~\eqref{eq:general_vqa_circuit}, we construct an ordered basis
\(\{b_1,\ldots,b_n\}\) of \(\mathfrak g\) from nested conjugates of the original generators
\(\mathbf a=\{a_1,\ldots,a_k\}\). We initialize
the construction by setting
\[
b_i=a_i,
\qquad 1\leq i\leq k.
\]

Fix \(\varepsilon>0\). For each \(j\geq1\), define the subset
\[
\mathcal C_j(\varepsilon)
:=
\left\{
\operatorname{Ad}_{e^{t_1a_{i_1}}\cdots e^{t_q a_{i_q}}}
(a_{i_0})
\;\middle|\;
\begin{array}{c}
0\leq q\leq j-1,\\
i_0,\ldots,i_q\in\{1,\ldots,k\},\\
|t_r|<\varepsilon,\quad 1\leq r\leq q
\end{array}
\right\}\subseteq \mathfrak g,
\]
where the product is understood as the identity when \(q=0\), and consider the subspace
\[
S_j(\varepsilon)
:=
\operatorname{span}_{\mathbb R}\mathcal C_j(\varepsilon)\subseteq\mathfrak g.
\]
Thus, \(\mathcal C_j(\varepsilon)\) contains conjugates evaluated at
all parameter values in the indicated intervals, rather than at a
single fixed choice of parameters. In particular,
\(
S_1(\varepsilon)=\mathcal F_1^{\mathbf a}.
\)

We first show that
\begin{equation}
\label{eq:filtration-contained-in-conjugate-span}
\mathcal F_j^{\mathbf a}\subseteq S_j(\varepsilon)
\qquad (j\geq1).
\end{equation}
Indeed, let \(c\in S_{j-1}(\varepsilon)\). Writing \(c\) as a finite
linear combination of elements of \(\mathcal C_{j-1}(\varepsilon)\)
and using the linearity of the adjoint action shows that
\[
\operatorname{Ad}_{e^{t a_i}}(c)\in S_j(\varepsilon)
\qquad\text{for every } |t|<\varepsilon.
\]
Since \(c\in S_{j-1}(\varepsilon)\subseteq S_j(\varepsilon)\), we
have, for every nonzero \(|t|<\varepsilon\),
\begin{equation}
    \label{eq:der_at_t=0}
\frac{\operatorname{Ad}_{e^{t a_i}}(c)-c}{t}
\in S_j(\varepsilon).
\end{equation}

The space \(S_j(\varepsilon)\) is a linear subspace of the finite-dimensional normed space \(\mathfrak g\), and is therefore closed. Taking the limit of \eqref{eq:der_at_t=0} as \(t\to0\) gives
\begin{equation}
\label{eq:commutator_belongs}
[a_i,c]
=
\left.\frac{d}{dt}\right|_{t=0}
\operatorname{Ad}_{e^{t a_i}}(c)
\in S_j(\varepsilon).
\end{equation}
The recursive definition of the filtration (see \eqref{eq:bracket_compatibility})
now proves~\eqref{eq:filtration-contained-in-conjugate-span} by induction on \(j\).

We next select the basis elements. Suppose that, through stage \(j-1\), we have selected exactly
\(\dim(\mathcal F_{j-1}^{\mathbf a})\) linearly independent elements within
\(\mathcal C_{j-1}(\varepsilon)\), and let \(W_{j-1}\) denote their span. Since
\[
W_{j-1}\subseteq S_{j-1}(\varepsilon)\subseteq S_j(\varepsilon)
\]
and
\(\mathcal F_j^{\mathbf a}\subseteq S_j(\varepsilon)\), we obtain

\begin{equation}
\dim\bigl(S_j(\varepsilon)/W_{j-1}\bigr)=
\dim(S_j(\varepsilon))-\dim (W_{j-1}) \geq
\dim(\mathcal F_j^{\mathbf a})
-\dim(\mathcal F_{j-1}^{\mathbf a}).
\end{equation}

As \(\mathcal C_j(\varepsilon)\) spans \(S_j(\varepsilon)\), its
image spans the quotient \(S_j(\varepsilon)/W_{j-1}\). We may
therefore select
\[
c_j
:=
\dim(\mathcal F_j^{\mathbf a})
-\dim(\mathcal F_{j-1}^{\mathbf a})
\]
elements of \(\mathcal C_j(\varepsilon)\) whose cosets are linearly
independent in \(S_j(\varepsilon)/W_{j-1}\). Appending these elements
to the previously selected collection produces a linearly independent
set of cardinality \(\dim(\mathcal F_j^{\mathbf a})\).

Repeating the preceding selection through stage \(\mathcal D_{\mathbf a}\) (see Definition~\ref{def:stabilization_depth}) produces \(n=\dim(\mathfrak g)\) linearly independent elements and hence a basis
\(\{b_1,\ldots,b_n\}\) of \(\mathfrak g\).

\begin{rmk}
\label{rmk:conjugation_advantage}
The argument of Ref.~\cite[Section~2]{dalessandro2009} also gives a
useful criterion for making individual selections. Suppose that \(W\)
is the span of the elements already selected and that an available
conjugate
\[
c\in W\cap\mathcal C_{j-1}(\varepsilon)
\]
satisfies \([a_i,c]\notin W\). Let
\(\pi_W:\mathfrak g\to\mathfrak g/W\) denote the quotient map.
Since \(c\in W\), the expansion
\[
\operatorname{Ad}_{e^{t a_i}}(c)
=
c+t[a_i,c]+O(t^2)
\]
gives
\[
\pi_W\!\left(\operatorname{Ad}_{e^{t a_i}}(c)\right)
=
t\,\pi_W([a_i,c])+O(t^2).
\]
Equivalently,
\[
\frac{1}{t}
\pi_W\!\left(\operatorname{Ad}_{e^{t a_i}}(c)\right)
\longrightarrow
\pi_W([a_i,c])\neq0
\qquad\text{as }t\to0.
\]
It follows that there exists \(\delta>0\) such that
\begin{equation}
\label{eq:conjugated_basis_element}
b_m:=\operatorname{Ad}_{e^{t a_i}}(c)\notin W
\qquad
\text{whenever }
0<|t|<\min\{\delta,\varepsilon\}.
\end{equation}
Thus every sufficiently small nonzero conjugation parameter supplies
a new linearly independent direction. 

The same conjugation structure gives an exact synthesis of every
selected flow. Indeed, if \(b_m\) is selected at stage \(j\), then
\(b_m\in\mathcal C_j(\varepsilon)\), so it has the form
\[
b_m=\operatorname{Ad}_{P_m}(a_{i_0}),
\qquad
P_m=e^{t_1a_{i_1}}\cdots e^{t_q a_{i_q}},
\qquad
q\leq j-1.
\]
Therefore, for every \(s\in\mathbb R\),
\[
e^{s b_m}=e^{sP_ma_{i_0}P_m^{-1}}=
P_m e^{s a_{i_0}}P_m^{-1} =
e^{t_1a_{i_1}}\cdots e^{t_q a_{i_q}}
e^{s a_{i_0}}
e^{-t_q a_{i_q}}\cdots e^{-t_1a_{i_1}}.
\]
In particular, if \(b_m=\operatorname{Ad}_{e^{t a_i}}(c)\), then
\[
e^{s b_m}
=
e^{t a_i}e^{s c}e^{-t a_i}.
\]
Although \(c\) need not be a selected basis element, its membership in
\(\mathcal C_{j-1}(\varepsilon)\) supplies an exact native
factorization of \(e^{s c}\). Hence \(e^{s b_m}\) is synthesized using
at most
\[
2q+1\leq2(j-1)+1\leq2j-1
\]
native generator flows, without Trotterization or
Baker--Campbell--Hausdorff approximations.
\end{rmk}

If \(b_m\) is selected at filtration stage \(j\), then
\(b_m\in\mathcal C_j(\varepsilon)\). Hence, as shown in
Remark~\ref{rmk:conjugation_advantage}, the corresponding flow
\(e^{s b_m}\) admits an exact factorization into at most \(2j-1\)
native generator flows. Since precisely
\[
c_j
:=
\dim(\mathcal F_j^{\mathbf a})
-
\dim(\mathcal F_{j-1}^{\mathbf a})
\]
basis elements are selected at stage \(j\), concatenating these
factorizations in the canonical-coordinate representation
\(\varphi_{\mathfrak g}\) shows that every
\(g\in U_{I,\varphi_{\mathfrak g}}\) can be represented using at most
\begin{equation}
\label{eq:total_exponentials}
\mathcal T_{\mathbf a}
:=
\sum_{j=1}^{\mathcal D_{\mathbf a}}(2j-1)c_j
\end{equation}
native exponential factors. Thus \(\mathcal T_{\mathbf a}\) is a
uniform upper bound furnished by the construction; cancellations or
shorter conjugating words may reduce the number of factors required
for a particular group element.

For any non-abelian Lie algebra $\mathfrak{g}$, this operation count satisfies the elementary bounds
\begin{equation}
    \label{eq:loose_bounds}
    \dim(\mathfrak{g}) \le \mathcal{T}_\mathbf{a} < \dim^2(\mathfrak{g}).
\end{equation}

Since $j \le \mathcal{D}_{\mathbf{a}}$ and $\sum\limits_j c_j = \dim(\mathfrak{g})$, \eqref{eq:total_exponentials} yields the refined upper bound
\begin{equation}
    \label{eq:worst_case_derivation}
    \mathcal{T}_{\mathbf{a}} \le (2\mathcal{D}_{\mathbf{a}} - 1)\dim\mathfrak{g}.
\end{equation}

These observations establish a stabilization depth dependent bound on local circuit synthesis complexity, summarized in the following theorem.

\begin{thm}
\label{thm:local_synthesis_depth_bound}
Let $\mathfrak{g} \subseteq \mathfrak{u}(W)$ be a dynamical Lie algebra generated by $\{a_1, \dots, a_k\}$ with stabilization depth $\mathcal{D}_{\mathbf{a}}$. Then every element $g$ in the identity neighborhood $U_{I, \varphi_{\mathfrak{g}}} \subseteq G$ \textup{(cf. \eqref{eq:local_diffeo})} admits a finite product-of-exponentials decomposition
\begin{equation}
    \label{eq:local_realization}
 g=  e^{\theta_L a_{j_L}} \cdots e^{\theta_1 a_{j_1}},
\end{equation}
where $a_{j_s} \in \{a_1, \dots, a_k\}$ for each $s \in \{1, \dots, \mathcal{T}_{\mathbf{a}}\}$, and the required depth $L\le \mathcal{T}_{\mathbf{a}}$ satisfies \eqref{eq:worst_case_derivation}. In particular, this bounds the local parameterization depth to $\mathcal{O}(\mathcal{D}_{\mathbf{a}} \dim \mathfrak{g})$.
\end{thm}


\subsection{Circuit Depth Bounds for Infinitesimal Evolutions of States}

Let $\ket{\psi} \in W$ be a unit vector ($\|\psi\| = 1$). The real dimension $m_\psi$ of the projective $G$-orbit $\mathcal O_{[\psi]}$ satisfies
\begin{equation}
\label{eq:orbit_dimension}
m_\psi \coloneqq \dim_{\mathbb{R}}\bigl(\mathcal O_{[\psi]}\bigr) = \dim_{\mathbb{R}}(\mathfrak{g}) - \dim_{\mathbb{R}}\!\bigl(\operatorname{stab}_{\mathfrak{g}}([\psi])\bigr),
\end{equation}
where
\begin{equation}
\label{eq:lie_algebra_stabilizer}
\operatorname{stab}_{\mathfrak{g}}([\psi]) \coloneqq \bigl\{A \in \mathfrak{g} \mid A \ket{\psi} \in i\mathbb R\ket{\psi}\bigr\}
\end{equation}
is the Lie algebra stabilizer of $[\psi]$ in $\mathfrak{g}$.

For the DLA filtration $\{\mathcal{F}_j^{\mathbf{a}}\}_{j\geq 1}$ generated by the original controls $\mathbf{a}=\{a_1,\ldots,a_k\}$
(see \eqref{eq:LieAlgFiltration}), define
\begin{equation}
\label{eq:psi_filtered_rank}
r_j(\psi)
:=
\dim_{\mathbb R}
\left\{
(I-|\psi\rangle\langle\psi|)A|\psi\rangle
\;\middle|\;
A\in\mathcal F_j^{\mathbf a}
\right\},
\qquad r_0(\psi):=0.
\end{equation}
Thus, $r_j(\psi)$ counts the number of linearly
independent infinitesimal directions in the tangent
space of the projective orbit $\mathcal O_{[\psi]}$
generated by the $j$-th filtered subspace of $\mathfrak{g}$. In particular,
\begin{equation}
    \label{eq:orb_filt_comp_dim_bound}
    r_j(\psi)
=
\dim_{\mathbb R}(\mathcal F_j^{\mathbf a})
-
\dim_{\mathbb R}(
\mathcal F_j^{\mathbf a}\cap
\operatorname{stab}_{\mathfrak g}([\psi])).
\end{equation}

Let
\begin{equation}
\mathcal D_{\mathbf{a},\psi}\
:=
\min\bigl\{
j\in \mathbb{N}\mid
r_j(\psi)=m_\psi
\bigr\}.
\end{equation}
be the \textit{$\psi$-stabilization depth} of the generating set $\mathbf{a}$, representing the minimal number of filtration steps required to fully span the tangent space of the projective orbit $\mathcal O_{[\psi]}$.

\begin{rmk}
    \label{rmk:orbit_bounds}
   It follows from \eqref{eq:orb_filt_comp_dim_bound} that for all $j \ge 1$:
    \begin{equation}
        \label{eq:trivial_bounds}
        r_j(\psi) \le \dim(\mathcal{F}_j^{\mathbf{a}}),
    \end{equation}

which, in turn, implies the state-independent uniform bounds

    \begin{equation}
        \label{eq:trivial_bounds2}
        \begin{aligned}
        m_\psi & \le \dim(\mathfrak{g})\\
         \mathcal{D}_{\mathbf{a},\psi} & \le \mathcal D_{\mathbf{a}}
        \end{aligned}
    \end{equation}
\end{rmk}

By accumulating the newly generated tangent directions at each filtration step, analogous to the derivation in \eqref{eq:worst_case_derivation}, we obtain bounds on the total number of exponentials, $\mathcal{T}_{\mathbf{a},\psi}$, required to prepare any target state within a sufficiently small neighborhood of $[\psi]$ in the orbit $\mathcal O_{[\psi]}$:

\begin{equation}
\label{eq:state_depth_bound}
m_\psi\le \mathcal T_{\mathbf{a},\psi}
\leq
\sum_{j=1}^{\mathcal{D}_{\mathbf{a},\psi}}
(2j-1)\bigl(r_j(\psi)-r_{j-1}(\psi)\bigr)
\leq
(2\mathcal{D}_{\mathbf{a},\psi}-1)m_\psi.
\end{equation}


The following theorem summarizes the primary results established in this subsection.

\begin{thm}
\label{thm:local_state_reachability}
Let $\mathfrak{g} \subseteq \mathfrak{u}(W)$ be a dynamical Lie algebra generated  by $\mathbf{a} = \{a_1, \dots, a_k\}$, let $G \subseteq U(W)$ be the corresponding connected analytic Lie group, and let $\ket{\psi}\in W$ be a reference state. There exists a neighborhood $\mathcal{V}_{[\psi]}$ of the ray $[\psi]$ in the projective state orbit $\mathcal O_{[\psi]}$ such that every state $\ket{\zeta}$ with $[\zeta]\in\mathcal{V}_{[\psi]}$ admits a finite product-of-exponentials realization

\begin{equation}
    \label{eq:state_poe_realization}
 \ket{\zeta}= e^{i\phi}U_{L}(\boldsymbol{\theta})\ket{\psi}=
    e^{i\phi}e^{\theta_L a_{j_L}} \cdots e^{\theta_1 a_{j_1}}\ket{\psi},
\end{equation}
where $\phi \in [0,2\pi]$, $a_{j_s} \in \{a_1, \dots, a_k\}$ for each $s \in \{1, \dots, L\}$, and the required depth $L = \mathcal{T}_{\mathbf{a},\psi}$ satisfies \eqref{eq:state_depth_bound}. In particular, this establishes an upper bound on the local parameterization depth of order $\mathcal{O}(\mathcal{D}_{\mathbf{a},\psi} m_\psi)$.
\end{thm} 

While producing a collection of circuits that reproduces every operator within an identity neighborhood $U_{I, \varphi_{\mathfrak{g}}}$ in the dynamical Lie group $G$ is sufficient to generate the corresponding neighborhood $\mathcal{V}_{[\psi]}$ of the projectivization of a  state $\ket{\psi}$ in $\mathcal O_{[\psi]}$, the upper bound on circuit depth appearing in \eqref{eq:state_depth_bound} can be noticeably smaller, as the following example demonstrates.

\begin{ex}
\label{ex:dimension_reduction}
Consider an $n$-qubit Hilbert space $W = \mathbb{C}^d$ with $d = 2^n$. The dimensions of the full special unitary Lie algebra $\mathfrak{su}(d)$ and the unit sphere of states $\mathbb{S}(\mathbb{C}^d)$ are
\begin{equation}
    \label{eq:dim_comparison}
    \dim (\mathfrak{su}(d)) = d^2 - 1, \qquad \dim_{\mathbb{R}} (\mathbb{P}(\mathbb{C}^d)) = 2d - 2.
\end{equation}
Hence, replacing the dimension of the full dynamical Lie algebra with the dimension of the state orbit reduces the governing dimension factor from order $\mathcal{O}(4^n)$ down to $\mathcal{O}(2^n)$.
\end{ex}

\section{Stabilization Depth Bounds for QAOA Architectures}
\label{sec:stab_depth_QAOA}

Next we compute explicit bounds on the stabilization depth $\mathcal{D}_{\mathbf{a}}$ for several prominent families of parameterized quantum circuits. Specifically, we analyze the dynamical Lie algebras associated with multi-angle  QAOA ans\"atze applied to various combinatorial optimization problems, as well as Grover-mixer QAOA (GM-QAOA) architectures. These concrete bounds characterize the bracket depth required to exhaust the underlying Lie algebra, thereby yielding explicit upper bounds on the local circuit depth required for expressivity near the identity of the group.

In Subsection~\ref{subsec:depth_bounds_examples}, we leverage these results to establish explicit circuit-depth upper bounds for local unitary synthesis and state preparation. A rigorous treatment and full technical proofs for these stabilization depth derivations are provided in Appendix~\ref{sec:depth_bounds_calculation}.

\begin{rmk}
\label{rmk:generator_vs_algebra_depth}
The stabilization depth $\mathcal{D}_{\mathbf{a}}$ is defined relative to a specific choice of generating set $\mathbf{a} \subset \mathfrak{g}$ rather than the Lie algebra $\mathfrak{g}$ itself (see Definition~\ref{def:stabilization_depth}). However, each circuit architecture considered in this section is equipped with a canonical, predefined set of elementary gate generators (e.g., individual Pauli terms or driver/problem Hamiltonian components). Consequently, when the generating set $\mathbf{a}$ is unambiguous from context, we adopt the slight abuse of notation $\mathcal{D}_{\mathfrak{g}} \equiv \mathcal{D}_{\mathbf{a}}$ to denote the stabilization depth of the corresponding dynamical Lie algebra.
\end{rmk}

\subsection{Dynamical Lie Algebras for the MaxCut Problem QAOA}
The Maximum Cut problem on a finite graph $\Gamma=(V,E)$ seeks a vertex partition maximizing the number of crossing edges. In the standard Quantum Approximate Optimization Algorithm framework, this combinatorial problem is mapped to an Ising spin system defined by the transverse-field mixer $B$ and problem Hamiltonian $H_P$:

\begin{equation}
\label{eq:OriginalHamiltonians}
B = \sum_{w \in V} X_w, \qquad H_P = \sum_{\{v,w\} \in E} Z_v Z_w.
\end{equation}
The associated standard dynamical Lie algebra is generated by these two collective operators:
\begin{equation}
\label{eq:StdDLAgenerators}
\mathfrak{g}_{\Gamma,\mathrm{std}} := \Big\langle i B, i H_P  \Big\rangle_{\mathrm{Lie}}.
\end{equation}

In multi-angle QAOA, each individual term in the mixer and problem Hamiltonians is assigned an independent variational parameter. The corresponding \emph{free} dynamical Lie algebra is formed by taking these individual operators as separate control fields:
\begin{equation}
\label{eq:FreeDLAgenerators}
\mathfrak{g}_{\Gamma,\mathrm{free}} := \Big\langle \{ i X_w \mid w \in V \} \cup \{ i Z_v Z_w \mid \{v,w\} \in E \} \Big\rangle_{\mathrm{Lie}}.
\end{equation}

\begin{rmk}
While the explicit structure and dimensions of standard DLAs are generally difficult to classify for arbitrary graph topologies, free DLAs admit full algebraic characterization. Specifically, Theorem~1 of \cite{KLFCCZ} provides a complete structural classification of $\mathfrak{g}_{\Gamma,\mathrm{free}}$. In particular, if $\Gamma$ is \emph{archetypal} (i.e., a connected graph that is neither bipartite nor a cycle), the free DLA is isomorphic to the direct sum of two special unitary Lie algebras:
\begin{equation}
\label{eq:FreeDLA_Archetypal}
\mathfrak{g}_{\Gamma,\mathrm{free}} \cong \mathfrak{su}(W_{\Gamma,\mathrm{even}}) \oplus \mathfrak{su}(W_{\Gamma,\mathrm{odd}}) \cong \mathfrak{su}(2^{n-1}) \oplus \mathfrak{su}(2^{n-1}),
\end{equation}
where $n = |V|$ is the order of $\Gamma$, and $W_{\Gamma,\mathrm{even}}$ (resp.\ $W_{\Gamma,\mathrm{odd}}$) denotes the subspace spanned by Hadamard basis states containing an even (resp.\ odd) number of $\ket{-}$ factors.
\end{rmk}

To highlight the structural differences between standard and free DLAs as well as their respective stabilization depths, we present numerical results for all non-isomorphic asymmetric graphs of order $n=6$ (see Table~\ref{table:SplittingStatisticsTable}).

\begin{table}[htbp]
\centering
\resizebox{0.53\linewidth}{!}{
\begin{tabular}{|c|c|c|c|c|c|c|c|c|}
\hline
$\Gamma$ & $1$ & $2$ & $3$ & $4$ & $5$ & $6$ & $7$ & $8$ \\
\hline
$\dim(\mathfrak g_{\Gamma,\mathrm{free}})$ & 2046 & 2046 & 2046 & 2046 & 2046 & 2046 & 2046 & 2046 \\
\hline
$\mathcal D_{\mathfrak g_{\Gamma,\mathrm{free}}}$ & 17 & 21 & 29 & 17 & 21 & 17 & 19 & 13 \\
\hline
$\dim(\mathfrak g_{\Gamma,\mathrm{std}})$ & 1984 & 1970 & 1983 & 1984 & 1984 & 1984 & 1984 & 1984 \\
\hline
$\mathcal D_{\mathfrak g_{\Gamma,\mathrm{std}}}$ & 19 & 21 & 29 & 19 & 21 & 17 & 19 & 17 \\
\hline
\end{tabular}
}
\caption{Comparison of dimensions and stabilization depths for free ($\mathfrak{g}_{\Gamma,\mathrm{free}}$) and standard ($\mathfrak{g}_{\Gamma,\mathrm{std}}$) dynamical Lie algebras across all eight non-isomorphic asymmetric graphs of order $n=6$.}
\label{table:SplittingStatisticsTable}
\end{table}

Complementing these empirical findings, the following theorem establishes analytical upper bounds on the stabilization depth of free DLAs across several canonical graph families.

\begin{thm}
\label{thm:free_DLA_filtration_classical}
Let $\Gamma$ be a graph belonging to one of the following families:

\begin{itemize}
\item a path graph $P_n$ (with $\mathfrak{g}_{P_n,\mathrm{free}} \cong \mathfrak{so}(2n)$),
\item a cycle graph $C_n$ (with $\mathfrak{g}_{C_n,\mathrm{free}} \cong \mathfrak{so}(2n) \oplus \mathfrak{so}(2n)$),
\item a complete graph $K_n$ (with $\mathfrak{g}_{K_n,\mathrm{free}} \cong \mathfrak{su}(2^{n-1}) \oplus \mathfrak{su}(2^{n-1})$).
\end{itemize}

The stabilization depth of the corresponding free dynamical Lie algebras satisfy
\begin{equation}
    \label{eq:free_DLA_filtr_bound_some_grpahs}
    \begin{aligned}
 \mathcal D_{\mathfrak{g}_{P_n,\mathrm{free}}}&\le2n-1\quad(n\ge2),\\
 \mathcal D_{\mathfrak{g}_{C_n,\mathrm{free}}}&\le2n-1\quad(n\ge3),\\
 \mathcal D_{\mathfrak{g}_{K_n,\mathrm{free}}}&\le2n-1\quad(n\ge2).
    \end{aligned}
\end{equation}
\end{thm}

\subsection{Reduced DLAs for the MaxCut Problem}
The MaxCut cost function exhibits a global $\mathbb{Z}_2$ (bit-flip) symmetry that permits fixing the classical state of an arbitrary reference vertex $v \in V$ (e.g., $x_v = 0$) without altering the maximum cut value, as the full solution set remains reconstructible via bitwise inversion. A comprehensive exposition of the corresponding symmetry-reduced QAOA formulations, including their dynamical Lie algebras, circuit properties, and algebraic structures, is provided in \cite{TBFS}; here, we briefly review the relevant setup. 

Quantum mechanically, imposing $x_v=0$ breaks this two-fold degeneracy and projects the $n$-qubit Hilbert space $(\mathbb{C}^2)^{\otimes n}$ onto an $(n-1)$-qubit subspace:

\begin{equation}
    \label{eq:reduced_hilbert}
    W_v = \mathbb{C}^2 \otimes \dots \otimes\mathbb{C}^2 \otimes\underbrace{\mathbb{C}|x_v=0\rangle}_{v\text{-th position}} \otimes\mathbb{C}^2\otimes \dots \otimes \mathbb{C}^2 \simeq (\mathbb{C}^2)^{\otimes (n-1)}.
\end{equation}

Under this reduction, Pauli $Z$-operator on vertex $v$ is replaced by its respective scalar eigenvalue ($Z_v \mapsto 1$). Consequently, edge interactions incident to $v$ simplify to single-qubit $Z$ operators. The reduced mixer and problem Hamiltonians acting on $V \setminus \{v\}$ become
\begin{align}
\label{eq:reduced_hams}
    H_M^v = \sum_{w \in V \setminus \{v\}} X_w, \qquad 
    H_P^v = \sum_{\substack{(w,w') \in E \\ w,w' \neq v}} Z_w Z_{w'} + \sum_{(v,w) \in E} Z_w.
\end{align}

The \emph{free} reduced dynamical Lie algebra is generated by taking the Lie closure of the individual local interaction terms in \eqref{eq:reduced_hams}:
\begin{equation}
\label{eq:reduced_free}
\mathfrak{g}^{\,v}_{\Gamma,\mathrm{free}} = \left\langle 
    \{ i X_w \mid w \neq v \}, \; 
    \{ i Z_w Z_{w'} \mid (w,w') \in E, \, w,w' \neq v \}, \; 
    \{ i Z_w \mid (v,w) \in E \}
\right\rangle_{\mathrm{Lie}}.
\end{equation}
In contrast, the \emph{standard} reduced DLA is generated directly by the full reduced Hamiltonians:
\begin{equation}
\label{eq:reduced_standard}
\mathfrak{g}^{\,v}_{\Gamma,\mathrm{std}} = \Big\langle
    i H_M^v, \; i H_P^v 
\Big\rangle_{\mathrm{Lie}}.
\end{equation}

To demonstrate the sensitivity of DLA dimensions and stabilization depths to vertex selection, we tabulate numerical data for all non-isomorphic asymmetric graphs on $n=6$ vertices. Results for the free and standard reduced DLAs are reported in Tables~\ref{AsymGraphsSixNodesFree} and~\ref{AsymGraphsSixNodesStandard}, respectively.

\begin{table}[htbp]
\begin{center}
\resizebox{0.78\linewidth}{!}{
\begin{tabular}{|c|c|c|c|c|c|c|c|}
\hline
$i$ &
$\dim\bigl(\mathfrak{g}_{\Gamma_i,\mathrm{free}}\bigr)$ & $\dim\bigl(\mathfrak{g}^{v_1}_{\Gamma_i,\mathrm{free}}\bigr)$ & $\dim\bigl(\mathfrak{g}^{v_2}_{\Gamma_i,\mathrm{free}}\bigr)$  &$\dim\bigl(\mathfrak{g}^{v_3}_{\Gamma_i,\mathrm{free}}\bigr)$ &
$\dim\bigl(\mathfrak{g}^{v_4}_{\Gamma_i,\mathrm{free}}\bigr)$ &$\dim\bigl(\mathfrak{g}^{v_5}_{\Gamma_i,\mathrm{free}}\bigr)$&$\dim\bigl(\mathfrak{g}^{v_6}_{\Gamma_i,\mathrm{free}}\bigr)$ \\
\hline
$1$ & $2046$ &  $1023$  & $1023$  & $1023$  & $1023$  & $1023$  & $258$\\
\hline
$2$ & $2046$ & $258$ &$1023$ &$1023$ & $1023$ &$1023$  &$258$ \\
\hline
$3$   & $2046$ & $258$ &$1023$ &$1023$  &$1023$  &$258$& $73$ \\
\hline
$4$  & $2046$ & $1023$  &$1023$ &$1023$  &$1023$  &$258$ & $1023$\\
\hline
$5$   & $2046$ & $1023$  &$1023$ &$1023$  &$1023$  &$258$ & $1023$ \\
\hline
$6$   &$2046$ & $1023$  &$1023$ &$1023$  &$1023$  &$1023$ & $1023$ \\
\hline
$7$   &$2046$ & $1023$  &$1023$ &$1023$  &$1023$  &$258$ & $1023$  \\
\hline
$8$  & $2046$& $1023$ & $1023$ & $1023$ & $1023$ & $1023$ & $1023$\\
\hline
\end{tabular}
}

\vspace{0.15in}

\resizebox{0.5\linewidth}{!}{\begin{tabular}{|c|c|c|c|c|c|c|c|}
\hline
$i$ & $\mathcal D_{\mathfrak g_{\Gamma_i,\mathrm{free}}}$& $\mathcal D_{\mathfrak g^{v_1}_{\Gamma_i,\mathrm{free}}}$  & $\mathcal D_{\mathfrak g^{v_2}_{\Gamma_i,\mathrm{free}}}$ & $\mathcal D_{\mathfrak g^{v_3}_{\Gamma_i,\mathrm{free}}}$ &
$\mathcal D_{\mathfrak g^{v_4}_{\Gamma_i,\mathrm{free}}}$  &  $\mathcal D_{\mathfrak g^{v_5}_{\Gamma_i,\mathrm{free}}}$ & $\mathcal D_{\mathfrak g^{v_6}_{\Gamma_i,\mathrm{free}}}$ \\
\hline
$1$ & $17$ & $21$ & $17$ & $15$ & $17$ & $17$ & $19$\\
\hline
$2$ & $21$ & $17$& $19$ & $19$& $19$& $20$ & $17$ \\
\hline
$3$   & $29$& $17$& $29$& $31$& $27$& $29$ & $17$ \\
\hline
$4$ &$17$ & $17$ & $17$& $15$& $17$ & $11$& $17$\\
\hline
$5$   & $21$ & $23$ & $20$& $19$& $24$& $12$& $31$  \\
\hline
$6$  & $17$& $19$& $15$& $16$& $14$& $14$& $19$ \\
\hline
$7$  & $19$& $17$& $17$& $17$& $18$& $9$& $20$\\
\hline
$8$  & $13$ & $13$ & $13$ & $11$ & $14$ & $13$ & $15$\\
\hline
\end{tabular}}
 \caption{Comparison of dimensions and stabilization depths for reduced free 
 dynamical Lie algebras across all non-isomorphic asymmetric graphs of order $6$.}
    \label{AsymGraphsSixNodesFree}
\end{center}
\end{table}

\begin{table}[htbp]
\begin{center}
\resizebox{0.78\linewidth}{!}{
\begin{tabular}{|c|c|c|c|c|c|c|c|}
\hline
$i$ &
$\dim\bigl(\mathfrak{g}_{\Gamma_i,\mathrm{std}}\bigr)$ & $\dim\bigl(\mathfrak{g}^{v_1}_{\Gamma_i,\mathrm{std}}\bigr)$ & $\dim\bigl(\mathfrak{g}^{v_2}_{\Gamma_i,\mathrm{std}}\bigr)$  &$\dim\bigl(\mathfrak{g}^{v_3}_{\Gamma_i,\mathrm{std}}\bigr)$ &
$\dim\bigl(\mathfrak{g}^{v_4}_{\Gamma_i,\mathrm{std}}\bigr)$ &$\dim\bigl(\mathfrak{g}^{v_5}_{\Gamma_i,\mathrm{std}}\bigr)$&$\dim\bigl(\mathfrak{g}^{v_6}_{\Gamma_i,\mathrm{std}}\bigr)$ \\
\hline
$1$ & $1984$ & $1023$  & $1023$  & $1023$  & $1023$  & $1023$  & $258$  \\
\hline
$2$ & $1970$ & $258$ &$1023$ &$1023$ & $1023$ &$1023$  &$258$  \\
\hline
$3$ & $1983$ & $258$ &$1023$ &$1023$  &$1023$  &$258$& $73$   \\
\hline
$4$ & $1984$ & $1023$  &$1023$ &$1023$  &$1023$  &$258$ & $1023$ \\
\hline
$5$ & $1984$ & $1023$  &$1023$ &$1023$  &$1023$  &$258$ & $1023$  \\
\hline
$6$ & $1984$ & $1023$  &$1023$ &$1023$  &$1023$  &$1023$ & $1023$  \\
\hline
$7$ & $1984$ & $1023$  &$1023$ &$1023$  &$1023$  &$258$ & $1023$  \\
\hline
$8$ & $1984$ & $1023$ & $1023$ & $1023$ & $1023$ & $1023$ & $1023$ \\
\hline
\end{tabular}
}

\vspace{0.15in}

\resizebox{0.5\linewidth}{!}{\begin{tabular}{|c|c|c|c|c|c|c|c|}
\hline
$i$ & $\mathcal D_{\mathfrak g_{\Gamma_i,\mathrm{std}}}$& $\mathcal D_{\mathfrak g^{v_1}_{\Gamma_i,\mathrm{std}}}$  & $\mathcal D_{\mathfrak g^{v_2}_{\Gamma_i,\mathrm{std}}}$ & $\mathcal D_{\mathfrak g^{v_3}_{\Gamma_i,\mathrm{std}}}$ &
$\mathcal D_{\mathfrak g^{v_4}_{\Gamma_i,\mathrm{std}}}$  &  $\mathcal D_{\mathfrak g^{v_5}_{\Gamma_i,\mathrm{std}}}$ & $\mathcal D_{\mathfrak g^{v_6}_{\Gamma_i,\mathrm{std}}}$ \\
\hline

$1$ & $19$ & $20$ & $16$ & $14$ & $17$ & $17$ & $19$ \\
\hline
$2$ & $21$ & $17$ & $18$ & $18$ & $18$ & $19$ & $17$ \\
\hline
$3$ & $29$ & $17$ & $28$ & $30$ & $26$ & $29$ & $17$ \\
\hline
$4$ & $19$ & $16$ & $17$ & $14$ & $16$ & $13$ & $16$ \\
\hline
$5$ & $21$ & $22$ & $20$ & $18$ & $24$ & $14$ & $30$ \\
\hline
$6$ & $17$ & $18$ & $14$ & $15$ & $15$ & $14$ & $20$ \\
\hline
$7$ & $19$ & $17$ & $17$ & $16$ & $17$ & $12$ & $19$ \\
\hline
$8$ & $17$ & $14$ & $14$ & $14$ & $14$ & $14$ & $14$ \\
\hline
\end{tabular}}
 \caption{Comparison of dimensions and stabilization depths for reduced  standard 
 dynamical Lie algebras across all non-isomorphic asymmetric graphs of order $6$.}
    \label{AsymGraphsSixNodesStandard}
\end{center}
\end{table}  

\begin{rmk}
\label{rmk:full_dla_dimension_six_nodes}
Since every reduced generator in Eqs.~\eqref{eq:reduced_free} and~\eqref{eq:reduced_standard} is traceless, each reduced DLA is a Lie subalgebra of $\mathfrak{su}(W_v) \cong \mathfrak{su}(32)$. Consequently, entries in Table~\ref{AsymGraphsSixNodesFree} with dimension $1023 = 2^{2(n-1)} - 1$ correspond to the full special unitary algebra $\mathfrak{su}(32)$, implying universal quantum controllability on the respective reduced subspace.
\end{rmk}

In the symmetry-reduced framework, Theorem~\ref{thm:free_DLA_filtration_classical} yields the following analogue for complete graphs.

\begin{thm}
\label{thm:reduced_free_DLA_filtration_classical}
Let $\Gamma = K_n$ be the complete graph on $n$ vertices, and let $v \in V$ be an arbitrary reference vertex. The reduced free dynamical Lie algebra $\mathfrak{g}^v_{\Gamma,\mathrm{free}}$ and its stabilization depth satisfy
\begin{equation}
\label{eq:reduced_free_DLA_filtr_bound_some_graphs}
        \mathfrak{g}^v_{\Gamma,\mathrm{free}} \cong \mathfrak{su}(2^{n-1}) \text{ and }
        \mathcal{D}_{\mathfrak{g}^v_{\Gamma,\mathrm{free}}} \le 2n-1.
\end{equation}
\end{thm}

\subsection{Free DLAs for the Maximum Independent Set Problem}
\label{subsec:MIS}

Let $\Gamma=(V,E)$ be a graph on $n$ vertices. An independent set in $\Gamma$ is a vertex subset containing no pair of adjacent vertices. Subsets of $V$ are identified with binary strings $x \in \{0,1\}^n$, where $x_v=1$ indicates that $v$ is included in the subset and $x_v=0$ indicates it is omitted.

The objective function for the Maximum Independent Set  problem is
\begin{equation}
    \label{eq:objective_function}
    F_{\mathrm{MIS}}(x) = \sum_{v \in V} x_v.
\end{equation}
Up to an additive constant and an overall sign factor, the corresponding problem Hamiltonian is encoded as
\begin{equation}
    \label{eq:MIS_problem_hamiltonian}
    H_{P,\mathrm{MIS}} = \sum_{v \in V} Z_v.
\end{equation}

Because adjacent vertices cannot simultaneously belong to an independent set, the feasible configurations form a restricted subset of binary strings:
\begin{equation}
    \label{eq:MIS_feasible_set}
    \mathcal{F} := \left\{ x \in \{0,1\}^n \;\middle|\; x_u x_v = 0 \text{ for all } \{u,v\} \in E \right\},
\end{equation}
inducing the feasible Hilbert subspace
\begin{equation}
    \label{eq:MIS_feasible_subspace}
    W_{\mathcal{F}} := \operatorname{span}\left\{ \ket{x} \;\middle|\; x \in \mathcal{F} \right\} \subseteq W.
\end{equation}

To enforce feasibility in constrained optimization problems, we employ a subspace-preserving mixer Hamiltonian that leaves the feasible subspace $W_{\mathcal{F}}$ invariant \cite{HHR, HWORVB, Saleem, TBS}. Let $\mathcal{N}_v := \{w \in V \mid \{v,w\} \in E\}$ denote the open neighborhood of vertex $v \in V$. The local controlled-$X$ mixer $H_{CX,v}$ acts on a computational basis state $\ket{x}$ according to the rule
\begin{equation}
    \label{eq:HCXv_action}
    H_{CX,v}\ket{x} = 
    \begin{cases}
        \ket{x^{(v)}}, & \text{if } x_w = 0 \text{ for all } w \in \mathcal{N}_v, \\[0.5ex]
        0, & \text{if } x_w = 1 \text{ for some } w \in \mathcal{N}_v,
    \end{cases}
\end{equation}
where $x^{(v)}$ denotes the bitstring obtained by flipping the $v$-th bit of $x$.

In terms of Pauli operators, each local mixer admits the explicit representation
\begin{equation}
    \label{eq:CX_mixer}
    \mathcal{P}_v \coloneqq \prod_{w \in \mathcal{N}_v} \frac{I + Z_w}{2}, \qquad
    H_{CX,v} \coloneqq X_v \mathcal{P}_v, \qquad
    H_{CX} \coloneqq \sum_{v \in V} H_{CX,v},
\end{equation}
where $\mathcal{P}_v$ is a projector onto the subspace in which all neighbors of $v$ are in the state $\ket{0}$. By construction, each $H_{CX,v}$ preserves the feasible subspace $W_{\mathcal{F}}$. Furthermore, since the problem Hamiltonian is diagonal in the computational basis, it preserves $W_{\mathcal{F}}$ as well.

Consequently, it is natural to define the restricted free dynamical Lie algebra associated with these generators on $W_{\mathcal{F}}$ as
\begin{equation}
    \label{eq:ReducedGeneratorsFreeDLAs}
    \mathfrak{g}_{\Gamma,\mathrm{free},\mathcal{F}} \coloneqq \Big\langle \Big\{ i H_{CX,v}\big|_{W_{\mathcal{F}}}, \; i Z_v\big|_{W_{\mathcal{F}}} \Big\}_{v \in V} \Big\rangle_{\mathrm{Lie}}.
\end{equation}

The following result establishes that this restricted algebra achieves full Lie-algebraic controllability on the feasible subspace, while providing an explicit upper bound on its stabilization depth that scales linearly with the number of vertices and the independence number.

\begin{thm}
\label{thm:restricted_free_DLA_controllability}
Let $\Gamma$ be a connected graph with $|V(\Gamma)| \ge 2$ vertices, and   independence number $\alpha(\Gamma)$. The restricted free dynamical Lie algebra $\mathfrak{g}_{\Gamma,\mathrm{free},\mathcal{F}}$ and its stabilization depth satisfy
\begin{equation}
\label{eq:mis_filtration_depth_bound}
    \mathfrak{g}_{\Gamma,\mathrm{free},\mathcal{F}} \cong \mathfrak{u}(W_{\mathcal{F}}) \text{ and } \mathcal{D}_{\mathfrak{g}_{\Gamma,\mathrm{free}, \mathcal{F}}} \le 8|V(\Gamma)| + 4\alpha(\Gamma) - 2.
\end{equation}
\end{thm}

\subsection{Dynamical Lie Algebras for the Grover-Mixer QAOA}

The Grover-mixer QAOA (GM-QAOA), introduced in~\cite{BE}, replaces standard local mixers with a highly nonlocal mixer Hamiltonian. The algorithm typically starts in a uniform superposition over all feasible configurations $\mathcal{F}$,
\begin{equation}
\label{eq:InitialState}
\ket{\xi} := \frac{1}{\sqrt{|\mathcal{F}|}} \sum_{x \in \mathcal{F}} \ket{x}.
\end{equation}
The Grover mixer $G_M$ is defined as the orthogonal projection (up to a negative sign) onto this initial state:
\begin{equation}
\label{eq:GroverMixer}
G_M := -\ket{\xi}\bra{\xi}.
\end{equation}

Let $d$ denote the number of distinct values attained by the objective function $F$ over $\mathcal{F}$. For each $j \in \{1, \dots, d\}$, let $\mathcal{F}_j \subseteq \mathcal{F}$ be the $j$-th level set of $F$, and define the normalized level-set state
\begin{equation}
\label{eq:LevelSetStates}
\ket{\xi_j} := \frac{1}{\sqrt{|\mathcal{F}_j|}} \sum_{x \in \mathcal{F}_j} \ket{x}.
\end{equation}
As established in~\cite[Section~V]{TNB}, the quantum evolution under GM-QAOA remains strictly confined to the $d$-dimensional invariant subspace $W_0 \subseteq W$ spanned by these level-set states:
\begin{equation}
\label{eq:GMQAOASubspace}
W_0 := \operatorname{span} \bigl\{ \ket{\xi_1}, \dots, \ket{\xi_d} \bigr\}.
\end{equation}

Restricting both the mixer $G_M$ and the problem Hamiltonian $H_P$ to $W_0$, we obtain the associated dynamical Lie algebra
\begin{equation}
\label{eq:GroverMixerDLA}
\mathfrak{g}_{\xi} := \left\langle i G_M\big|_{W_0}, \, i H_P\big|_{W_0} \right\rangle_{\mathrm{Lie}}.
\end{equation}

Theorem~III.1 of~\cite{TNB} shows that $\mathfrak{g}_{\xi}$ is the full unitary algebra on this subspace:
\begin{equation}
\label{eq:GroverMixerLieAlgebra}
\mathfrak{g}_{\xi} \cong \mathfrak{u}(W_0) \cong \mathfrak{u}(d).
\end{equation}

Consequently, GM-QAOA is fully controllable on $W_0$, whose dimension equals the number $d$ of distinct objective-function values on the feasible set. 

\subsection{Analytical Depth Bounds for Specific Ans\"atze}
\label{subsec:depth_bounds_examples}

Combining the stabilization depths $\mathcal{D}_{\mathfrak{g}}$ derived for the DLA families in the preceding subsections with the general depth bounds given in \eqref{eq:worst_case_derivation} and \eqref{eq:state_depth_bound}, we establish explicit upper bounds on the circuit depths required for local unitary synthesis and state preparation. Specifically, these bounds guarantee the local generation of unitaries within a neighborhood of the identity in the dynamical Lie group $G$, as well as the preparation of states within a sufficiently small neighborhood of the ray orbit of the initial state $\ket{\psi}$. The analytical results for Lie algebra dimensions, orbit dimension bounds, stabilization depths, and the corresponding sufficient circuit depths $\mathcal{T}_{\mathfrak{g}}$ and $\mathcal{T}_{\mathfrak{g},\psi}$ across these models are summarized in Table~\ref{tab:examples_depths}.

\begin{table}[htbp]
\begin{center}
\resizebox{1.\linewidth}{!}{
\begin{tabular}{|c|c|c|c|c|c|c|}
\hline
Circuit family
& DLA, $\mathfrak g$
& $\dim(\mathfrak g)$
& $\dim(\mathcal O_{[\psi]})$ 
& $\mathcal D_{\mathfrak g}$ 
& Sufficient $\mathcal T_\mathfrak{g}$
& Sufficient $\mathcal T_{\mathfrak{g},\psi}$\\
\hline

FMC, $\mathcal P_n$
& $\mathfrak{so}(2n)$
& $n(2n-1)$
& $\le n(2n-1)$
& $\le 2n-1$
& $n(4n-3)(2n-1)$
& $n(4n-3)(2n-1)$\\
\hline

FMC, $\mathcal C_n$
& $\mathfrak{so}(2n)^{\oplus2}$
& $2n(2n-1)$
& $\le 2n(2n-1)$
& $\le 2n-1$
& $2n(4n-3)(2n-1)$
& $2n(4n-3)(2n-1)$\\
\hline

FMC, $K_n$
& $\mathfrak{su}(2^{n-1})^{\oplus2}$
& $2^{2n-1}-2$
& $\le 2^{n+1}-3$
& $\le 2n-1$
& $(4n-3)(2^{2n-1}-2)$
& $(4n-3)(2^{n+1}-3)$\\
\hline

Red. FMC, $K_n$
& $\mathfrak{su}(2^{n-1})$
& $2^{2n-2}-1$
& $ 2^{n}-2$
& $\le 2n-1$
& $(4n-3)(2^{2n-2}-1)$
& $(4n-3)(2^{n}-2)$ \\
\hline

FMIS, $n\ge3$
& $\mathfrak{u}(|F|)$
& $|\mathcal F|^2$
& $ 2|\mathcal F|-2$
& $\le 8n+4m-2$
& $(16n+8m-5)|\mathcal F|^2$
& $(16n+8m-5)(2|\mathcal F|-2)$\\
\hline

GM-QAOA 
& $\mathfrak{u}(d)$
& $d^2$
& $ 2d-2$
& $\le d^2$
& $(2d^2-1)d^2$
& $(2d^2-1)(2d-2)$\\

\hline
\end{tabular}%
}
\caption{Sufficient numbers of original-control exponentials for local dynamical Lie group element synthesis ($\mathcal{T}_{\mathfrak{g}}$) and projective state orbit preparation ($\mathcal{T}_{\mathfrak{g},\psi}$) across various QAOA ans\"atze. Here, FMC and FMIS denote free (multi-angle) MaxCut and Maximum Independent Set QAOA variants, respectively, and GM-QAOA denotes Grover-Mixer QAOA. The parameters $n$ and $m$ represent the number of vertices and independence number of the underlying graph. The given values represent upper bounds derived from stabilization depths $\mathcal{D}_{\mathfrak{g}}$, dynamical Lie algebra dimensions $\dim(\mathfrak{g})$, and projective orbit dimensions $\dim(\mathcal{O}_{[\psi]})$, and are not necessarily optimal circuit depths.}
\label{tab:examples_depths}
\end{center}
\end{table}

\section{Global Synthesis on Full Unitary Invariant Sectors}
\label{sec:global_full_unitary}

In this section, we pass from local synthesis near the identity and the initial-state ray to uniform global synthesis on $\mathfrak{g}$-invariant subspaces \(V\subseteq W\). We assume throughout that the image of the dynamical Lie algebra on \(V\) is
\begin{equation}
    \label{eq:g_restricts_to_unitary}
    \mathfrak{g}|_V = \mathfrak{u}(V) \qquad \text{or} \qquad \mathfrak{g}|_V = \mathfrak{su}(V).
\end{equation}
Thus, \emph{full} in this section refers specifically to full unitary or special-unitary controllability on \(V\), rather than to pure-state controllability in its more general sense.

In this regime, global depth bounds can be obtained by reducing the decomposition problem to a two-level unitary synthesis procedure.

\subsection{Two-Stage Approach to Global Unitary Synthesis}
\label{subsec:two_stage_unitary_synthesis}

Let $D \coloneqq \dim(V)$ and fix an orthonormal basis $\{|1\rangle, \ldots, |D\rangle\}$ for $V$. For $1 \le r < s \le D$, the standard Hermitian generators for the two-level $\mathfrak{su}(2)$ sub-blocks are defined by
\begin{equation}
\label{eq:two_level_generators_recap}
    X_{rs} \coloneqq E_{rs} + E_{sr}, \qquad Y_{rs} \coloneqq -i(E_{rs} - E_{sr}), \qquad Z_{rs} \coloneqq E_{rr} - E_{ss},
\end{equation}
where $E_{rs} \coloneqq |r\rangle\langle s|$ denotes the elementary matrix unit. The global synthesis procedure operates via the following two-stage paradigm.

\medskip
\textbf{Step 1. Synthesis of elementary generator flows.}
Starting from the dynamical Lie algebra generators, we first construct exact implementations of the single-parameter flows
\begin{equation}
    \label{eq:basic_su(n)_flows}
    e^{itY_{rs}} \quad (\text{or } e^{itX_{rs}}) \qquad \text{and} \qquad e^{itZ_{rs}},
\end{equation}

for all $1 \le r < s \le D$. In the full unitary case $\mathfrak{g}|_V = \mathfrak{u}(V)$, this construction is extended to include the individual diagonal phase flows $e^{itE_{rr}}$.

\medskip
\textbf{Step 2. Application of global unitary decomposition.}
Once arbitrary two-level $SU(2)$ operations are accessible, we assemble them into an arbitrary element of $U(D)$ or $SU(D)$. Specifically, Spengler, Huber, and Hiesmayr~\cite{spengler2010composite} provided an explicit composite parameterization of $U(D)$ directly in terms of exponentials appearing in~\eqref{eq:basic_su(n)_flows}. Namely, every $g \in U(D)$ can be factored as
\begin{equation}
\label{eq:composite_parameterization}
g = \prod_{j=1}^{D-1} \prod_{s=j+1}^D \left( e^{i\widehat{\theta}_{j,s} E_{ss}} e^{i\theta_{j,s} Y_{js}} \right) \cdot \prod_{l=1}^{D} e^{i\widehat{\theta}_{l} E_{ll}},
\end{equation}
for suitable parameters $\widehat{\theta}_{j,s}, \widehat{\theta}_{l} \in [0, 2\pi]$ and $\theta_{j,s}\in\left[0,\frac{\pi}{2}\right]$.

To formulate this quantitatively, suppose every diagonal unitary on $V$ (modulo a scalar phase in the $SU(D)$ case) can be synthesized using at most $K_{\mathrm{diag}}$ native pulses, and every $e^{itY_{rs}}$ flow  using at most $K_Y$ pulses. The total parameterized circuit depth of global synthesis is then bounded by
\begin{equation}
\label{eq:general_full_unitary_bound}
L_{\mathrm{glob}} \le \binom{D}{2} \bigl(K_Y + K_{\mathrm{diag}}\bigr) + K_{\mathrm{diag}}.
\end{equation}

\begin{rmk}
\label{rmk:alternative_X_global_bound}
An equivalent global synthesis bound can be formulated using the generators $X_{rs} = E_{rs} + E_{sr}$ from \eqref{eq:two_level_generators_recap}. Using the diagonal unitary operator
\begin{equation}
\label{eq:R_diagonal_def}
R_{rs} \coloneqq e^{-\frac{i\pi}{4}(E_{rr} - E_{ss})},
\end{equation}
one obtains the relation
\begin{equation}
\label{eq:Y_from_X_diagonal_conjugation}
Y_{rs} = R_{rs} X_{rs} R_{rs}^{-1}, \qquad \text{and consequently,} \qquad e^{itY_{rs}} = R_{rs} e^{itX_{rs}} R_{rs}^{-1}.
\end{equation}

To determine the resulting circuit cost, set $M \coloneqq \binom{D}{2}$ and express the composite parameterization \eqref{eq:composite_parameterization} schematically as
\[
U = A_1 e^{it_1 Y_1} A_2 e^{it_2 Y_2} \cdots A_M e^{it_M Y_M} A_{\mathrm{f}},
\]
where $A_1, \ldots, A_M, A_{\mathrm{f}}$ are diagonal unitaries. Substituting \eqref{eq:Y_from_X_diagonal_conjugation} yields
\begin{equation}
\label{eq:X_decomp}
U = (A_1 R_{r_1s_1}) e^{it_1 X_{r_1s_1}} (R_{r_1s_1}^{-1} A_2 R_{r_2s_2}) e^{it_2 X_{r_2s_2}} \cdots (R_{r_{M-1}s_{M-1}}^{-1} A_M R_{r_Ms_M}) e^{it_M X_{r_Ms_M}} (R_{r_Ms_M}^{-1} A_{\mathrm{f}}).
\end{equation}
As each parenthesized factor is a product of diagonal matrices, it remains strictly diagonal. The full product therefore consists of $M$ two-level $X$-flows interleaved with $M + 1$ diagonal unitaries.

\end{rmk}

The following general theorem establishes explicit circuit depth bounds for synthesizing arbitrary group elements and state-to-state transitions when the restricted dynamical Lie algebra is $\mathfrak{u}(D)$ or $\mathfrak{su}(D)$. These bounds are parameterized by the primitive native synthesis costs $K_X$ (for off-diagonal generator flows) and $K_{\mathrm{diag}}$ (for diagonal unitaries). The formal proof is provided in Appendix~\ref{app:global_synthesis}.

In the special unitary case, the diagonal factors in
\eqref{eq:Y_from_X_diagonal_conjugation} may be chosen within
\(\mathrm{SU}(V)\). Indeed, if \(A_1,\ldots,A_{M+1}\) are the
diagonal factors and \(U\in\mathrm{SU}(V)\), choose scalars
\(\lambda_r\) satisfying
\[
\lambda_r^D=\det(A_r),
\qquad
\prod_{r=1}^{M+1}\lambda_r=1,
\]
where the branches of the \(D\)-th roots are chosen so that the second
condition holds. Then \(\widetilde A_r:=\lambda_r^{-1}A_r\) is diagonal
and belongs to \(\mathrm{SU}(V)\), while replacing every \(A_r\) by
\(\widetilde A_r\) leaves the total product unchanged.

\begin{thm}
\label{thm:uniform_orbit_generation}
Let $V \cong \mathbb{C}^D$ be a $D$-dimensional Hilbert space equipped with an orthonormal basis $\{\ket{e_1}, \dots, \ket{e_D}\}$. Suppose the restricted dynamical Lie algebra satisfies $\mathfrak{g}\big|_V = \mathfrak{u}(D)$ or $\mathfrak{su}(D)$. Let $K_X$ denote a uniform depth bound for synthesizing any two-level flow $e^{itX_{rs}}$ for all parameters $t \in \mathbb{R}$ and indices $1 \le r < s \le D$, and let $K_{\mathrm{diag}}$ denote a uniform depth bound for synthesizing an arbitrary diagonal unitary on $V$ (or arbitrary diagonal unitary within \(\mathrm{SU}(V)\) if  \(\mathfrak g\big|_V=\mathfrak{su}(V)\)). Then the following bounds hold:
\begin{enumerate}
    \item Any unitary operator $U$ on $V$ (or $U \in \mathrm{SU}(V)$ when $\mathfrak{g}\big|_V = \mathfrak{su}(D)$) can be synthesized with a global depth $L_{\mathrm{glob}}$ bounded by
    \begin{equation}
    \label{eq:general_full_unitary_bound_X}
    L_{\mathrm{glob}} \le \frac{D(D-1)}{2}\cdot \bigl(K_X + K_{\mathrm{diag}}\bigr) + K_{\mathrm{diag}}.
    \end{equation}

    \item From any computational basis state $\ket{e_r}$, an arbitrary normalized state $\ket{\phi} \in V$ can be synthesized, up to a global phase, with a circuit depth $L_{e_r \to \phi}$ bounded by
    \begin{equation}
    \label{eq:basis_orbit_generation_bound}
    L_{e_r \to \phi} \le (D-1)K_X + K_{\mathrm{diag}}.
    \end{equation}

    \item Any pair of normalized states $\ket{\psi}, \ket{\phi} \in V$ can be connected, up to a global phase, with a circuit depth $L_{\mathrm{state}}$ bounded by
    \begin{equation}
    \label{eq:uniform_state_to_state_bound}
    L_{\mathrm{state}} \le 2(D-1)K_X + 2K_{\mathrm{diag}}.
    \end{equation}
\end{enumerate}
\end{thm}

\begin{rmk}
\label{rmk:complexity_comparison}
Relating the system dimension to the DLA dimension via $D = \mathcal{O}(\sqrt{\dim(\mathfrak{g})})$, where $\dim(\mathfrak{g}) = D^2$ for $\mathfrak{u}(D)$ and $D^2 - 1$ for $\mathfrak{su}(D)$, and letting $\mathcal{K}_{\max} \coloneqq \max(K_X, K_{\mathrm{diag}})$ denote the maximum primitive synthesis cost bound, Table~\ref{tab:complexity_comparison} summarizes the asymptotic upper bounds on circuit depths for local versus global synthesis of unitary operators and quantum states.

Both local and global unitary synthesis scale directly with $\dim(\mathfrak{g})$, with their fundamental distinction lying in their operational depth parameters: local-vicinity unitary synthesis is governed by the generator stabilization depth $\mathcal{D}_{\mathbf{a}}$ (see \eqref{eq:worst_case_derivation}), whereas global unitary synthesis is governed by the primitive cost bound $\mathcal{K}_{\max}$. Similarly, for quantum state generation, local-vicinity synthesis is governed by the state orbit specific stabilization depth and local orbit dimension (see \eqref{eq:state_depth_bound}), whereas the global uniform bound scales as $\mathcal{K}_{\max}\sqrt{\dim(\mathfrak{g})}$.
\begin{table}[htbp]
\centering
\begin{tabular}{|c|c|c|}
\hline
\textbf{Regime} & \textbf{Unitary Synthesis} & \textbf{State-to-State} \\ \hline
\textbf{Local} & 
$\mathcal{O}(\mathcal{D}_{\mathbf{a}} \dim(\mathfrak{g}))$ & 
$\mathcal{O}(\mathcal{D}_{\mathbf{a},\psi} m_\psi)$  \\
\hline
\textbf{Global} & 
$\mathcal{O}(\mathcal{K}_{\max}\dim(\mathfrak{g}))$ & 
$\mathcal{O}( \mathcal{K}_{\max}\sqrt{\dim(\mathfrak{g})})$ \\ \hline
\end{tabular}
\caption{Asymptotic upper bounds on circuit depths for local versus global synthesis of unitary operators and quantum states via dynamical Lie algebra generator flows.}
\label{tab:complexity_comparison}
\end{table}
As a concrete illustration, Theorem~\ref{thm:MIS_connected_global_bound} establishes explicit global synthesis bounds for multi-angle MIS QAOA, where the primitive synthesis cost bound evaluates to $\mathcal{K}_{\max} = \mathcal{O}(D \Delta_\Gamma)$ with $\Delta_\Gamma$ denoting the toggle group diameter and $D = |\mathcal{F}|$ representing the feasible independent-set subspace dimension.
\end{rmk}

\subsection{Circuit-Level Synthesis via Standard Gate Decompositions}
\label{subsec:circuit_level_synthesis}

In certain regimes, the foregoing global synthesis approach can be further refined by leveraging standard circuit compilation architectures. Specifically, the universal unitary synthesis results of Shende et al.~\cite{shende2006synthesis} establish that an arbitrary operator in $\mathrm{SU}(2^n)$, up to an overall phase, can be implemented using $\mathcal{O}(4^n)$ CNOT gates and $\mathcal{O}(4^n)$ single-qubit rotations, while its state-preparation construction uses only \(\mathcal O(2^n)\) such gates.

Translating this gate-level decomposition into our framework allows us to bound global unitary and state-to-state transition depths directly in terms of the primitive native implementation costs for single-qubit gates ($K_{\mathrm{1q}}$) and CNOT gates ($K_{\mathrm{CX}}$).

\begin{thm}
\label{thm:uniform_orbit_generation2}
Let $V \cong (\mathbb{C}^2)^{\otimes q}$ be a $D = 2^q$-dimensional Hilbert space equipped with an orthonormal basis $\{\ket{e_1}, \dots, \ket{e_D}\}$ corresponding to an $n$-qubit system ($D = 2^n$). Suppose the restricted dynamical Lie algebra satisfies $\mathfrak{g}\big|_V = \mathfrak{u}(V)$ or $\mathfrak{su}(V)$, and assume that every primitive single-qubit gate is synthesized by a native circuit of length at most $K_{\mathrm{1q}}$ and every CNOT gate by one of length at most $K_{\mathrm{CX}}$. 

Then, every projective unitary $[U] \in \mathrm{PU}(V)$ can be synthesized with a circuit of depth
\begin{equation}
\label{eq:general_full_unitary_bound_circuit}
L_{\mathrm{glob}}^{\mathrm{proj}}
\le
\left(4^q - 3 \cdot 2^{q-1}\right) K_{\mathrm{1q}}
+
\left(
\frac{3}{4}4^q - \frac{3}{2}2^q
\right) K_{\mathrm{CX}}.
\end{equation}
Furthermore, any pair of normalized states $\ket{\psi}, \ket{\phi} \in V$ can be connected, up to a global phase, with a circuit of depth
\begin{equation}
\label{eq:uniform_state_to_state_bound_circuit}
L_{\mathrm{state}}
\le
\left(2^{q+2} - 4\right) K_{\mathrm{1q}}
+
\left(2^{q+2} - 4q - 4\right) K_{\mathrm{CX}}.
\end{equation}
\end{thm}

\subsection{Application to Pauli-Generated Dynamical Lie Algebras}
\label{subsec:pauli_generated_dlas}

Next, we specialize the preceding circuit-level synthesis results to the setting where the dynamical Lie algebra is generated by a finite collection of imaginary Pauli strings. This bridges the abstract bounds with explicit, constructive gate sequences implementable within variational quantum algorithms.

\begin{thm}
\label{thm:Pauli_global_synthesis}
Let \(W=(\mathbb{C}^2)^{\otimes n}\) be a qubit Hilbert space with $N=\dim(W)=2^n$, and suppose that the dynamical Lie algebra
\[
\mathfrak{g}=\langle iP_1,\ldots,iP_m\rangle_{\mathrm{Lie}},
\qquad
P_r\in\{I,X,Y,Z\}^{\otimes n},
\]
is generated by a finite collection of imaginary Pauli strings and satisfies $\mathfrak{g}=\mathfrak{u}(W) $ or $\mathfrak{g}=\mathfrak{su}(W)$. Let $\mathcal D_{\mathbf a}$ be the stabilization depth of the generating set $\mathbf a=\{iP_1,\ldots,iP_m\}$. Then every nonidentity Pauli flow on \(W\) can be synthesized using a variational quantum circuit of depth at most
\begin{equation}
\label{eq:general_Pauli_flow_cost}
K_P \le 2\mathcal D_{\mathbf a}-1.
\end{equation}
Furthermore, the following global bounds hold:
\begin{enumerate}
    \item Any unitary operator $U$ on $W$ (or $U \in \mathrm{SU}(W)$ when $\mathfrak{g} \simeq \mathfrak{su}(W)$) can be synthesized with a global depth $L_{\mathrm{glob}}$ bounded by
    \begin{equation}
    \label{eq:general_full_unitary_bound_circuitPauli}
    L_{\mathrm{glob}} \le 3(2\mathcal D_{\mathbf a}-1)\left(\frac{7}{4}\cdot4^{n}-3\cdot2^{n-1}
    \right).
    \end{equation}

    \item Any pair of normalized states $\ket{\psi}, \ket{\phi} \in W$ can be connected, up to a global phase, with a circuit depth $L_{\mathrm{state}}$ bounded by
    \begin{equation}
    \label{eq:uniform_state_to_state_bound_circuitPauli}
    L_{\mathrm{state}} 
    \le
    12 (2\mathcal D_{\mathbf a}-1)\left( 2^{n+1} - n - 2 \right) .
    \end{equation}
\end{enumerate}
\end{thm}

\begin{cor}
\label{cor:reduced_free_maxcut_K_n}
For the reduced multi-angle MaxCut QAOA on the complete graph $K_n$ ($n \ge 2$), the free reduced dynamical Lie algebra constitutes the full unitary algebra on the reduced Hilbert space (see Theorem~\ref{thm:reduced_free_DLA_filtration_classical}). Thus the reduced space carries \(q=n-1\) logical qubits, and
Theorem~\ref{thm:Pauli_global_synthesis} provides the explicit upper bounds:
\begin{align}
L_{\mathrm{glob}}^{\mathrm{proj}}
&\le 3(4n-3)\left(\frac{7}{4}\cdot 4^{n-1} - 3 \cdot 2^{n-1}\right), \\
L_{\mathrm{state}}
&\le 12(4n-3)\left(2^n - n - 1\right).
\end{align}
In particular, the corresponding asymptotic depth complexities scale as $\mathcal{O}(n 4^n)$ and $\mathcal{O}(n 2^n)$, respectively.
\end{cor}

\begin{rmk}
\label{rmk:unreduced_vs_reduced_DLAs}
Unlike their unreduced counterparts, symmetry-reduced free MaxCut QAOA instances often achieve full special-unitary, and hence projective operator, controllability on the reduced Hilbert space. In the unreduced framework, the free DLA for  \emph{archetypal}  graphs (connected graphs that are neither bipartite nor cyclic) is semisimple and isomorphic to $\mathfrak{su}(2^{n-1}) \oplus \mathfrak{su}(2^{n-1})$ \cite[Theorem~1]{KLFCCZ}; because it preserves the two invariant parity sectors, it fails to be transitive on the full $n$-qubit Hilbert space. By contrast, the symmetry-reduced framework eliminates this degeneracy, yielding a simple Lie algebra isomorphic to the full special unitary algebra $\mathfrak{su}(2^{n-1})$ on the $(n-1)$-qubit reduced space whenever the reduced graph is archetypal (see \cite[Theorem~IV.14]{TBFS}). Consequently, Theorem~\ref{thm:Pauli_global_synthesis} is directly applicable to these reduced systems once the  stabilization depth $\mathcal{D}_{\mathbf{a}}$ is determined.
\end{rmk}

\subsection{Global Synthesis for Free Flip-or-Stay MIS QAOA}
\label{subsubsec:MIS_global_matrix_units}

In this subsection, we apply the orbit-generation bounds from Theorem~\ref{thm:uniform_orbit_generation} to the free DLA for MIS QAOA on a connected graph $\Gamma = (V, E)$, obtaining concrete global bounds on circuit synthesis.

We retain the notation established in Section~\ref{subsec:MIS}. However, rather than working directly with the restricted free DLA $\mathfrak{g}_{\Gamma,\mathrm{free},\mathcal{F}}$ from \eqref{eq:ReducedGeneratorsFreeDLAs}, we analyze the closely related algebra
\begin{equation}
    \label{eq:ReducedGeneratorsFreeDLAs_hat}
    \widehat{\mathfrak{g}}_{\Gamma,\mathrm{free},\mathcal{F}} \coloneqq \Big\langle \Big\{ i \widehat{H}_{CX,v}\big|_{W_{\mathcal{F}}}, \; i Z_v\big|_{W_{\mathcal{F}}} \Big\}_{v \in V} \Big\rangle_{\mathrm{Lie}},
\end{equation}
where $\widehat{H}_{CX,v}\big|_{W_{\mathcal{F}}}$ denotes the restriction of the local controlled-X mixer operator to the feasible subspace $W_{\mathcal{F}}$. The action of $\widehat{H}_{CX,v}$ on a computational basis state $\ket{x}$ (for $x \in \{0,1\}^n$) is given by 
\begin{equation}
    \label{eq:HCXv_hat_action}
    \widehat{H}_{CX,v}\ket{x} \coloneqq H_{CX,v}\ket{x} + (I - \mathcal{P}_v)\ket{x} =
    \begin{cases}
        |x^{(v)}\rangle, & \text{if } x_w = 0 \text{ for all } w \in \mathcal{N}_v, \\[0.5ex]
        \ket{x}, & \text{if } x_w = 1 \text{ for some } w \in \mathcal{N}_v,
    \end{cases}
\end{equation}
where $x^{(v)}$ denotes the bitstring $x$ with the $v$-th bit flipped (cf.~\eqref{eq:HCXv_action} and \eqref{eq:CX_mixer}). As shown in \cite[Theorem~IV.2]{TBS}, this modification still results in the generation of the full unitary algebra:
\begin{equation}
\label{eq:algebra_isomorphism}
    \widehat{\mathfrak{g}}_{\Gamma,\mathrm{free}, \mathcal{F}} \cong \mathfrak{u}(W_{\mathcal{F}}).
\end{equation}

Throughout this subsection, \emph{native pulse} refers to the
Flip-or-Stay control family
\[
\widehat{\mathbf a}_{\mathrm{MIS}}
:=
\left\{
i\widehat H_{CX,v}\big|_{W_{\mathcal F}},
\,iZ_v\big|_{W_{\mathcal F}}
\;\middle|\;
v\in V
\right\}.
\]
Thus the bounds below are  stated in the hatted
Flip-or-Stay pulse alphabet. 

We next observe that each local mixer stabilizes the set $\mathcal{F}$ of feasible bitstrings, thereby inducing a well-defined combinatorial action directly on $\mathcal{F}$. Specifically, for each vertex $v \in V$, the local mixer defines a set-theoretic toggle map $\tau_v \colon \mathcal{F} \to \mathcal{F}$ given by
\begin{equation}
    \label{eq:set_toggle_def}
    \tau_v(x) \coloneqq 
    \begin{cases}
        x^{(v)}, & \text{if } x_w = 0 \text{ for all } w \in \mathcal{N}_v, \\[0.5ex]
        x, & \text{if } x_w = 1 \text{ for some } w \in \mathcal{N}_v.
    \end{cases}
\end{equation}
As $\tau_v$ either leaves a bitstring $x \in \mathcal{F}$ invariant or swaps an isolated pair $x \leftrightarrow x^{(v)}$, each $\tau_v$ acts as an involution on the feasible set $\mathcal{F}$:
\begin{equation}
    \label{eq:mixer_set_involution}
    \tau_v^2 = \mathrm{id}_{\mathcal{F}} \qquad \text{for all } v \in V.
\end{equation}
The collection of these local toggles generates a permutation subgroup of the symmetric group $\mathrm{Sym}(\mathcal{F})$, known as the \emph{independent-set toggle group}:
\begin{equation}
    \label{eq:toggle_group}
    \mathcal{T}(\Gamma) \coloneqq \langle \tau_v \rangle_{v \in V} \subseteq \mathrm{Sym}(\mathcal{F}).
\end{equation}

Importantly, the involutory nature of $\widehat{H}_{CX,v}\big|_{W_{\mathcal{F}}}$ (see \eqref{eq:mixer_set_involution}) implies that exponentiating the local mixer flow at parameter $t = \pi/2$ yields
\begin{equation}
\label{eq:toggle_pulse_identity}
e^{i\frac{\pi}{2} \widehat{H}_{CX,v}\big|_{W_{\mathcal{F}}}} = \cos\left(\frac{\pi}{2}\right) I_{W_{\mathcal{F}}} + i \sin\left(\frac{\pi}{2}\right) \widehat{H}_{CX,v}\big|_{W_{\mathcal{F}}} = i \widehat{H}_{CX,v}\big|_{W_{\mathcal{F}}}.
\end{equation}
Thus, up to an overall global phase factor $i$, executing the action of a single discrete toggle permutation $\tau_v$ on $W_{\mathcal{F}}$ requires exactly one native continuous pulse evolution under the Hamiltonian $\widehat{H}_{CX,v}\big|_{W_{\mathcal{F}}}$.

To quantify the pulse complexity of synthesizing arbitrary permutations within the toggle group $\mathcal{T}(\Gamma) \coloneqq \langle \tau_v \rangle_{v \in V} \subseteq \mathrm{Sym}(\mathcal{F})$, we define the toggle group diameter $\Delta_\Gamma$ as the diameter of its Cayley graph with respect to the generating set in \eqref{eq:toggle_group}:
\begin{equation}
\label{eq:MIS_toggle_diameter_general}
\Delta_\Gamma \coloneqq \operatorname{diam} \operatorname{Cay}\bigl( \mathcal{T}(\Gamma), \{\tau_v \mid v \in V\} \bigr).
\end{equation}
Consequently, every permutation $g \in \mathcal{T}(\Gamma)$ admits a word decomposition $g = \tau_{v_k} \cdots \tau_{v_1}$ of length $k \le \Delta_\Gamma$. On the Hilbert space $W_{\mathcal{F}}$, the corresponding unitary operator $\widehat{g} \in \mathrm{U}(W_{\mathcal{F}})$ is synthesized as
\begin{equation}
\label{eq:toggle_circuit_synthesis}
\widehat{g} = \widehat{H}_{CX,v_k} \cdots \widehat{H}_{CX,v_1} = e^{-i \frac{k\pi}{2}} \prod_{j=0}^{k-1} e^{i\frac{\pi}{2} \widehat{H}_{CX,v_{k-j}}},
\end{equation}
yielding a continuous mixer circuit implementation with native pulse depth not exceeding $\Delta_\Gamma$.

\begin{thm}
\label{thm:MIS_connected_global_bound}
Let $\Gamma = (V,E)$ be a connected graph with feasible independent-set subspace $W_{\mathcal{F}}$ of dimension $D = |\mathcal{F}| \ge 4$, and let $\Delta_\Gamma$ denote the toggle group diameter defined in \eqref{eq:MIS_toggle_diameter_general}. Then:
\begin{enumerate}
    \item Every unitary \(U\in\mathrm{U}(W_{\mathcal F})\) admits an exact native Flip-or-Stay pulse synthesis with total depth \(L_{\mathrm{glob}}(\Gamma)\) bounded by 
    \begin{equation}
    \label{eq:MIS_connected_global_bound}
    L_{\mathrm{glob}}(\Gamma) \le  \Delta_\Gamma(D-1)(4D^2-3D+2)+2D^3-D^2.
    \end{equation}
    \item  For any pair of normalized quantum states $|\psi\rangle, |\phi\rangle \in W_{\mathcal{F}}$, there exists a native continuous pulse circuit connecting $|\psi\rangle$ to $|\phi\rangle$ (up to a global phase) with total depth $L_{\mathrm{state}}(\Gamma)$ bounded by
    \begin{equation}
    \label{eq:MIS_connected_global_state_to_state_bound}
    L_{\mathrm{state}}(\Gamma) \le 12D^2\Delta_\Gamma + 6D^2 - 16D\Delta_\Gamma + 4\Delta_\Gamma - 4.
    \end{equation}
\end{enumerate}
In particular, as functions of the feasible dimension $D = |\mathcal{F}|$ and toggle diameter $\Delta_\Gamma$, the global depth complexities satisfy the asymptotic upper bounds:
\begin{equation}
\label{eq:MIS_connected_global_asymptotic}
L_{\mathrm{glob}}(\Gamma) = \mathcal{O}\left( D^3 \Delta_\Gamma \right), \qquad L_{\mathrm{state}}(\Gamma) = \mathcal{O}\left(D^2\Delta_\Gamma \right).
\end{equation}
\end{thm}

\begin{rmk}
    \label{rmk:local_vs_global_comparison}
    It is instructive to compare these global bounds with the local-vicinity depth complexities summarized in Table~\ref{tab:examples_depths}. Specifically, while global unitary and state-to-state synthesis scale as $\mathcal{O}(D^3 \Delta_\Gamma)$ and $\mathcal{O}(D^2 \Delta_\Gamma)$, respectively, local synthesis bounds are governed by the DLA stabilization depth and orbit dimensions—such as $\mathcal{O}(\mathcal{D}_{\mathbf{a}} D^2)$ for local unitaries and $\mathcal{O}(\mathcal{D}_{\mathbf{a},\psi} m_\psi)$ for local states. This trade-off underscores how global pulse schedules incorporate an explicit topological dependence on the toggle group diameter $\Delta_\Gamma$.
\end{rmk}

To see how these theoretical bounds manifest in a concrete setting, we next analyze the complete graph family.

\begin{ex}
\label{ex:complete_graph_toggle_diameter}
Let $\Gamma = K_n$ be the complete graph on $n\ge 3$ vertices. The feasible independent sets consist of the empty set and all singletons:
\begin{equation}
\mathcal{F}(K_n) = \bigl\{ \varnothing, \{1\}, \ldots, \{n\} \bigr\}, \qquad D \coloneqq |\mathcal{F}(K_n)| = n + 1.
\end{equation}
Identifying the empty set $\varnothing$ with the symbol $0$ and each singleton $\{v\}$ with the symbol $v \in \{1, \dots, n\}$, the feasible set maps bijectively to $\{0, 1, \dots, n\}$. 

In the complete graph \(K_n\), for any vertex \(v \in V(K_n)\), every independent set distinct from \(\varnothing\) and \(\{v\}\)  contains a vertex adjacent to \(v\). Therefore, the toggle group element \(\tau_v\) associated with vertex \(v\) acts as a simple transposition that swaps the empty set, encoded by \(0\), with the singleton set, encoded by \(v\):
\begin{equation}
\tau_v = (0 \, v) \in \mathrm{Sym}(\mathcal{F}(K_n)), \qquad 1 \le v \le n.
\end{equation}
Consequently, the independent-set toggle group is generated by the following transpositions:
\begin{equation}
\mathcal{T}(K_n) = \big\langle (0 \, 1), (0 \, 2), \ldots, (0 \, n) \big\rangle \cong S_{n+1}.
\end{equation}
The corresponding toggle Cayley graph
\begin{equation}
\operatorname{Cay}\Big( \mathcal{T}(K_n), \{\tau_1, \ldots, \tau_n\} \Big) \cong \mathrm{ST}_{n+1}
\end{equation}
is the classical star graph $\mathrm{ST}_{n+1}$ on $n+1$ symbols. By a well-known result of Akers and Krishnamurthy \cite[Theorem~7]{akers1989group}, the exact diameter of the star graph $\mathrm{ST}_{n+1}$ is
\begin{equation}
\label{eq:complete_graph_toggle_diameter}
\Delta_{K_n} = \left\lfloor \frac{3(n+1-1)}{2} \right\rfloor = \left\lfloor \frac{3n}{2} \right\rfloor = \mathcal{O}(n).
\end{equation}

Substituting $D = n + 1$ and $\Delta_{K_n} = \mathcal{O}(n)$ into Theorem~\ref{thm:MIS_connected_global_bound} yields the following explicit polynomial circuit-depth scaling for free MIS QAOA on complete graphs:
\begin{equation}
\label{eq:complete_graph_generic_global_bound}
L_{\mathrm{glob}}(K_n) = \mathcal{O}(n^4), \qquad L_{\mathrm{state}}(K_n) = \mathcal{O}(n^3).
\end{equation}
\end{ex}

\section{Finite-depth Exposure of Descent Directions}
\label{subsec:finite_depth_descent}

The finite-depth preparation results established in the preceding sections guarantee the existence of parametrized circuits that reproduce elements within a neighborhood $\mathcal{V}_{[\psi]}$ of a ray $[\psi]$ in the projective orbit $\mathcal O_{[\psi]}$ (see Theorem~\ref{thm:local_state_reachability}). However, they do not provide an explicit mechanism for determining concrete parameter values that realize specific states in this neighborhood. In this section, we take a first step toward resolving this constructive problem by connecting parameter identification to the infinitesimal framework developed in Section~\ref{sec:infinitesimal}. Specifically, we show that any finite-depth circuit preparing a non-optimal state can be augmented by a bounded-length identity block, ensuring that the expanded parametrization explicitly exhibits either a first-order descent direction for the loss function or a direction of negative curvature.

Before presenting our main result, we formalize the framework for circuit extensions. Let $\ket{\xi}$ be an initial reference state, and let 
\begin{equation}
    \rho_p = \lvert\psi(\boldsymbol{\theta}_p)\rangle \langle\psi(\boldsymbol{\theta}_p)\rvert
\end{equation}
be the density matrix corresponding to the state $\lvert\psi(\boldsymbol{\theta}_p)\rangle = U(\boldsymbol{\theta}_p)\lvert\xi\rangle$ prepared by a depth-$p$ parameterized quantum circuit $U(\boldsymbol{\theta}_p)$. Throughout the remainder of this section, we assume that the observable (problem Hamiltonian) $H_P$ satisfies $iH_P \in \mathfrak{g}$, where $\mathfrak{g}$ is the dynamical Lie algebra, and that the associated Lie group $G = \exp(\mathfrak{g})$ is compact.

An \emph{identity block} $V(\boldsymbol{x})$ is a parameterized unitary operation satisfying $V(\boldsymbol{0}) = I$, generated by a sequence of at most $\mathcal{T}_{\mathfrak{g}}$ elementary evolutions under generators of $\mathfrak{g}$. Appending $V(\boldsymbol{x})$ yields the augmented loss function
\begin{equation}
    \label{eq:loss_function}
    \ell(\boldsymbol{x}, \boldsymbol{\theta}_p) := \operatorname{Tr}\!\left[ V(\boldsymbol{x}) \rho_p V^\dagger(\boldsymbol{x}) H_P \right],
\end{equation}
which recovers the baseline energy at the origin: $\ell(\boldsymbol{0}, \boldsymbol{\theta}_p) = \ell(\boldsymbol{\theta}_p)$.

The theorem below formalizes the finite-depth exposure guarantee for arbitrary non-optimal states. For complete derivations and technical proofs, we refer the reader to Appendix~\ref{sec:finite_depth_descent_proofs}.

\begin{thm}[Finite-depth exposure of descent directions]
\label{thm:finite_depth_exposure_descent}
Assume that the $G$-orbit $\mathcal O_\psi$ of the initial state $\ket{\psi}$ intersects the ground eigenspace of $H_P$, and suppose the energy prepared by the current circuit is strictly non-optimal:
\[
    \ell_{\boldsymbol\theta_p}(\rho, H_P) > \lambda_{\min}(H_P),
\]
where $\rho = \ket{\psi}\bra{\psi}$ is the initial state density matrix and  $\ell_{\boldsymbol\theta_p}(\rho, H_P) = \operatorname{Tr}\!\left[ U(\boldsymbol\theta_p) \rho U^\dagger(\boldsymbol\theta_p) H_P \right]$ denotes the current expectation value. Then, there exists an identity block $V(\boldsymbol{x})$ requiring at most $\mathcal{T}_{\mathfrak{g}}$ elementary generator evolutions such that after appending $V(\boldsymbol{x})$, exactly one of the following holds for the augmented loss function $\ell_{(\boldsymbol{x}, \boldsymbol{\theta}_p)}(\rho, H_P)$ associated with the extended unitary $U(\boldsymbol{x}, \boldsymbol{\theta}_p) =  V(\boldsymbol{x})U(\boldsymbol{\theta}_p)$:

\begin{enumerate}
    \item \textbf{First-Order Descent:} The parameter gradient at the origin is non-vanishing,
    \begin{equation}
    \label{eq:nonzero_parameter_gradient}
        \left. \nabla_{\boldsymbol x} \ell_{(\boldsymbol x, \boldsymbol\theta_p)}(\rho, H_P) \right|_{\boldsymbol x = \boldsymbol 0} \neq \boldsymbol 0.
    \end{equation}

    \item \textbf{Second-Order Escape:} The gradient vanishes at $\boldsymbol x = \boldsymbol 0$, but there exists a direction $\boldsymbol v \in \mathbb R^m$ exhibiting negative second-order curvature:
    \begin{equation}
    \label{eq:negative_parameter_curvature}
        \left. \frac{d^2}{dt^2} \right|_{t=0} \ell_{(t\boldsymbol v, \boldsymbol\theta_p)}(\rho, H_P) < 0.
    \end{equation}
\end{enumerate}

Consequently, every neighborhood of $\boldsymbol 0$ contains parameters $\boldsymbol x$ yielding strict energy improvement:
\begin{equation}
\label{eq:strict_improvement_after_extension}
    \ell_{(\boldsymbol x, \boldsymbol\theta_p)}(\rho, H_P) < \ell_{\boldsymbol\theta_p}(\rho, H_P).
\end{equation}
The identity block extension requires at most $\mathcal T_{\mathfrak g}$ elementary generator evolutions.
\end{thm}



\begin{rmk}[Computational complexity of the escape block]
\label{rmk:escape_block_complexity}
Recall that the identity block depth satisfies  $\mathcal T_{\mathbf{a}} = \mathcal{O}(\mathcal{D}_{\mathbf{a}} \dim \mathfrak{g}) \le\mathcal O(\dim(\mathfrak g)^2)$ (see Theorem~\ref{thm:local_synthesis_depth_bound}), where $\mathcal{D}_{\mathbf{a}}$ is the filtration depth of the generator pool. Consequently, for Lie-algebraic architectures with polynomially bounded DLA dimension $\dim(\mathfrak g) = \operatorname{poly}(n)$,  Theorem~\ref{thm:finite_depth_exposure_descent} guarantees that any suboptimal stationary point can be escaped by appending a local identity block of depth at most $\operatorname{poly}(n)$.
\end{rmk}

\subsection{Relation to Existing Landscape and Circuit Extension Results}
\label{sec:comparison_overparameterization}

Theorem~\ref{thm:finite_depth_exposure_descent} connects naturally to existing strategies for escaping local minima in Quantum Approximate Optimization Algorithm landscapes, adaptive control architectures, and quantum optimization theory.

\textbf{Comparison with QAOA Layer-Extension Frameworks.}
A prominent approach to breaking out of suboptimal stationary points involves inserting additional gates into an existing circuit. Sack, Medina, Kueng, and Serbyn \cite{sack2023recursive} start from a local minimum of a depth-$p$ QAOA circuit energy function and insert two zero-angle gates in $2p+1$ prescribed ways. The resulting depth-$(p+1)$ circuits prepare the same state and act as stationary points of the enlarged loss function. When the enlarged Hessians are non-degenerate, each of these $2p+1$ stationary points possesses exactly one negative Hessian eigenvalue, exposing a direction of strict energy improvement \cite[Theorems~1,2]{sack2023recursive}. Medina and Serbyn \cite{medina2024recursive} subsequently refined this mechanism by constructing an explicit escape vector without needing Hessian diagonalization \cite[Sec.~III.A]{medina2024recursive}.

\textbf{Key Distinctions and Generality.}
While sharing the operational intuition of identity-initialized circuit extensions ($V(\boldsymbol{0}) = I$), Theorem~\ref{thm:finite_depth_exposure_descent} extends beyond these QAOA-specific results in several fundamental ways:

\begin{enumerate}
    \item \textbf{Generality of States and Architectures.} 
   Theorem~\ref{thm:finite_depth_exposure_descent} places no structural restrictions on the baseline ansatz $U(\boldsymbol{\theta}_p)$ and applies to \emph{any} reachable state $\rho_p$ satisfying $\ell_{\boldsymbol{\theta}_p}(\rho_p, H_P) > \lambda_{\min}(H_P)$ (i.e., not a global minimum). It does not require the baseline parameters $\boldsymbol{\theta}_p$ to form a non-degenerate local minimum; the conclusion also covers arbitrary stationary points, thereby guaranteeing that no suboptimal stationary state traps the optimization process at finite depth.

    \item \textbf{Local Parameter Isolation.} Strict local improvement $\ell_{(\boldsymbol{x}, \boldsymbol{\theta}_p)}(\rho, H_P) < \ell_{\boldsymbol{\theta}_p}(\rho, H_P)$ is achieved by varying \emph{only} the newly appended parameters $\boldsymbol{x}$ near $\boldsymbol{0}$, leaving the trained baseline parameters $\boldsymbol{\theta}_p$ entirely untouched.

    \item \textbf{Extension Depth vs.\ Structural Scope.} Whereas QAOA extensions insert two zero-angle gates to form a single extra layer, our identity block has length $L \le \mathcal T_{\mathfrak g} = \mathcal{O}(\mathcal{D}_{\mathbf{a}} \dim \mathfrak{g})$ built from the full control pool. Allowing an extension block longer than a single QAOA layer buys universal applicability across arbitrary Lie-algebraic control spaces $\mathfrak g$.

\end{enumerate}

\section{Summary and Outlook}
\label{sec:conclusion}

As this work connects quantum control theory, dynamical Lie algebras, orbit geometry, and quantum-circuit synthesis across several variational architectures, we organize the summary around the central questions addressed by our results. This question-and-answer format also clarifies their broader implications and the problems that remain open.

\medskip
\noindent\textit{What is the correct notion of reachability for a variational circuit?}
The accessible states form the dynamical group orbit of the initial state, rather than the entire cyclic invariant subspace generated by the initial state under the DLA action. This distinction can be substantial: an orbit of only polynomial dimension may contain exponentially many linearly independent optima. Orbit geometry therefore provides both the appropriate reachability criterion and a precise description of its limitations.

\medskip
\noindent\textit{When does asymptotic reachability imply exact
finite-depth preparation?}
We show that this implication holds whenever the relevant projective orbit is closed. In particular, a nondegenerate eigenstate of one of the initial control operators has a closed projective orbit even when the full dynamical group is not closed inside the ambient unitary group. For QAOA employing an irreducible stoquastic mixer, this converts asymptotic convergence  into exact ground state preparation at some finite depth. 

\medskip
\noindent\textit{Which circuit depth is sufficient?}
Locally, the commutator filtration translates the rate at which native controls generate new Lie-algebra and orbit directions into explicit bounds on the required number of native pulses. Globally, when the restricted DLA is full or special unitary, we develop complementary constructions based on diagonal and two-level synthesis, alongside logical-gate compilation. The Pauli-generated DLA case links these global costs directly to the filtration depth, whereas the MIS toggle construction provides a fully explicit realization from the original controls. Together, these results elevate qualitative controllability into quantitative resource guarantees, while clearly distinguishing the cost of arbitrary unitary synthesis from the typically more efficient state-to-state transfer.

\medskip
\noindent\textit{What are the implications for variational
optimization?}
Under the stated compactness and reachable-ground state assumptions, the local synthesis construction produces a bounded-depth, identity-initialized circuit extension block that exposes either first-order descent or a negative-curvature direction at every suboptimal circuit output. Thus, the algebra and geometry of the reachable orbit constrain not only expressivity but also the local optimization landscape.

\medskip
\noindent\textit{What remains open?}
The principal questions are quantitative and algorithmic:
how tight the depth bounds are, how large the local synthesis
neighborhoods can be chosen, and how the required circuit parameters
can be found efficiently. Incorporating hardware connectivity,
locality, and noise may lead to substantially sharper
architecture-dependent bounds. Resolving these questions would bring
the structural guarantees developed here closer to practical
finite-depth circuit design and optimization.

\bibliographystyle{plain}
\bibliography{References}

\appendix
\section{Proofs of the Results in Sections \ref{sec:reachability_2} and \ref{sec:performance_guarantees}}
\label{sec:proofs_reachability_2_and_performance_guarantees}

This section collects the technical proofs for the theoretical results presented in Sections~\ref{sec:reachability_2} and~\ref{sec:performance_guarantees}. We begin by establishing the classification of projectively transitive Lie algebras (Theorem~\ref{thm:projective_transitivity_classification}), followed by the topological closedness criteria for dynamical group orbits.

\begin{proof}[Proof of Theorem \ref{thm:projective_transitivity_classification}]

Projective transitivity implies that the representation of \(\mathfrak{g}\) on \(V\) is irreducible: any proper nonzero invariant subspace would define a proper invariant projective subvariety containing the orbit, contradicting transitivity. Moreover, since \(V\) $V$ is an irreducible representation of a reductive Lie algebra $\mathfrak{g}$, Schur's lemma implies that the center $\mathfrak{z}(\mathfrak{g})$ acts on $V$ via scalar operators, and thus acts trivially on the projective space $\mathbb{P}(V)\cong \mathbb{CP}^{d-1}$. Consequently, the effective action on $\mathbb{P}(V)$ factors through the semisimple commutator ideal $\mathfrak{g}_{\mathrm{ss}} = [\mathfrak{g}, \mathfrak{g}]$. In turn, the connected analytic subgroup $G_{\mathrm{ss}} \subset \mathrm{U}(d)$ corresponding to $\mathfrak{g}_{\mathrm{ss}}$ is closed and compact (see, e.g., \cite[Proposition~7.9]{Knapp}).

By the classification of connected compact linear groups acting transitively on complex projective space \cite[Table~1]{Oni}, any connected compact subgroup of $\mathrm{U}(d)$ acting transitively on $\mathbb{CP}^{d-1}$ has a Lie algebra that is, up to unitary conjugation, one of the following:

\[
\mathfrak{su}(d),\,
\mathfrak{u}(d),
\]
or, when $d=2m$,
\[
\mathfrak{sp}(m),\,
\mathfrak{sp}(m)\oplus\mathfrak u(1),
\]
with the indicated defining representations. This establishes the necessity of the listed Lie algebras.

Conversely, $\mathrm{SU}(d)$ and $\mathrm{U}(d)$ act transitively on the unit sphere in $\mathbb{C}^d$, and consequently transitively on $\mathbb{CP}^{d-1}$. Similarly, the compact symplectic group $\mathrm{Sp}(m)$ acts transitively on the unit sphere of its defining representation $\mathbb{C}^{2m}$, and thus transitively on the corresponding projective space $\mathbb{CP}^{2m-1}$. Adjoining the central scalar factor $\mathrm{U}(1)$ leaves the projective orbits unchanged. Therefore, all listed Lie algebras admit transitive actions on complex projective space.

\end{proof}

\begin{proof}[Proof of Theorem \ref{thm:closed_dynamical_lie_group}]
Let
\(
K:=\overline{G}
\)
denote the topological closure of $G$ in $\mathrm{U}(W)$, and let $\mathfrak{k}$ be its Lie algebra. Since $G$ is connected, $K$ is a compact, connected Lie group. 

Since $G$ is generated by the exponentials of elements in $\mathfrak{g}$, we have $\operatorname{Ad}_g(\mathfrak{g}) = \mathfrak{g}$ for all $g \in G$. By the density of $G$ in $K$ and the continuity of the adjoint action, it follows that
\[
\operatorname{Ad}_U(\mathfrak g)=\mathfrak g
\qquad
\text{for every }U\in K.
\]

Differentiating this relation yields $[\mathfrak{k}, \mathfrak{g}] \subseteq \mathfrak{g}$, establishing that $\mathfrak{g}$ is an ideal of $\mathfrak{k}$.

By the standard structure theory of compact Lie algebras, $\mathfrak{k}$ decomposes as
\[
\mathfrak{k} = \mathfrak{z} \oplus \mathfrak{k}_1 \oplus \cdots \oplus \mathfrak{k}_r,
\]
where $\mathfrak{z} = Z(\mathfrak{k})$ is the center of $\mathfrak{k}$ and each $\mathfrak{k}_\ell$ is a compact simple ideal. Since $\mathfrak{g} \trianglelefteq \mathfrak{k}$ is an ideal, it respects this decomposition, meaning there exist a central subspace $\mathfrak{c} \subseteq \mathfrak{z}$ and an index subset $S \subseteq \{1, \ldots, r\}$ such that
\begin{equation}
\label{eq:ideal_decomposition_in_closure}
    \mathfrak{g} = \mathfrak{c} \oplus \bigoplus_{\ell \in S} \mathfrak{k}_\ell.
\end{equation}

The analytic subgroup $G_{\mathrm{ss}}$ corresponding to the semisimple ideal
\[
\mathfrak{g}_{\mathrm{ss}} := \bigoplus_{\ell \in S} \mathfrak{k}_\ell
\]
is closed and compact (see, e.g., \cite[Proposition~7.9]{Knapp}). Consequently, it suffices to show that the analytic subgroup associated with the central subspace $\mathfrak{c}$ is also closed.

Consider the canonical projection onto the abelianization,
\[
\pi \colon K \longrightarrow K / [K, K].
\]
Since $K$ is compact and connected, $K / [K, K]$ is a compact torus. At the Lie algebra level, bracket terms vanish under the differential $d\pi$. As $\{iH_1, \ldots, iH_k\}$ generate $\mathfrak{g}$, we have
\begin{equation}
\label{eq:abelianized_DLA}
    d\pi(\mathfrak{g}) = \operatorname{span}_{\mathbb{R}} \bigl\{ d\pi(iH_1), \ldots, d\pi(iH_k) \bigr\}.
\end{equation}

Let
\[
\Lambda := \ker \!\left( \exp \colon \operatorname{Lie}\bigl(K / [K, K]\bigr) \longrightarrow K / [K, K] \right)
\]
be the integral lattice of this torus. From the periodicity condition in \eqref{eq:generator_periodicity}, we obtain
\[
\exp \bigl( \tau_j d\pi(iH_j) \bigr) = \pi \bigl( e^{i \tau_j H_j} \bigr) = I,
\]
which implies $\tau_j d\pi(iH_j) \in \Lambda$ for each $j$. Thus, $d\pi(\mathfrak{g})$ is spanned over $\mathbb{R}$ by vectors in the lattice $\Lambda$, making it a rational subspace of the toral Lie algebra $\operatorname{Lie}(K / [K, K])$. It consequently integrates to a closed subtorus of $K / [K, K]$.

Equivalently, the central subalgebra $\mathfrak{c}$ in \eqref{eq:ideal_decomposition_in_closure} integrates to a closed torus $T_{\mathfrak{c}} \subseteq K$. Denoting by $G_{\mathrm{ss}}$ the closed compact subgroup associated with $\mathfrak{g}_{\mathrm{ss}}$, the product
\[
G_{\mathrm{ss}} T_{\mathfrak{c}}
\]
is a compact, connected subgroup of $K$ with Lie algebra $\mathfrak{g}_{\mathrm{ss}} \oplus \mathfrak{c} = \mathfrak{g}$. By uniqueness of the connected analytic subgroup associated with a given Lie subalgebra, it follows that
\[
G = G_{\mathrm{ss}} T_{\mathfrak{c}}.
\]
Being the product of two compact subgroups, $G$ is compact and therefore closed in $\mathrm{U}(W)$.

\end{proof}


\begin{proof}[Proof of Theorem \ref{thm:closure_to_finite_word_reachability}]
Under the first additional assumption, $\pi(G)$ is a closed subgroup of the compact group $\mathrm{PU}(W)$, and hence its orbit $G\cdot[s]=\pi(G)\cdot[s]$ is closed:

\[
    \overline{G\cdot[s]} = G\cdot[s].
\]
Combined with \eqref{eq:target_in_projective_orbit_closure}, this implies $[t]\in G\cdot[s]$.

Next, we verify the sufficiency of the second assumption. We first show that this nondegeneracy condition guarantees that the projective orbit $G\cdot[s]$ is closed, even though $\pi(G)$ itself need not be closed.

Since $\mathfrak{g}$ is a reductive Lie subalgebra of the compact Lie algebra $\mathfrak{u}(W)$, it decomposes into semisimple and central components:
\[
\mathfrak{g} = \mathfrak{g}_{\mathrm{ss}} \oplus \mathfrak{z}, \qquad \mathfrak{g}_{\mathrm{ss}} = [\mathfrak{g},\mathfrak{g}], \qquad \mathfrak{z} = Z(\mathfrak{g}).
\]
Let $G_{\mathrm{ss}}$ and $Z_0$ denote the connected analytic subgroups corresponding to $\mathfrak{g}_{\mathrm{ss}}$ and $\mathfrak{z}$, respectively. At the group level,
\[
G = G_{\mathrm{ss}}Z_0,
\]
where $G_{\mathrm{ss}}$ and $Z_0$ commute.

For any $A \in \mathfrak{z}$, the centrality of $A$ together with the inclusion $iH_r \in \mathfrak{g}$ implies
\[
[A,H_r] = 0.
\]
Consequently, if $H_r\ket{s} = \lambda\ket{s}$, then
\[
H_r A\ket{s} = A H_r\ket{s} = \lambda A\ket{s}.
\]
Thus, $A\ket{s}$ belongs to the $\lambda$-eigenspace of $H_r$. As this eigenspace is assumed to be one-dimensional, there exists a real scalar $\alpha_A \in \mathbb{R}$ such that
\[
A\ket{s} = i\alpha_A\ket{s},
\]
which upon exponentiation yields
\[
e^A\ket{s} = e^{i\alpha_A}\ket{s}.
\]
Therefore, the connected central subgroup $Z_0$ acts on $[s]$ purely by phase multiplication, leaving the projective class $[s]$ invariant.

Furthermore, as $Z_0$ commutes with $G_{\mathrm{ss}}$, for every $g_{\mathrm{ss}}\in G_{\mathrm{ss}}$ and $z\in Z_0$ there exists a phase $\theta(z)\in\mathbb{R}$ satisfying
\[
z g_{\mathrm{ss}}\ket{s} = g_{\mathrm{ss}} z\ket{s} = e^{i\theta(z)} g_{\mathrm{ss}}\ket{s}.
\]
It follows that the full projective orbit reduces to the semisimple orbit:
\begin{equation}
\label{eq:PF_orbit_semisimple_reduction}
G\cdot[s] = G_{\mathrm{ss}}\cdot[s].
\end{equation}

Since its Lie algebra $\mathfrak{g}_{\mathrm{ss}}$ is compact and semisimple, the connected analytic subgroup $G_{\mathrm{ss}}$ is compact \cite[Proposition~7.9]{Knapp}. Being the continuous image of a compact group, the orbit $G_{\mathrm{ss}} \cdot [s]$ is compact, and therefore closed in $\mathbb{P}(W)$. Together with \eqref{eq:PF_orbit_semisimple_reduction}, this establishes the statement in \eqref{eq:exact_projective_finite_word}.

\end{proof}

\begin{cor}
\label{cor:continuous_objective_attainment}
Under either hypothesis of Theorem~\ref{thm:closure_to_finite_word_reachability}, let
\[
    F \colon \mathbb{P}(W) \longrightarrow \mathbb{R}
\]
be a continuous function. Then both
\[
    \inf_{[v] \in G \cdot [s]} F([v])
    \qquad \text{and} \qquad
    \sup_{[v] \in G \cdot [s]} F([v])
\]
are attained by finite parametrized circuits generated by \(H_1, \ldots, H_k\).
\end{cor}

\begin{proof}
Under either hypothesis, the orbit \(G \cdot [s]\) is a closed subset of the compact space \(\mathbb{P}(W)\), and is therefore compact. Consequently, the continuous function \(F\) attains its minimum and maximum on this orbit. By Theorem~\ref{thm:closure_to_finite_word_reachability}, every point of the orbit is realized by a finite word in the generating one-parameter subgroups.
\end{proof}

\begin{cor}
\label{cor:exact_ground_energy_attainment}
Let \(C\) be a Hamiltonian on \(W\), and suppose that the hypotheses of Theorem~\ref{thm:closure_to_finite_word_reachability} hold for \(\ket{s}\). If
\begin{equation}
\label{eq:asymptotic_ground_energy}
    \inf_{[v] \in G \cdot [s]}\bra{v} C \ket{v} = \lambda_{\min}(C),
\end{equation}
then a ground state of \(C\) can be prepared (up to a global phase) by a finite parametrized circuit generated by \(H_1, \ldots, H_k\).
\end{cor}

\begin{proof}
We apply Corollary~\ref{cor:continuous_objective_attainment} to the Rayleigh quotient of \(C\), which defines a continuous function on \(\mathbb{P}(W)\). A state vector achieves the value \(\lambda_{\min}(C)\) if and only if it belongs to the ground eigenspace of \(C\). Since the infimum is attained on the orbit, the corresponding ground state is reachable via a finite parametrized circuit.
\end{proof}

We conclude this section with the proof of Corollary~\ref{cor:projectively_semisimple_exact_reachability}.

\begin{proof}[Proof of Corollary~\ref{cor:projectively_semisimple_exact_reachability}]
The connected subgroup corresponding to the derived Lie algebra \([\mathfrak{g}, \mathfrak{g}]\) is compact, whereas the central summand \(\mathfrak{c}\) acts trivially on the projective space \(\mathbb{P}(W)\). Consequently, every projective orbit is compact and, hence, closed. The result follows directly from Theorem~\ref{thm:closure_to_finite_word_reachability}.
\end{proof}

Finally, Theorem~\ref{thm:infimum_attainability} is established by noting that, since $iH_P$ is in the dynamical Lie algebra $\mathfrak{g}$, it commutes with its center $\mathfrak{z}(\mathfrak{g})$; hence, the center acts trivially on the expectation value:
\[
\operatorname{Tr}(H_P g \rho_s g^\dagger) 
= \operatorname{Tr}(g H_P \rho_s g^\dagger) 
= \operatorname{Tr}(H_P \rho_s) 
\quad \text{for any } g \in \exp(\mathfrak{z}(\mathfrak{g})).
\]
As the connected analytic subgroup $G_{\mathrm{ss}}$ corresponding to the semisimple part of the DLA is compact, the statement follows (see the proof of Theorem~\ref{thm:closure_to_finite_word_reachability}).

\section{A Fixed-Particle \texorpdfstring{$XY$}{XY} Family with Exponentially Many Exactly Reachable Optima}
\label{app:fixed_particle_XY}

In this section, we expand upon the construction presented in Section~\ref{subsec:fixed_particle_XY_example}, providing the requisite algebraic background and rigorously establishing properties stated therein.

In the computational basis of the $k$-excitation sector $W_k$, the restricted mixer Hamiltonian $H_M\big|_{W_k}$ coincides with the negative adjacency matrix of the $k$-token graph $F_k(P_n)$ of the path graph $P_n$. The vertices of $F_k(P_n)$ correspond to $k$-element subsets $S \subseteq \{1,\ldots,n\}$, with an edge connecting two subsets $S$ and $T$ whenever $T$ is obtained from $S$ by hopping a single excitation across an edge of $P_n$. Because the underlying path $P_n$ is connected, its token graph $F_k(P_n)$ is connected as well. Consequently, $H_M\big|_{W_k}$ is a real symmetric, irreducible matrix whose off-diagonal entries are all nonpositive ($0$ or $-1$).

By the Perron--Frobenius theorem, its lowest eigenvalue is non-degenerate, guaranteeing a unique normalized ground state $\ket{\xi_k}$ with strictly positive coefficients in the computational basis. Recall that, when restricted to the $k$-excitation sector $W_k$, the ground space $W_{P,\min}$ of the problem Hamiltonian $H_P = \frac{I-Z_1}{2}$ is spanned by all $k$-particle computational basis states corresponding to configurations in which the first site is unoccupied (see~\eqref{eq:fixed_weight_problem_ground_space}).


Let $\{e_1, \ldots, e_n\}$ denote the standard basis of $\mathbb{C}^n$, and let $E_{ab} \coloneqq \ket{a}\!\bra{b} \in \operatorname{End}(\mathbb{C}^n)$ denote the standard matrix units. Under the Jordan--Wigner transformation and the canonical vector space isomorphism
\begin{equation}
\label{eq:W_k}
    W_k \cong \bigwedge^k \mathbb{C}^n,
\end{equation}
the $n$-qubit Hamiltonians in~\eqref{eq:fixed_weight_XY_pair} take the second-quantized form
\begin{equation}
\label{eq:fixed_weight_second_quantization}
    H_M\big|_{W_k} = -\mathrm{d}\Gamma_k(A_{P_n}),
    \qquad
    H_P\big|_{W_k} = \mathrm{d}\Gamma_k(E_{11}),
\end{equation}
where
\begin{equation}
\label{eq:path_adjacency_matrix}
    A_{P_n} \coloneqq \sum_{j=1}^{n-1} \left(E_{j,j+1} + E_{j+1,j}\right) \in \operatorname{End}(\mathbb{C}^n)
\end{equation}
is the single-particle adjacency matrix of the open path graph $P_n$, and 
\[
\mathrm{d}\Gamma_k \colon \mathfrak{gl}(n,\mathbb{C}) \to \mathfrak{gl}\left(W_k\right)
\]
denotes the $k$-particle second-quantization operator (i.e., the differential of the $k$-th exterior power representation; see, e.g., \cite{FH}). Explicitly, for any operator $B \in \operatorname{End}(\mathbb{C}^n)$, its action on elementary wedge products is given by the derivation rule
\begin{equation}
\label{eq:dGamma_action}
    \mathrm{d}\Gamma_k(B)\left(v_1 \wedge \cdots \wedge v_k\right) = \sum_{m=1}^{k} v_1 \wedge \cdots \wedge (B v_m) \wedge \cdots \wedge v_k.
\end{equation}
As $\mathrm{d}\Gamma_k$ is a Lie algebra representation, it preserves Lie brackets:
\begin{equation}
\label{eq:dGamma_homomorphism}
    \left[\mathrm{d}\Gamma_k(A),\, \mathrm{d}\Gamma_k(B)\right] = \mathrm{d}\Gamma_k\left([A, B]\right) \qquad \forall A, B \in \mathfrak{u}(n).
\end{equation}


Recall that a quantum system consisting of a single particle constrained to $n$ discrete modes has a single-particle Hilbert space $\mathcal{H}_1 \cong \mathbb{C}^n$ spanned by localized basis states $\{\ket{1}, \ket{2}, \dots, \ket{n}\}$. 
The Lie algebra of all skew-Hermitian operators acting on the $n$-dimensional single-particle Hilbert space forms the full unitary Lie algebra $\mathfrak{u}(n)$. Physical generators such as the adjacency matrix $A_{P_n}$ (see \eqref{eq:path_adjacency_matrix}) and the local site energy operator $E_{11} = \ket{1}\bra{1}$ act directly on $\mathbb{C}^n$. Under repeated Lie brackets, these two local operators generate the full single-particle unitary Lie algebra:
\begin{equation}
\label{eq:endpoint_path_single_particle_DLA}
\mathfrak{g}_1 \coloneqq \left\langle iA_{P_n}, iE_{11}\right\rangle_{\mathrm{Lie}} \cong \mathfrak{u}(n).
\end{equation}
To establish this, define the elementary basic matrix units for $j \in \{1, \dots, n-1\}$:
\[
S_j = i(E_{j,j+1} + E_{j+1,j}), \qquad
A_j = E_{j,j+1} - E_{j+1,j}.
\]
Expressing the tunneling generator as $iA_{P_n} = \sum\limits_{r=1}^{n-1} S_r$, we proceed by induction. Initiating with $D_1 = iE_{11}$, suppose that $iE_{jj}$, $S_j$, and $A_j$ have been generated up to index $j$. The nested commutators
\[
\left[iE_{jj}, iA_{P_n} - \sum_{r<j} S_r\right] = -A_j, 
\quad
[iE_{jj}, A_j] = S_j, 
\quad
[S_j, A_j] = 2i(E_{j+1,j+1} - E_{jj})
\]
successively yield $iE_{jj+1}$, $S_{j+1}$, and $A_{j+1}$. Finally, taking commutators of adjacent off-diagonal matrix units generates all remaining off-diagonal generators, confirming that $\langle iA_{P_n}, iE_{11}\rangle_{\mathrm{Lie}} = \mathfrak{u}(n)$.

The corresponding dynamical Lie group acts transitively on the single-particle state space, establishing full single-particle controllability.
Furthermore, for any excitation number $1 \le k \le n-1$, the $k$-particle second-quantization representation $\mathrm{d}\Gamma_k$ of $\mathfrak{u}(n)$ on $W_k \cong \bigwedge^k \mathbb{C}^n$ is faithful, implying that the corresponding dynamical Lie algebra $\mathfrak{g}_{n,k}$ coincides with the exact image of the single-particle algebra $\mathfrak{g}_1 = \mathfrak{u}(n)$:
\begin{equation}
\label{eq:fixed_weight_DLA}
    \mathfrak{g}_{n,k} \coloneqq \left\langle i H_M\big|_{W_k}, i H_P\big|_{W_k}\right\rangle_{\operatorname{Lie}}
    =\left\langle
id\Gamma_k A_{P_n} ,id\Gamma_k E_{11}\right\rangle_{\mathrm{Lie}} \cong
    \mathrm{d}\Gamma_k\!\left(\mathfrak{u}(n)\right)
 \cong \mathfrak{u}(n).
\end{equation}

In particular, its real dimension satisfies
\begin{equation}
\label{eq:fixed_weight_DLA_dim}
    \dim_{\mathbb{R}}(\mathfrak{g}_{n,k}) = \dim_{\mathbb{R}}(\mathfrak{u}(n)) = n^2.
\end{equation}
For $k\propto n$, this quadratic scaling is exponentially smaller than that of the full ambient space of skew-Hermitian operators $\mathfrak{u}(W_k)$, which has dimension $\binom{n}{k}^2$.

Recall that the complex Grassmannian $\operatorname{Gr}(k,n)$ is the complex manifold of $k$-dimensional linear subspaces of $\mathbb{C}^n$:
\begin{equation}
\label{eq:fixed_weight_grassmann_def}
    \operatorname{Gr}(k,n) \coloneqq \left\{ V \subseteq \mathbb{C}^n \;\middle|\; \dim_{\mathbb{C}}(V) = k \right\}.
\end{equation}
The Grassmannian embeds into the projectivization of the $k$-excitation subspace $\mathbb{P}(W_k) = \mathbb{P}\left(\bigwedge^k \mathbb{C}^n\right)$ via the classical Plücker embedding \cite[Chapter~3.2]{EisenbudHarris2016}:
\begin{equation}
\label{eq:Plucker_embedding}
    \iota_{\mathcal{P}} \colon \operatorname{Gr}(k,n) \hookrightarrow \mathbb{P}(W_k), \qquad \iota_{\mathcal{P}}\left(\operatorname{span}_\mathbb C\{v_1, \ldots, v_k\}\right) = [v_1 \wedge \cdots \wedge v_k].
\end{equation}
The image $\iota_{\mathcal{P}}(\operatorname{Gr}(k,n))$ consists precisely of the rays of \emph{decomposable} $k$-tensors, physically corresponding to single Slater determinants of $k$ occupied single-particle orbitals. As the mixer restriction \(H_M\big|_{W_k} = -d\Gamma_k(A_{P_n})\) is nondegenerate, its unique $k$-particle ground state is given by the wedge product of the $k$ lowest-energy single-particle orbitals. In particular, the projective ground state \([\xi_k]\) lies within the Plücker image of the Grassmannian \(\operatorname{Gr}(k,n)\). Furthermore, since the natural action of $\mathrm{U}(n)$ on $\operatorname{Gr}(k,n)$ is transitive, the dynamical Lie group $G_{n,k}$ orbit of any state $[\psi] \in \iota_{\mathcal{P}}(\operatorname{Gr}(k,n))$ coincides with the full Grassmannian image:
\begin{equation}
\label{eq:fixed_weight_grassmann_orbit}
    \mathcal{O}_{[\psi]} \coloneqq G_{n,k} \cdot [\psi] = \iota_{\mathcal{P}}(\operatorname{Gr}(k,n)).
\end{equation}

Since the dynamical Lie group acts transitively on $\operatorname{Gr}(k,n)$ with stabilizer isomorphic to the maximal block-diagonal subgroup $\mathrm{U}(k) \times \mathrm{U}(n-k)$, the orbit can be identified with the homogeneous space $\mathrm{U}(n) / \left(\mathrm{U}(k) \times \mathrm{U}(n-k)\right)$. Its real manifold dimension is given by the difference of Lie group dimensions:
\begin{equation}
\label{eq:fixed_weight_grassmann_orbit_dimn}
    \dim_{\mathbb{R}}(\mathcal{O}_{[\psi]}) = \dim_{\mathbb{R}}(\mathrm{U}(n)) - \dim_{\mathbb{R}}(\mathrm{U}(k)) - \dim_{\mathbb{R}}(\mathrm{U}(n-k)) = n^2 - k^2 - (n-k)^2 = 2k(n-k),
\end{equation}
which scales at most quadratically in $n$ ($\dim_{\mathbb{R}}(\mathcal{O}_{[\psi]})\le n^2/2$). On the other hand, the dimension of the Hilbert space $W_k$ is $\binom{n}{k}$, and the ground space $W_{P,\min}$ of the problem Hamiltonian in \eqref{eq:fixed_weight_problem_ground_dimension} has exponential dimension $\binom{n-1}{k}$ when $k \propto n$.

\section{Derivation of Stabilization Depths for QAOA Architectures}
\label{sec:depth_bounds_calculation}

This section provides the rigorous analytical derivation of the stabilization depth bounds for dynamical Lie algebras arising from the QAOA architectures defined in Section~\ref{sec:stab_depth_QAOA}, formally verifying the assertions made therein.

\begin{proof}[Proof of Theorem \ref{thm:free_DLA_filtration_classical}]

\textbf{Path and Cycle Graphs.}
According to Table~V of \cite{KLFCCZ}, the bases of the corresponding free DLAs with respective generating sets 
\[
\begin{aligned}
    \mathbf{a}_{P_n,\mathrm{free}}&= \left\{
        iX_s,\, iZ_j Z_{j+1}
        \mid s \in \{1,\dots,n\},\, j \in \{1,\dots,n-1\} 
    \right\}\\
    \mathbf{a}_{C_n,\mathrm{free}}&= \left\{
        iX_s,\, iZ_j Z_{j+1}, iZ_1Z_n  
        \mid s \in \{1,\dots,n\},\, j \in \{1,\dots,n-1\} 
    \right\},
\end{aligned}
\]
consist of the single-site generators $iX_s$ (for $s \in \{1, \dots, n\}$) alongside multi-qubit Pauli strings comprising a contiguous block of $X$ operators on the interior sites $\{b+1, \dots, b+k-1\}$, bounded at the endpoints $b$ and $b+k$ by pairs of operators from $\{Z, Y\}$, with identity operators implicitly understood to be assigned to the remaining positions. These composite basis elements are indexed by their starting position $b$ and length parameter $k$:

\begin{equation}
    \label{eq:path_cycle_basis_strings}
    \begin{aligned}
        P^{b,k}_{ZZ}&=Z_bX_{b+1}\dots X_{b+k-1} Z_{b+k},\\
        P^{b,k}_{ZY}&=Z_bX_{b+1}\dots X_{b+k-1} Y_{b+k},\\
        P^{b,k}_{YZ}&=Y_bX_{b+1}\dots X_{b+k-1} Z_{b+k},\\
        P^{b,k}_{YY}&=Y_bX_{b+1}\dots X_{b+k-1} Y_{b+k},
    \end{aligned}
\end{equation}
where $1 \le b < b+k \le n$ for the path graph, and $1 \le b < b+k \le 2n$ for the cycle graph (with indices evaluated modulo $n$).

Any such string can be generated iteratively from the core two-local generators. For instance, $P^{b,k}_{ZZ}$ is constructed via the following sequence of commutators:
\begin{equation}
    \label{eq:path_generation_chain}
    \begin{aligned}
        [X_{b+1},Z_bZ_{b+1}]&\propto Z_bY_{b+1},\\
        [Z_{b+1}Z_{b+2},Z_bY_{b+1}]&\propto Z_bX_{b+1}Z_{b+2},\\
        [X_{b+2},Z_bX_{b+1}Z_{b+2}]&\propto Z_bX_{b+1}Y_{b+2},\\
        [Z_{b+2}Z_{b+3},Z_bX_{b+1}Y_{b+2}]&\propto Z_bX_{b+1}X_{b+2}Z_{b+3},\\
        &\;\;\vdots
    \end{aligned}
\end{equation}
Tracking the filtration depth through this chain reveals that $P^{b,k}_{ZZ}\in \mathcal  F_{2k-1}^{\mathfrak \mathbf{a}_{P_n,\mathrm{free}}}$. Similarly, by applying up to two additional commutations with single-qubit $X$-generators at the boundaries, we find that
\[
P^{b,k}_{YZ},\, P^{b,k}_{ZY}\in \mathcal F_{2k}^{\mathbf{a}_{P_n,\mathrm{free}}}, \qquad P^{b,k}_{YY}\in \mathcal F_{2k+1}^{\mathbf{a}_{P_n,\mathrm{free}}}.
\]
Because the maximum path length of a non-trivial contiguous block within $n$ sites is bounded by $n-1$, the maximum filtration level required to span the algebra cannot exceed $2n-1$. Therefore, $\mathcal D_{\mathfrak \mathbf{a}_{P_n,\mathrm{free}}} \le 2n-1$ and $\mathcal D_{\mathfrak \mathbf{a}_{C_n,\mathrm{free}}} \le 2n-1$, completing the proof.

\medskip

\textbf{Complete Graphs.} We denote by $x(P)$, $y(P)$, $z(P)$, and $i(P)$ the number of $X$-, $Y$-, $Z$-, and identity factors in a Pauli string $P$, respectively, so that
\begin{equation}
    \label{eq:total_num_of_gates}
    x(P) + y(P) + z(P) + i(P) = n.
\end{equation}
By Theorem~1 in \cite{KLFCCZ}, the free DLA $\mathfrak g_{K_n,\mathrm{free}}$ admits a basis consisting of Pauli strings $P$ satisfying
\begin{equation}
\label{eq:KnBasisConditions}
\begin{cases}
    y(P) + z(P) \equiv 0 \pmod{2}, \\
    x(P) \neq n, \\
    i(P) \neq n.
\end{cases}
\end{equation}
We establish degree bounds for every such basis element $P$ based on  its constituents.

\medskip
\textbf{Case 1: strings with no $X$ factors ($x(P) = 0$).}
Let $s := y(P) + z(P) \ge 2$ denote the support size of non-$X$ factors (which is even, $s \in 2\mathbb Z_{>0}$). We prove by induction on $s$ that $iP \in \mathcal F^{\mathbf{a}_{K_n,\mathrm{free}}}_{2s-1}$ (i.e., $\deg(iP) \le 2s-1$).

For the base case $s=2$, two-body strings of type $ZZ$, $ZY/YZ$, and $YY$ have generation degrees $1$, $2$, and at most $3$, respectively.

Assume $s \ge 4$ is even and that the claim holds for all valid Pauli strings with non-$X$ support strictly smaller than $s$. 

We select a pivot site $j \in \operatorname{supp}(P)$ as follows: if $y(P) \ge 1$, choose $j$ such that $P_j = Y$; otherwise ($y(P) = 0$), select any site $j$ in the support, so $P_j = Z$. Next, choose two distinct sites $a, b \in \operatorname{supp}(P) \setminus \{j\}$ and define $P'$ by replacing $P_a$ and $P_b$ with identity operators $I_a$ and $I_b$, so that $s(P') = s - 2 \ge 2$.

For each $v \in \{a, b\}$, define the auxiliary operator
\begin{equation}
A_v \coloneqq 
\begin{cases}
iP_v Z_j, & \text{if } P_j = Y_j, \\
iP_v Y_j, & \text{if } P_j = Z_j.
\end{cases}
\end{equation}
We verify that $\deg(A_v) \le 2$ for all $v \in \{a, b\}$ as follows:
\[
iZ_aY_j \propto  [iX_j, iZ_aZ_j], \qquad iZ_bY_j \propto [iX_j, iZ_bZ_j],
\]

Evaluating the iterated Lie bracket confirms that $iP$ is recovered (up to a non-zero scalar factor) via the nested commutator:
\begin{equation}
    iP \propto [A_b, [A_a, iP']].
\end{equation}
Applying the inductive hypothesis $\deg(iP') \le 2(s-2) - 1$ together with the filtration bracket compatibility property~\eqref{eq:bracket_compatibility}, we obtain
\begin{align}
    \deg(iP) &\le \deg(iP') + \deg(A_a) + \deg(A_b) \nonumber \\
             &\le (2(s-2) - 1) + 2 + 2 \nonumber \\
             &= 2s - 5 + 4 = 2s - 1,
\end{align}
which completes the inductive step and establishes $iP \in \mathcal F_{2s-1}$.

\textbf{Case 2: strings with $X$ factors and $y(P) + z(P) \ge 2$.}
Let $s := x(P) + y(P) + z(P)$ denote the total support size of $P$. We prove by induction on $x(P) \ge 1$ that $\deg(iP) \le 2s - 1$. The base case $x(P) = 0$ is established by Case~1.

For $x(P) \ge 1$, select an $X$ site $a$ and a $Y/Z$ site $b \in \operatorname{supp}(P)$. Define $P'$ by replacing $X_a$ with $Y_a$ and $P_b$ with $I_b$. The resulting string $P'$ has one fewer $X$ factor ($x(P') = x(P) - 1$) while preserving the even parity of $Y/Z$ factors ($y(P') + z(P') = y(P) + z(P)$), giving a reduced support size of $s(P') = s - 1$.

Define the auxiliary two-body operator $B \coloneqq i Z_a P_b$. If $P_b = Z$, $B = i Z_a Z_b$ is a fundamental generator ($\deg(B) = 1$); if $P_b = Y$, $B = i Z_a Y_b$ has degree $\deg(B)=2$. Evaluating the Lie bracket $[B, iP']$ confirms that $iP$ is recovered (up to a non-zero scalar) via:
\begin{equation}
    iP \propto [i Z_a P_b, iP'].
\end{equation}
Applying the induction hypothesis $\deg(iP') \le 2s(P') - 1$ together with the filtration bracket compatibility property~\eqref{eq:bracket_compatibility}, we obtain
\[
    \deg(iP) \le \deg(iP') + \deg(i Z_a P_b) 
             \le (2s(P') - 1) + 2 
             = 2(s - 1) - 1 + 2 = 2s - 1,
\]
completing the inductive step.

\textbf{Case 3: Pure $X$ strings ($y(P) = z(P) = 0$).} 
Let $t := x(P)$ denote the number of $X$ factors. By the basis conditions~\eqref{eq:KnBasisConditions}, $t < n$, so $1 \le t \le n - 1$.  If $t = 1$, $P = X_a$ is a single-qubit generator, so $\deg(iP) = 1 \le 2n - 1$. In case $t \ge 2$, since $t < n$, there exists at least one unoccupied site $c \notin \operatorname{supp}(P)$ ($P_c = I$). Select two distinct sites $a, b \in \operatorname{supp}(P)$ and construct the auxiliary string $P'$ by replacing $X_a$ and $X_b$ with $Y_a$ and $Y_b$, respectively. The string $P'$ has $x(P') = t - 2$  and $y(P') = 2, z(P') = 0$, yielding a total support size of $s(P') = t$.

Since $P'$ \(P'\) falls under Case 1 if \(t=2\) (mixed string with $y(P') + z(P') = 2$), and under Case 2 otherwise, its degree is bounded by $\deg(iP') \le 2s(P') - 1 = 2t - 1$. Evaluating the double Lie bracket with the fundamental generators $i Z_c Z_a$ and $i Z_c Z_b$ (each of degree $1$) confirms that $iP$ is recovered via:
\begin{equation}
    iP \propto [i Z_c Z_b, [i Z_c Z_a, iP']].
\end{equation}
Applying the filtration bracket compatibility property~\eqref{eq:bracket_compatibility}, we obtain

\[
\deg(P)\le \deg(P')+1+1\le 2(x(P')+y(P')+z(P'))-1+2=2x(P)+1\le 2(n-1)+1\le 2n-1
\]

Since every basis element of $\mathfrak g_{K_n,\mathrm{free}}$ falls into one of these three cases, every basis string $P$ satisfies $\deg(iP) \le 2n - 1$, establishing the stabilization bound $\mathcal D_{\mathfrak g_{K_n,\mathrm{free}}} \le 2n - 1$.

\end{proof}

\begin{rmk}
\label{rmk:support_lower_bound}
The structural properties of the generators can be exploited to establish a straightforward linear lower bound on the filtration level of any given Pauli string. For any two Pauli strings $p$ and $q$, the support of their commutator is contained within the union of their individual supports, $\mathrm{supp}([P, Q]) \subseteq \mathrm{supp}(P) \cup \mathrm{supp}(Q)$. As the generators of $\mathfrak g_{\Gamma,\mathrm{free}}$ are at most $2$-local, commuting a Pauli string $P$ with a generator string $Q$ (where $|\mathrm{supp}(Q)| \le 2$) can increase the cardinality of the support of $P$ by at most $|\mathrm{supp}(Q)| - 1 \le 1$. Consequently, if a basis element $P$ belongs to the minimal filtration component $\mathcal F_{m}^{\mathfrak g_{\Gamma,\mathrm{free}}}$, its filtration index must satisfy $m \ge k - 1$.
\end{rmk}

\begin{proof}[Proof of Theorem \ref{thm:reduced_free_DLA_filtration_classical}]
For \(n=2\), the reduced generating set is \(\{iX_1,iZ_1\}\), so its algebra is \(\mathfrak{su}(2)\) and its filtration stabilizes
at degree \(2\leq2n-1\). Henceforth we assume \(n\geq3\). Since vertex $v$ is connected to every other vertex in $K_n$, its neighborhood is $\mathcal N_v = V(\Gamma) \setminus \{v\}$, containing $n-1$ vertices. Consequently, every single-site operator $i Z_w$ for $w \in V(\Gamma_v)$ is a fundamental generator of the reduced Lie algebra $\mathfrak g^v_{K_n,\mathrm{free}}$. Combining these with the generators $i X_w $ yields the full single-site local algebra $\mathfrak{su}(2)$ at each vertex $w \in V(\Gamma_v)$ within the second filtration component.

To establish the depth bound, let $\mathcal Q \neq I$ be a non-trivial string on the $n-1$ remaining qubits. We analyze the generation degree $\deg(i\mathcal Q)$ across three exhaustive cases:

\medskip
\textbf{Case 1: standard parity strings.}
If $\mathcal Q$ satisfies the basis conditions~\eqref{eq:KnBasisConditions} on the complete graph $K_{n-1}$, then by Theorem~\ref{thm:free_DLA_filtration_classical} applied to $n-1$ vertices, its generation degree is bounded by
\begin{equation}
    \deg(i\mathcal Q) \le 2(n-1) - 1 = 2n - 3 \le 2n - 1.
\end{equation}

\medskip
\textbf{Case 2: all-$X$ string.}
Consider the full $X$-string $P = X_1 X_2 \cdots X_{n-1}$ on $n-1$ sites. Define the auxiliary string
\begin{equation}
    \label{eq:auxiliary_P_string}
    L \coloneqq X_1 \cdots X_{n-3} Y_{n-2} Y_{n-1}.
\end{equation}
Since $y(L) + z(L) = 2$ is even, Theorem~\ref{thm:free_DLA_filtration_classical} on $K_{n-1}$ yields $\deg(i L) \le 2n - 3$. Taking successive commutators of $i L$ with the single-site generators $i Z_{n-2}$ and $i Z_{n-1} $ recovers $P$ up to a non-zero scalar:
\begin{equation}
    i X_1 X_2 \cdots X_{n-1} \propto [i Z_{n-1}, [i Z_{n-2}, i L]].
\end{equation}
As each $i Z_k$ has degree $1$, we obtain:
\begin{equation}
    \deg(i X_1 X_2 \cdots X_{n-1}) \le \deg(i L) + 1 + 1 \le (2n - 3) + 2 = 2n - 1.
\end{equation}

\medskip

\textbf{Case 3: odd-parity strings.}
Finally, suppose that $y(\mathcal Q) + z(\mathcal Q)$ is odd, so that $\mathcal Q$ violates the first condition in~\eqref{eq:KnBasisConditions}. Choose a site $w' \in V(\Gamma_v)$ where $\mathcal Q_{w'} \in \{Y, Z\}$, and define a modified Pauli string $\mathcal Q'$ by setting
\begin{equation}
    \mathcal Q'_w \coloneqq
    \begin{cases}
    X, & w = w', \\
    \mathcal Q_w, & w \neq w'.
    \end{cases}
\end{equation}

By construction, $y(\mathcal Q') + z(\mathcal Q') = y(\mathcal Q) + z(\mathcal Q) - 1$, which is even. Applying Theorem~\ref{thm:free_DLA_filtration_classical} to $K_{n-1}$ yields $\deg(i\mathcal Q') \le 2(n-1) - 1 = 2n - 3$.

To recover $i\mathcal Q$, we take the Lie bracket of $i\mathcal Q'$ with an appropriate single-site element $iZ_{w'}$ if $\mathcal Q_{w'} = Y$ and $iY_{w'}$ if $\mathcal Q_{w'} = Z$.  Given that $\deg(i Z_{w'}) = 1$ and $\deg(i Y_{w'}) = 2$, the subadditivity of degrees under the Lie bracket implies

\begin{equation}
    \deg(i\mathcal Q) \le \deg(i\mathcal Q') + 2 \le (2n - 3) + 2 = 2n - 1.
\end{equation}

\medskip
As the traceless Pauli strings on $n-1$ qubits form a basis of $\mathfrak{su}(2^{n-1})$ and every basis string is generated at degree at most $2n - 1$, we conclude that
\begin{equation}
    \mathcal D_{\mathfrak g^v_{K_n,\mathrm{free}}} \le 2n - 1.
\end{equation}
\end{proof}

\begin{proof}[Proof of Theorem~\ref{thm:restricted_free_DLA_controllability}]
We establish the asserted bound on filtration depth by carefully tracing the steps in the proof of Ref.~\cite[Theorem~IV.2]{TBS}. Let $\mathcal{L}_j$ denote the $j$-th filtration space generated by the native controls according to~\eqref{eq:LieAlgFiltration}, which satisfies the graded bracket property $[\mathcal{L}_p, \mathcal{L}_q] \subseteq \mathcal{L}_{p+q}$ (see \eqref{eq:bracket_compatibility}).

For any distinct feasible configurations $x, y \in \mathcal{F}$, define the standard skew-Hermitian basis elements of $\mathfrak{u}(W_\mathcal{F})$ as
\[
A_{x,y} := E_{x,y} - E_{y,x}, \qquad B_{x,y} := i(E_{x,y} + E_{y,x}).
\]
The localization procedure established in Ref.~\cite[Theorem~IV.2]{TBS} first yields
\[
iZ_v Q_{\mathcal{N}_v}\big|_{W_\mathcal{F}} \in \mathcal{L}_3,
\]
where, for any subset $S \subseteq V$, we define the corresponding projector
\begin{equation}
\label{eq:zero_projector_subset}
Q_S := \prod_{u \in S} \frac{I + Z_u}{2}, \qquad Q_{\varnothing} := I
\end{equation}
(see Ref.~\cite[Eq.~(B5)]{TBS}). The associated propagation identity guarantees that if vertices $r$ and $w$ are adjacent and $C \subseteq V \setminus \{r\}$, then
\[
iZ_r Q_C \in \mathcal{L}_j \quad\Longrightarrow\quad iZ_w Q_{(C \setminus \{w\}) \cup \mathcal{N}_w} \in \mathcal{L}_{j+2}.
\]

Fix $v \in V(\Gamma)$. A closed depth-first traversal of a spanning tree starting and ending at $v$ visits every vertex, forming a walk of length $2(n-1)$. Applying the propagation identity along this walk gives
\[
iZ_v Q_{V \setminus \{v\}}\big|_{W_\mathcal{F}} = iE_{0,0} - iE_{e_v,e_v} \in \mathcal{L}_{3 + 2\cdot2(n-1)} = \mathcal{L}_{4n-1}.
\]
Consequently,
\begin{equation}
\label{eq:MIS_interm_depth}
A_{0,e_v} = -\tfrac{1}{2}\bigl[iZ_v Q_{V \setminus \{v\}}\big|_{W_\mathcal{F}}, i\widehat{H}_{CX,v}\big|_{W_{\mathcal{F}}}\bigr] \in \mathcal{L}_{4n}, 
\qquad
B_{0,e_v} = \tfrac{1}{2}[iZ_v, A_{0,e_v}] \in \mathcal{L}_{4n+1}.
\end{equation}

We proceed by induction on $k = |x|$ to prove that for every nonempty independent set $x$ of size $k$,
\[
A_{0,x}, B_{0,x} \in \mathcal{L}_{d_k}, \qquad d_k := 4n + 2k - 1.
\]
The base case $k = 1$ is established by~\eqref{eq:MIS_interm_depth}. For $k \ge 2$, choose $v \in x$ and let $y = x \setminus \{v\}$. The mixer matching relations from Ref.~\cite[Eq.~(B24)]{TBS} provide the identities
\[
C_A := [B_{0,y}, i\widehat{H}_{CX,v}\big|_{W_{\mathcal{F}}}] = -A_{0,x} - A_{y,e_v}, 
\qquad
C_B := [A_{0,y}, i\widehat{H}_{CX,v}\big|_{W_{\mathcal{F}}}] = B_{0,x} - B_{y,e_v}.
\]
Since both pairs $(0,x)$ and $(y,e_v)$ exhibit a $Z_v$-eigenvalue difference of $2$, we have
\[
[iZ_v, A_{p,q}] = 2B_{p,q}, \qquad [iZ_v, B_{p,q}] = -2A_{p,q}
\]
for these pairs, allowing us to invert the relations as
\[
A_{0,x} = -\tfrac{1}{2}C_A - \tfrac{1}{4}[iZ_v, C_B], 
\qquad
B_{0,x} = \tfrac{1}{2}C_B - \tfrac{1}{4}[iZ_v, C_A].
\]
By the induction hypothesis, $C_A, C_B \in \mathcal{L}_{d_{k-1}+1}$, which places these expressions in $\mathcal{L}_{d_{k-1}+2} = \mathcal{L}_{d_k}$.

Set \(m:=\alpha(\Gamma)\) and \(\delta:=d_m=4n+2m-1\). Then every vacuum transition $A_{0,x}, B_{0,x}$ resides in $\mathcal{L}_\delta$. Taking commutators of two such transitions generates $A_{x,y}$ and $B_{x,y}$ for all distinct nonzero $x, y \in \mathcal{F}$. Furthermore, the relation
\[
\tfrac{1}{2}[A_{0,x}, B_{0,x}] = i(E_{0,0} - E_{x,x})
\]
implies that $\mathfrak{su}(W_\mathcal{F}) \subseteq \mathcal{L}_{2\delta}$.

Finally, for any edge $\{v, w\}$, the generator combination $i(Z_v + Z_w)\big|_{W_\mathcal{F}} \in \mathcal{L}_1$ has a strictly positive trace. Indeed, since $\Gamma$ is connected and has at least two vertices, $\operatorname{Tr}_{W_{\mathcal F}}(Z_v) > 0$ for every vertex $v$: removing $v$ defines a non-surjective one-to-one map from independent sets containing $v$ to those omitting it. Consequently, $\operatorname{Tr}_{W_{\mathcal F}}(Z_v + Z_w) > 0$. As its traceless component is a linear combination of diagonal operators already contained in $\mathcal{L}_{2\delta}$, the identity operator satisfies $iI_{W_\mathcal{F}} \in \mathcal{L}_{2\delta}$ as well, yielding
\[
\mathcal{L}_{2\delta} = \mathfrak{u}(W_\mathcal{F}), \qquad D_{\mathfrak{g}_{\Gamma,\mathrm{free},\mathcal{F}}} \le 2\delta = 8n + 4m - 2.
\]
\end{proof}

\section{Theoretical Derivations of Bounds on Circuit Depth for Orbit Generation and Global Element Synthesis}
\label{app:global_synthesis}
In this appendix, we prove Theorems~\ref{thm:uniform_orbit_generation}, \ref{thm:uniform_orbit_generation2}, \ref{thm:Pauli_global_synthesis}, and~\ref{thm:MIS_connected_global_bound}. 

For Theorem~\ref{thm:uniform_orbit_generation}, the factorization in Eqs.~\eqref{eq:composite_parameterization}--\eqref{eq:X_decomp} comprises $\binom{D}{2}$ two-level flows and $\binom{D}{2} + 1$ diagonal factors, thereby establishing Eq.~\eqref{eq:general_full_unitary_bound_X}. Next, we prove the basis-to-state bound \eqref{eq:basis_orbit_generation_bound} for the reference state $\ket{e_1}$; the corresponding result for an arbitrary reference state $\ket{e_r}$ follows immediately by relabeling the basis.

\begin{proof}[Proof of Theorem~\ref{thm:uniform_orbit_generation}]
The global unitary bound in~\eqref{eq:general_full_unitary_bound_X} follows directly from the factorizations given in Eq.~\eqref{eq:composite_parameterization} through~\eqref{eq:X_decomp}. It remains to prove assertions~2 and~3. We first establish the basis-to-state transition synthesis bound~\eqref{eq:basis_orbit_generation_bound} for $|e_1\rangle$; the result for any arbitrary basis state follows by relabeling. Expressing the target state as
\[
|\phi\rangle = \sum_{j=1}^{D} c_j |e_j\rangle, \qquad p_j \coloneqq |c_j|,
\]
we define the cumulative norm sequence via
\[
\rho_j \coloneqq \sqrt{\sum_{k=1}^{j} p_k^2}, \qquad 1 \le j \le D.
\]
Normalization of $|\phi\rangle$ implies $\rho_D = 1$. For each index $j = D, D-1, \ldots, 2$, select the rotation angle $\theta_j \in [0, \pi/2]$ according to
\[
\cos(\theta_j) \coloneqq \begin{cases} \dfrac{\rho_{j-1}}{\rho_j} & \text{if } \rho_j > 0, \\ 1 & \text{if } \rho_j = 0, \end{cases} \qquad 
\sin(\theta_j) \coloneqq \begin{cases} \dfrac{p_j}{\rho_j} & \text{if } \rho_j > 0, \\ 0 & \text{if } \rho_j = 0. \end{cases}
\]

Starting from the state $|e_1\rangle$, we sequentially apply the $D-1$ single-parameter two-level rotations $e^{-i\theta_j X_{1j}}$ in descending order $j = D, D-1, \ldots, 2$. Because $X_{1j} = |e_1\rangle\langle e_j| + |e_j\rangle\langle e_1|$ acts nontrivially only on $\operatorname{span}\{|e_1\rangle, |e_j\rangle\}$, the rotation at step $j$ transforms the amplitude on $|e_1\rangle$ from $\rho_j$ to $\rho_j \cos(\theta_j) = \rho_{j-1}$, while depositing the amplitude $-i \rho_j \sin(\theta_j) = -i p_j$ onto $|e_j\rangle$. All previously populated components $|e_k\rangle$ ($k > j$) remain unaffected. Consequently, after applying all $D-1$ rotations, the state becomes
\[
p_1 |e_1\rangle - i \sum_{j=2}^{D} p_j |e_j\rangle.
\]
A single diagonal unitary transformation adjusts the phases of each basis component simultaneously, mapping this vector directly to $|\phi\rangle$ (up to a global phase). The total circuit depth is therefore at most $(D-1)K_X + K_{\mathrm{diag}}$, establishing \eqref{eq:basis_orbit_generation_bound}. 

To prove assertion~3, let $C_\psi$ and $C_\phi$ be unitary synthesis circuits satisfying
\[
C_\psi \ket{e_1} = e^{i\alpha} |\psi\rangle, \qquad C_\phi \ket{e_1} = e^{i\beta} \ket{\phi},
\]
each produced by a parameterized  circuit with depth bounded by $(D-1)K_X + K_{\mathrm{diag}}$. Since the native pulse operations generate single-parameter sub-groups, the inverse circuit $C_\psi^{-1} = C_\psi^\dagger$ is obtained by reversing the sequence of pulses and negating their corresponding parameters. Hence, $\operatorname{depth}(C_\psi^{-1}) = \operatorname{depth}(C_\psi)$. Combining these circuits yields
\[
\left(C_\phi C_\psi^{-1}\right) |\psi\rangle = e^{i(\beta - \alpha)} |\phi\rangle,
\]
which implies the combined depth bound
\[
L_{\mathrm{state}} \le \operatorname{depth}(C_\psi^{-1}) + \operatorname{depth}(C_\phi) \le 2(D-1)K_X + 2K_{\mathrm{diag}},
\]
completing the proof of \eqref{eq:uniform_state_to_state_bound}.
\end{proof}

\begin{proof}[Proof of Theorem~\ref{thm:uniform_orbit_generation2}]
Let \(s_q\) and \(c_q\) denote, respectively, the numbers of
one-qubit gates and CNOT gates in the unoptimized quantum Shannon
decomposition of a \(q\)-qubit unitary. The recursive decomposition
contains four \((q-1)\)-qubit unitaries and three uniformly controlled
rotations. Each such rotation uses \(2^{q-1}\) one-qubit rotations and
\(2^{q-1}\) CNOTs. Hence
\[
s_q\le 4s_{q-1}+3\cdot 2^{q-1},
\qquad
c_q\le 4c_{q-1}+3\cdot 2^{q-1},
\]
with \(s_1=1\) and \(c_1=0\). Solving these recurrences gives
\[
s_q\le 4^q-3\cdot2^{q-1},
\qquad
c_q\le \frac34\,4^q-\frac32\,2^q.
\]
This proves \eqref{eq:general_full_unitary_bound_X}.

For state preparation, at the stage involving \(k\) remaining qubits,
the disentangling construction of
Ref.~\cite[Theorems~8--9]{shende2006synthesis} uses at most \(2^k\)
one-qubit rotations and \(2^k-2\) CNOTs. Including two rotations for
the final one-qubit state gives
\[
s_q^{\mathrm{prep}}
\le
2+\sum_{k=2}^{q}2^k
=
2^{q+1}-2,
\]
and
\[
c_q^{\mathrm{prep}}
\le
\sum_{k=2}^{q}(2^k-2)
=
2^{q+1}-2q-2.
\]
A state-to-state circuit is obtained by composing the inverse
preparation circuit for \(\ket{\psi}\) with the preparation circuit
for \(\ket{\phi}\). Doubling these counts gives
\[
s_q^{\mathrm{state}}\le 2^{q+2}-4,
\qquad
c_q^{\mathrm{state}}\le 2^{q+2}-4q-4,
\]
which proves \eqref{eq:uniform_state_to_state_bound}.
\end{proof}

\subsection{Verification of Theorems~\ref{thm:Pauli_global_synthesis} and~\ref{thm:MIS_connected_global_bound}}
\label{subsec:MIS_global_proofs}

In this subsection, we present the constructive proofs and analytical verifications for the global circuit depth bounds established in Theorem~\ref{thm:Pauli_global_synthesis} for Pauli-generated full DLAs and Theorem~\ref{thm:MIS_connected_global_bound} for multi-angle MIS QAOA. 

\begin{proof}[Proof of Theorem~\ref{thm:Pauli_global_synthesis}]
For a nonidentity Pauli string \(P\), define its filtration depth by
\[
\delta(P)
:=
\min\{j\mid iP\in\mathcal F_j^{\mathbf a}\}.
\]
The space \(\mathcal F_j^{\mathbf a}\) is spanned by left-nested Lie words of length at most \(j\) in the native generators. Because every nonzero commutator of Pauli strings is proportional to a single Pauli string, and distinct Pauli strings are linearly independent, the inclusion \(iP\in\mathcal F_{\delta(P)}^{\mathbf a}\) implies that some nonzero nested Lie word of length at most \(\delta(P)\) is proportional to \(iP\).

If \(S\) and \(Q\) are anticommuting Hermitian Pauli strings satisfying $SQ+QS=0$, then
\[
\operatorname{Ad}_{e^{\frac{\pi}{4}iS}}(iQ)
=
e^{\frac{\pi}{4}iS}(iQ)e^{-\frac{\pi}{4}iS}
=
-SQ.
\]
It follows recursively that \(iP\) can be written in the form
\[
iP
=
\pm\operatorname{Ad}_{V}(iP_r),
\qquad
V=e^{\frac{\pi}{4}iP_{r_1}}\cdots e^{\frac{\pi}{4}iP_{r_{\ell-1}}},
\qquad
\ell\le\delta(P),
\]
for some native Pauli generator \(P_r \in \mathbf a\). Consequently, the targeted flow decomposes as
\[
e^{itP}
=
V e^{\pm itP_r}V^{-1},
\]
which incurs at most
\[
2\delta(P)-1\le 2\mathcal D_{\mathbf a}-1
\]
primitive exponential steps, establishing \eqref{eq:general_Pauli_flow_cost}.

In particular, since any single-qubit unitary $U\in SU(2)$ can be parameterized via Euler rotations as
\begin{equation}
    \label{eq:1q_euler_decomposition}
    U = e^{i\frac{\alpha}{2}X} e^{i\frac{\beta}{2}Y} e^{i\frac{\gamma}{2}Z},
\end{equation}
each constituent single-qubit rotation requires at most $2\mathcal{D}_{\mathbf{a}}-1$ native generator steps. Accounting for the three rotation axes yields the uniform bound
\begin{equation}
\label{eq:K_1q_bound_pauli}
K_{\mathrm{1q}}
\le 3(2\mathcal D_{\mathbf a}-1) = 6\mathcal D_{\mathbf a}-3.
\end{equation}

Similarly, utilizing the decomposition of a CNOT gate into Pauli string exponentials up to a global phase,
\begin{equation}
\label{eq:cnot_pauli_string_decomposition}
\left[\mathrm{CX}_{c\to t}\right] = \left[e^{-i\pi(I_c\otimes X_t)/4}
e^{-i\pi(Z_c\otimes I_t)/4}
e^{ i\pi(Z_c\otimes X_t)/4}\right],
\end{equation}
each of the three constituent two-qubit Pauli string flows can likewise be conjugated into native generators with depth at most $2\mathcal{D}_{\mathbf{a}}-1$, giving
\begin{equation}
\label{eq:K_cx_bound_pauli}
K_{\mathrm{CX}}
\le 3(2\mathcal D_{\mathbf a}-1) = 6\mathcal D_{\mathbf a}-3.
\end{equation}

Substituting $K_{\mathrm{1q}}$ and $K_{\mathrm{CX}}$ into the general circuit-level bounds of Theorem~\ref{thm:uniform_orbit_generation2} directly yields the total depth bounds.
\end{proof}

We next establish the proof of Theorem~\ref{thm:MIS_connected_global_bound} by breaking down the construction into three core technical phases: exact synthesis of diagonal operators, toggle-controlled extraction of single matrix units, and obtaining the resulting global and state-to-state depth bounds.

\subsubsection{Exact Synthesis of Diagonal Operators}
\label{subsubsec:MIS_diagonal_synthesis}

We now establish an explicit depth bound for synthesizing arbitrary diagonal unitaries on the feasible subspace $W_{\mathcal{F}}$. Let $D \coloneqq |\mathcal{F}| \ge 4$, and assume the underlying graph $\Gamma = (V,E)$ is connected.

For any vertex $v \in V$, consider the local Pauli-$Z$ operator restricted to $W_{\mathcal{F}}$, denoted by $Z_v |_{W_{\mathcal{F}}}$. Partition the feasible set into bitstrings where the bit at position indexed by $v$ is empty vs.\ occupied:
\begin{equation}
    \mathcal{F}_v^0 \coloneqq \{x \in \mathcal{F} \mid x_v = 0\}, \qquad \mathcal{F}_v^1 \coloneqq \{x \in \mathcal{F} \mid x_v = 1\}.
\end{equation}
The matrix trace of $Z_v |_{W_{\mathcal{F}}}$ counts the difference in subset sizes:
\begin{equation}
    \label{eq:Zv_trace_positive}
    \operatorname{Tr}_{W_{\mathcal{F}}}(Z_v) = |\mathcal{F}_v^0| - |\mathcal{F}_v^1| > 0.
\end{equation}
Indeed, flipping the $v$th bit gives a one-to-one map from $\mathcal{F}_v^1$ into $\mathcal{F}_v^0$. However, this map fails to be surjective whenever $v$ has a neighbor $w \in \mathcal{N}_v$, as the single-vertex independent set $\{w\} \in \mathcal{F}_v^0$ cannot lie in its image.

Let $\mu_v \coloneqq \frac{1}{D} \operatorname{Tr}_{W_{\mathcal{F}}}(Z_v) > 0$ denote the normalized trace, and define the traceless centered operator
\begin{equation}
    \label{eq:Zv_traceless}
    Z_{v,0} \coloneqq Z_v |_{W_{\mathcal{F}}} - \mu_v I_{W_{\mathcal{F}}}, \qquad \operatorname{Tr}_{W_{\mathcal{F}}}(Z_{v,0}) = 0.
\end{equation}
For any permutation $P \in \mathcal{T}(\Gamma)$, denote its unitary matrix representation on $W_{\mathcal{F}}$ by $\widehat{P} \in \mathrm{U}(W_{\mathcal{F}})$. We form the traceless diagonal difference operator
\begin{equation}
    \label{eq:BP_def}
    B_P \coloneqq Z_v |_{W_{\mathcal{F}}} - \widehat{P} Z_v |_{W_{\mathcal{F}}} \widehat{P}^\dagger = Z_{v,0} - \widehat{P} Z_{v,0} \widehat{P}^\dagger,
\end{equation}
and define the real linear span
\begin{equation}
    \label{eq:Ev_subspace}
    \mathcal{E}_v \coloneqq \operatorname{span}_{\mathbb{R}} \{ B_P \mid P \in \mathcal{T}(\Gamma) \} \subseteq \mathfrak{d}_0(W_{\mathcal{F}}),
\end{equation}
where $\mathfrak{d}_0(W_{\mathcal{F}}) \cong \mathbb{R}^{D-1}$ denotes the space of all traceless diagonal Hermitian operators on $W_{\mathcal{F}}$.

By a result of Striker \cite[Corollary~3.24]{striker2018rowmotion}, for every connected graph $\Gamma$, the toggle group satisfies
\begin{equation}
    \label{eq:MIS_toggle_alternating}
    A_{\mathcal{F}} \subseteq \mathcal{T}(\Gamma) \subseteq \mathrm{Sym}(\mathcal{F}),
\end{equation}
where $A_{\mathcal{F}}$ and $\mathrm{Sym}(\mathcal{F})$ denote the alternating and symmetric groups on $\mathcal{F}$, respectively. 

For every $Q \in \mathcal{T}(\Gamma)$, one has
\[
\widehat{Q} B_P \widehat{Q}^\dagger = B_{QP} - B_Q \in E_v,
\]
which proves that $E_v$ is a $\mathcal{T}(\Gamma)$-invariant subspace (subrepresentation) of $\mathfrak{d}_0(W_{\mathcal{F}})$. 

To show that $E_v = \mathfrak{d}_0(W_{\mathcal{F}})$, first note that $E_v \neq \{0\}$: the diagonal operator $Z_{v,0}$ is nonconstant because the feasible configurations $\varnothing$ and $\{v\}$ have opposite $Z_v$-eigenvalues, and the transitive action of $\mathcal{T}(\Gamma)$ cannot fix a nonconstant diagonal vector. Since the subspace $\mathfrak{d}_0(W_{\mathcal{F}})$ is isomorphic to the standard real representation of $A_D$ (where $D = |\mathcal{F}|$), and this representation is irreducible for all $D \ge 4$, it follows that the subspace $\mathcal{E}_v$ generated by orbit differences coincides with the entire space:
\[
E_v = \mathfrak{d}_0(W_{\mathcal{F}}).
\]

We can therefore select $D-1$ permutations $P_1, \dots, P_{D-1} \in \mathcal{T}(\Gamma)$ such that $\{B_{P_1}, \dots, B_{P_{D-1}}\}$ forms a basis for $\mathfrak{d}_0(W_{\mathcal{F}})$. Since $\operatorname{Tr}_{W_{\mathcal{F}}}(Z_v) \neq 0$, appending $Z_v |_{W_{\mathcal{F}}}$ yields a full basis for all diagonal Hermitian operators $\mathfrak{d}(W_{\mathcal{F}}) \cong \mathbb{R}^D$:
\begin{equation}
    \operatorname{span}_{\mathbb{R}} \big\{ Z_v |_{W_{\mathcal{F}}}, \, B_{P_1}, \, \dots, \, B_{P_{D-1}} \big\} = \mathfrak{d}(W_{\mathcal{F}}).
\end{equation}

Thus, any diagonal Hamiltonian $H \in \mathfrak{d}(W_{\mathcal{F}})$ expands uniquely as
\begin{equation}
    \label{eq:H_diagonal_expansion}
    H = c_0 Z_v |_{W_{\mathcal{F}}} + \sum_{k=1}^{D-1} c_k \Big( Z_v |_{W_{\mathcal{F}}} - \widehat{P}_k Z_v |_{W_{\mathcal{F}}} \widehat{P}_k^\dagger \Big), \qquad c_0, c_1, \dots, c_{D-1} \in \mathbb{R}.
\end{equation}
Since all diagonal operators mutually commute, exponentiating Eq.~\eqref{eq:H_diagonal_expansion} yields an exact product factorization for an arbitrary diagonal unitary $e^{itH}$:
\begin{equation}
    \label{eq:diagonal_unitary_factorization}
    e^{itH} = e^{i t \left( c_0 + \sum\limits_{k=1}^{D-1} c_k \right) Z_v |_{W_{\mathcal{F}}}} \cdot \prod_{k=1}^{D-1} \widehat{P}_k e^{ -i t c_k Z_v |_{W_{\mathcal{F}}}} \widehat{P}_k^\dagger, \qquad t \in \mathbb{R}.
\end{equation}
Although a native toggle word realizes $\widehat{P}_k$ only up to an overall phase, this phase factor cancels out identically under conjugation, ensuring that the phase ambiguity of the toggle implementation does not affect the exactness of the diagonal factorization in~\eqref{eq:diagonal_unitary_factorization}.

By \eqref{eq:MIS_toggle_diameter_general}, each unitary permutation operator $\widehat{P}_k$ is implementable (up to a global phase) to depth at most $\Delta_\Gamma$ using native mixer pulses. Each conjugated factor $\widehat{P}_k e^{-itc_k Z_v} \widehat{P}_k^\dagger$ therefore requires at most $2\Delta_\Gamma + 1$ native operations ($1$ diagonal pulse drive under $Z_v$ surrounded by $2$ toggle sub-circuits of depth $\Delta_\Gamma$). Combining these with the single un-conjugated $Z_v$ pulse gives a bound on the exact diagonal synthesis complexity.

\begin{prop}
\label{prop:MIS_exact_diagonal_cost}
Let $\Gamma$ be a connected graph with $D = |\mathcal{F}| \ge 4$. Every diagonal unitary operator $U_{\mathrm{diag}} \in \mathrm{U}(W_{\mathcal{F}})$ (including its global phase) can be synthesized exactly using at most
\begin{equation}
    \label{eq:MIS_diagonal_cost}
    K_{\mathrm{diag}} \le 1 + (D-1)(2\Delta_\Gamma + 1)
\end{equation}
native pulse operations.
\end{prop}

\subsubsection{Extraction of Single Two-Level Matrix Units}
\label{subsubsec:MIS_matrix_unit_extraction}

Having established exact diagonal synthesis, we now show how to isolate an individual two-level matrix unit $X_{rs}$ with $r,s\in \mathcal F$ (see \eqref{eq:two_level_generators_recap}).

Fix a vertex $v \in V$. Because $\tau_v$ acts as a set-theoretic involution on $\mathcal{F}$, its action on the basis of $W_{\mathcal{F}}$ decomposes into a disjoint union of transpositions and fixed points:
\begin{equation}
\label{eq:MIS_toggle_matching}
\widehat{H}_{CX,v}\big|_{W_{\mathcal{F}}} = Q_v + \sum_{\{r,s\} \in M_v} X_{rs}, \qquad X_{rs} \coloneqq |r\rangle\langle s| + |s\rangle\langle r|,
\end{equation}
where $M_v$ is a non-empty matching of feasible strings transposed by $\tau_v$, and $Q_v \coloneqq \sum_{w \in \mathrm{Fix}(\tau_v)} |w\rangle\langle w|$ is the diagonal projector onto the fixed subspace of $\widehat{H}_{CX,v}\big|_{W_{\mathcal{F}}}$.

Consider a pair of strings $\{a,b\} \in M_v$ swapped by $\tau_v$. Because the edges of the matching $M_v$ are pairwise disjoint, we can construct a diagonal sign operator $S \in \mathrm{U}(W_{\mathcal{F}})$:
\begin{equation}
    S \coloneqq \sum_{x \in \mathcal{F}} \varepsilon_x |x\rangle\langle x|, \qquad \varepsilon_x \in \{\pm 1\},
\end{equation}
such that the pair signs satisfy
\begin{equation}
    \label{eq:sign_assignment}
    \varepsilon_a \varepsilon_b = +1, \qquad \text{and} \qquad \varepsilon_r \varepsilon_s = -1 \quad \text{for every } \{r,s\} \in M_v \setminus \{\{a,b\}\}.
\end{equation}
Since $S = S^\dagger = S^{-1}$, conjugating $X_{rs}$ by $S$ yields $S X_{rs} S^{-1} = \varepsilon_r \varepsilon_s X_{rs}$. Applied to $\widehat{H}_{CX,v}\big|_{W_{\mathcal{F}}}$, this sign assignment flips the direction of every matching term except the target $X_{ab}$:
\begin{equation}
    S \widehat{H}_{CX,v}\big|_{W_{\mathcal{F}}} S^{-1} = Q_v + X_{ab} - \sum_{\{r,s\} \in M_v \setminus \{\{a,b\}\}} X_{rs}.
\end{equation}
Averaging $\widehat{H}_{CX,v}\big|_{W_{\mathcal{F}}}$ with its conjugated counterpart yields
\begin{equation}
\label{eq:MIS_refocusing_identity}
\frac{1}{2} \left( \widehat{H}_{CX,v}\big|_{W_{\mathcal{F}}} + S \widehat{H}_{CX,v}\big|_{W_{\mathcal{F}}} S^{-1} \right) = Q_v + X_{ab}.
\end{equation}

Crucially, the operators $\widehat{H}_{CX,v}\big|_{W_{\mathcal{F}}}$ and $S \widehat{H}_{CX,v}\big|_{W_{\mathcal{F}}} S^{-1}$ commute. Indeed, both are block-diagonal with respect to the orthogonal decomposition
\begin{equation}
    W_{\mathcal{F}} = \bigoplus_{w \in \mathrm{Fix}(\tau_v)} \operatorname{span}_{\mathbb C}\{|w\rangle\} \;\oplus \bigoplus_{\{r,s\} \in M_v} \operatorname{span}_{\mathbb C}\{|r\rangle, |s\rangle\}.
\end{equation}
On each two-dimensional matching subspace $\operatorname{span}_{\mathbb C}\{|r\rangle, |s\rangle\}$, $\widehat{H}_{CX,v}\big|_{W_{\mathcal{F}}}$ restricts to the Pauli-$X$ matrix, while $S \widehat{H}_{CX,v}\big|_{W_{\mathcal{F}}} S^{-1}$ restricts to either $+X$ (on $\{a,b\}$) or $-X$ (on $M_v \setminus \{\{a,b\}\}$). Because $X$ commutes with $\pm X$, the sum-of-generators exponentiates to a product of commuting unitaries:
\begin{equation}
e^{-i \frac{t}{2} \widehat{H}_{CX,v}\big|_{W_{\mathcal{F}}}} e^{ -i \frac{t}{2} S \widehat{H}_{CX,v}\big|_{W_{\mathcal{F}}} S^{-1}} = e^{-i t (Q_v + X_{ab})}.
\end{equation}

As the diagonal projector $Q_v$ and the off-diagonal generator $X_{ab}$ have disjoint support on $W_{\mathcal{F}}$, they commute ($[Q_v, X_{ab}] = 0$). We can therefore isolate the target two-level flow $e^{-it X_{ab}}$ as
\begin{equation}
\label{eq:MIS_matrix_unit_flow}
e^{-it X_{ab}} = e^{it Q_v} e^{ -i \frac{t}{2} \widehat{H}_{CX,v}\big|_{W_{\mathcal{F}}}} S e^{-i \frac{t}{2} \widehat{H}_{CX,v}\big|_{W_{\mathcal{F}}}} S^{-1},
\end{equation}
where we used the identity $e^{-i \frac{t}{2} S \widehat{H}_{CX,v}\big|_{W_{\mathcal{F}}} S^{-1}} = S e^{-i \frac{t}{2} \widehat{H}_{CX,v}\big|_{W_{\mathcal{F}}}} S^{-1}$.

\begin{prop}
\label{prop:MIS_matrix_unit_cost}
Let \(\Gamma = (V, E)\) be a connected graph with \(D = |\mathcal{F}| \ge 4\). For any vertex \(v \in V\) and every toggle edge \(\{a, b\} \in M_v\), the two-level flow \(e^{-itX_{ab}}\) (\(t \in \mathbb{R}\)) can be synthesized exactly using at most
\begin{equation}
\label{eq:MIS_matrix_unit_cost_v}
K_X^{(v)}
\le 3K_{\mathrm{diag}} + 2
\le
6\Delta_\Gamma D + 3D - 6\Delta_\Gamma + 2
\end{equation}
native pulses.

Consequently, for any pair of distinct feasible configurations \(r, s \in \mathcal{F}\), the corresponding two-level flow \(e^{-itX_{rs}}\) can be synthesized exactly using at most
\begin{equation}
  \label{eq:MIS_matrix_unit_cost} 
  K_X\le K_X^{(v)} + 2\Delta_\Gamma \le 3K_{\mathrm{diag}} + 2 + 2\Delta_\Gamma \le
6\Delta_\Gamma D + 3D - 4\Delta_\Gamma + 2
\end{equation}

native pulses.
\end{prop}

\begin{proof}
Fix \(v\in V\) and a toggle edge \(\{a,b\}\in M_v\).
We employ the realization of $e^{-itX_{ab}}$ from~\eqref{eq:MIS_matrix_unit_flow}.
The diagonal unitary \(e^{itQ_v}\) requires at most
\(K_{\mathrm{diag}}\) native pulses. Each of the two evolutions under
\(\widehat H_{CX,v}\) requires one native mixer pulse, while \(S\) and
\(S^{-1}\) each require at most \(K_{\mathrm{diag}}\) pulses.
Therefore,
\[
K_X^{(v)}
\leq
K_{\mathrm{diag}}+1+K_{\mathrm{diag}}+1+K_{\mathrm{diag}}
=
3K_{\mathrm{diag}}+2.
\]
Substituting the upper bound for \(K_{\mathrm{diag}}\) from~\eqref{eq:MIS_diagonal_cost} yields the bound in~\eqref{eq:MIS_matrix_unit_cost_v}.

Now let $r, s \in \mathcal{F}$ be distinct feasible bitstrings. By~\eqref{eq:MIS_toggle_alternating}, the toggle group contains the alternating group:
\[
\mathcal{A}_\mathcal{F} \subseteq \mathcal{T}(\Gamma).
\]
Since $D \ge 4$, the alternating group $\mathcal{A}_\mathcal{F}$ acts transitively on ordered pairs of distinct elements in $\mathcal{F}$. Hence, there exists a permutation $P \in \mathcal{T}(\Gamma)$ such that
\[
P(a) = r, \qquad P(b) = s.
\]
Denoting by $\widehat{P}\in U(W_\mathcal{F})$ the corresponding permutation operator, we obtain
\[
\widehat{P} X_{ab} \widehat{P}^{-1} = X_{rs},
\]
and consequently,
\begin{equation}
\label{eq:arbitrary_matrix_unit_conjugation}
e^{-itX_{rs}} = \widehat{P} \, e^{-itX_{ab}} \, \widehat{P}^{-1}.
\end{equation}
By definition of the toggle-group diameter $\Delta_\Gamma$, both $\widehat{P}$ and $\widehat{P}^{-1}$ can be implemented using at most $\Delta_\Gamma$ local mixer generators. 
It follows that
\[
K_X \le K_X^{(v)} + 2\Delta_\Gamma.
\]
Combining this inequality with~\eqref{eq:MIS_matrix_unit_cost_v}  yields~\eqref{eq:MIS_matrix_unit_cost}.

\end{proof}

\subsubsection{Global Synthesis Complexity and State-to-State Bounds}
\label{subsubsec:MIS_global_bounds}

By substituting the explicit primitive costs $K_X$ (Proposition~\ref{prop:MIS_matrix_unit_cost}) and $K_{\mathrm{diag}}$ (Proposition~\ref{prop:MIS_exact_diagonal_cost}) into the general pulse-complexity bounds of Eq.~\eqref{eq:general_full_unitary_bound_X} and Theorem~\ref{thm:uniform_orbit_generation}, we establish the global synthesis and state-to-state transition bounds asserted in Theorem~\ref{thm:MIS_connected_global_bound}.

\section{Finite-Depth Exposure of Descent Directions: Detailed Treatment}
\label{sec:finite_depth_descent_proofs}

Let $\mathfrak{g} \subseteq \mathfrak{u}(W)$ be the dynamical Lie algebra generated by the available skew-Hermitian control operators, and let $G$ denote the associated connected dynamical Lie group. We assume that the problem Hamiltonian satisfies $iH_P \in \mathfrak{g}$. Working in the density operator formalism eliminates unphysical global phase ambiguities; we further assume that  $G$ is a compact Lie group.

For a normalized initial state $\ket{\xi}$, we recall that $\rho = \ket{\xi}\!\bra{\xi}$ is the initial density matrix. The corresponding \emph{reachable state orbit} under $G$ is the adjoint orbit
\begin{equation}
    \label{eq:adjoint_orbit}
    \mathcal{O}_\rho := \{ g \rho g^\dagger : g \in G \}.
\end{equation}
A parameterized quantum circuit specifies a smooth map $U(\boldsymbol{\theta}) \in G$, which induces the circuit state endpoint map
\[
\Phi_\rho(\boldsymbol{\theta}) := U(\boldsymbol{\theta}) \rho U^\dagger(\boldsymbol{\theta}) \in \mathcal{O}_\rho.
\]

The circuit loss function measures the expectation value of the problem Hamiltonian $H_P$:
\begin{equation}
\label{eq:loss_finite_depth_descent}
     \ell_{\boldsymbol{\theta}}\!\left(\rho, H_P\right) 
    := \bra{\psi(\boldsymbol{\theta})} H_P \ket{\psi(\boldsymbol{\theta})} 
    = \operatorname{Tr}\!\left( H_P U(\boldsymbol{\theta}) \rho U^\dagger(\boldsymbol{\theta}) \right),
\end{equation}
where $\ket{\psi(\boldsymbol{\theta})} = U(\boldsymbol{\theta})\ket{\xi}$. The loss function factors through the state orbit as the composition
\[
\ell_{\boldsymbol{\theta}}\!\left(\rho, H_P\right) 
  = (h \circ \Phi_\rho)(\boldsymbol{\theta}),
\]
where $h : \mathcal{O}_\rho \longrightarrow \mathbb{R}$ is the linear height function defined by 

\begin{equation}
    \label{eq:height_function}
    h(A) \coloneqq \operatorname{Tr}(H_P\cdot A).
\end{equation}

\subsection{The loss landscape on the reachable orbit}

\begin{prop}[Absence of suboptimal orbit minima]
\label{prop:no_suboptimal_orbit_minima}
Assume that the adjoint orbit $\mathcal O_\rho$ of a pure-state projector $\rho$ (see \eqref{eq:adjoint_orbit}) contains a projector $\ket{\phi}\!\bra{\phi}$ onto a normalized ground state of $H_P$:
\begin{equation}
\label{eq:orbit_intersects_ground_space_descent}
    \min_{A \in \mathcal O_\rho} h(A) = \bra{\phi}H_P\ket{\phi}=\lambda_{\min}(H_P).
\end{equation}
Then $h$ has no suboptimal local minima on $\mathcal O_\rho$. Specifically:
\begin{enumerate}[label=(\roman*)]
    \item Every local minimum $A \in \mathcal O_\rho$ is a global minimum, corresponding to a rank-one projector onto a ground state of $H_P$ with $h(A) = \lambda_{\min}(H_P)$.
    \item Every critical point $A \in \mathcal O_\rho$ with $h(A) > \lambda_{\min}(H_P)$ possesses a tangent direction $Y \in T_A \mathcal O_\rho$ of strictly negative second variation:
    \begin{equation}
        \operatorname{Hess}(h|_{\mathcal O_\rho})_A(Y, Y) < 0.
    \end{equation}
\end{enumerate}
\end{prop}

\begin{proof}[Proof of Proposition~\ref{prop:no_suboptimal_orbit_minima}]
Let $\mathfrak h = i\mathfrak g$ denote the Hermitian form of the dynamical Lie algebra, and let $\operatorname{proj}_{\mathfrak h}$ be the Hilbert--Schmidt orthogonal projection onto $\mathfrak h$. Define the compact connected adjoint orbit 
\[
    \mathcal A := \operatorname{Ad}_G\bigl(\operatorname{proj}_{\mathfrak h}\rho\bigr) \subset \mathfrak h.
\]
By $G$-equivariance of $\operatorname{proj}_{\mathfrak h}$ and inclusion of stabilizers $G_\rho \subseteq G_{\operatorname{proj}_{\mathfrak h}\rho}$, the map
\begin{equation}
\label{eq:state_orbit_to_adjoint_orbit}
    q: \mathcal O_\rho \longrightarrow \mathcal A,
    \qquad
    q(A) = \operatorname{proj}_{\mathfrak h} A,
\end{equation}
is a well-defined surjective submersion. Because $H_P \in \mathfrak h$, the energy functional factors through $q$ as $h(A) = f(q(A))$, where
\begin{equation}
\label{eq:energy_factors_through_projection}
    f: \mathcal A \longrightarrow \mathbb R,
    \qquad
    f(M) = \operatorname{Tr}(M H_P).
\end{equation}

The function $f$ is the component of the moment map for the Hamiltonian $G$-action on the orbit $\mathcal A$ along the line spanned by $H_P \in \mathfrak h$. By \cite[Lemma~2.2]{atiyah1982convexity}, any such moment-map component is a Morse--Bott function whose critical manifolds all have even Morse indices. Since critical manifolds of index zero correspond to local minima and distinct sublevel set components can merge only at critical manifolds of index one (which are absent here), $f$ possesses a unique local-minimum connected component on $\mathcal A$, which is necessarily the global minimum set \cite[Lemma~2.1]{atiyah1982convexity}. Because all Morse--Bott indices of $f$ are even, any critical point lying outside this global minimum set cannot have index zero and must instead have index $\lambda \ge 2$, meaning the Hessian has at least two negative eigenvalues. Consequently, every non-global critical point $M \in \mathcal A$  admits a tangent direction $Z \in T_M \mathcal A$ such that $\operatorname{Hess}(f)_M(Z, Z) < 0$.

As $q$ is a smooth submersion, its differential $dq_A: T_A \mathcal O_\rho \to T_{q(A)} \mathcal A$ is surjective at every point $A \in \mathcal O_\rho$. Pullback by a submersion preserves local minima and critical points \cite[Proposition~9.6]{Boumal2023}. If $A \in \mathcal O_\rho$ is a critical point of $h$ with $h(A) > \lambda_{\min}(H_P)$, then $q(A)$ is a non-global critical point of $f$. Choosing $Y \in T_A \mathcal O_\rho$ such that $dq_A(Y) = Z \in T_{q(A)} \mathcal A$, the second-order chain rule at a critical point ($df_{q(A)} = 0$) yields
\[
    \operatorname{Hess}(h|_{\mathcal O_\rho})_A(Y, Y) = \operatorname{Hess}(f)_{q(A)}(Z, Z) < 0.
\]
Finally, assumption \eqref{eq:orbit_intersects_ground_space_descent} identifies the global minimum value on $\mathcal O_\rho$ with $\lambda_{\min}(H_P)$, completing the proof.
\end{proof}

At a noncritical orbit point $\sigma$, an explicit loss function descent generator direction is
\begin{equation}
\label{eq:projected_orbit_gradient}
    X_\sigma := \operatorname{proj}_{\mathfrak g} \bigl([\sigma, H_P]\bigr),
\end{equation}
where the projection uses the Hilbert--Schmidt inner product $\langle X, Y\rangle_{\mathrm{HS}} = -\operatorname{Tr}(XY)$ on skew-Hermitian matrices. For the one-parameter unitary flow $\sigma(t) = e^{tX_\sigma} \sigma e^{-tX_\sigma}$, the corresponding loss function $\ell_t(\sigma, H_P) = \operatorname{Tr}(H_P \sigma(t))$ satisfies
\begin{equation}
\label{eq:strict_orbit_descent_detailed}
\frac{d}{dt}\bigg\rvert_{t=0} \ell_t(\sigma, H_P)
= -\|X_\sigma\|_{\mathrm{HS}}^2 < 0.
\end{equation}
Moreover, $X_\sigma = 0$ if and only if $\sigma$ is a critical point of the orbit loss under $G$.

\subsubsection{Lifting orbit descent to circuit parameters}

Let $m = \dim_{\mathbb R}(\mathfrak g)$ denote the dimension of the dynamical Lie algebra. The construction from Section~\ref{sec:infinitesimal} provides a smooth local family of parameterized unitaries
\begin{equation}
\label{eq:identity_block_elementary_factors}
    V(\boldsymbol x)
       =\prod_{\ell=1}^{L}
          e^{\vartheta_\ell(\boldsymbol x)a_{j_\ell}},
    \qquad
    L\leq\mathcal T_{\mathfrak g},
\end{equation}
defined on a neighborhood of $\boldsymbol 0 \in \mathbb R^m$, where the $a_{j_\ell}$ are skew-Hermitian control generators and
\begin{equation}
\label{eq:identity_block_full_rank}
    V(\boldsymbol0)=I,
    \qquad
    (dV)_{\boldsymbol0}(\mathbb R^m)=\mathfrak g.
\end{equation}

Fix a circuit $U(\boldsymbol\theta_p)$ of depth $p$, and set 
\[
\rho_p \coloneqq U(\boldsymbol\theta_p) \rho U^\dagger(\boldsymbol\theta_p).
\]
For the augmented circuit 
\begin{equation}
    \label{eq:augmented_circuit}
    U(\boldsymbol x, \boldsymbol\theta_p) = V(\boldsymbol x) U(\boldsymbol\theta_p),
\end{equation}
 the loss function becomes
\begin{equation}
\label{eq:augmented_identity_block_objective}
    \ell_{(\boldsymbol x,\boldsymbol\theta_p)}
       \!\left(\rho,H_P\right)
       =\operatorname{Tr}\!\left[
          V(\boldsymbol x)\rho_p
          V^\dagger(\boldsymbol x)H_P
        \right].
\end{equation}
In particular,
\begin{equation}
\label{eq:identity_extension_preserves_energy}
    \ell_{(\boldsymbol0,\boldsymbol\theta_p)}
       \!\left(\rho,H_P\right)
       =\ell_{\boldsymbol\theta_p}\!\left(\rho,H_P\right),
\end{equation}
demonstrating that appending the identity block initially preserves the state and energy.

\begin{proof}[Proof of Theorem~\ref{thm:finite_depth_exposure_descent}]
Consider the state endpoint map $\Psi_p : \mathbb R^m \to \mathcal O_\rho$ given by
\[
    \Psi_p(\boldsymbol x) = V(\boldsymbol x) \rho_p V^\dagger(\boldsymbol x).
\]
By virtue of \eqref{eq:identity_block_full_rank}, its differential at the origin satisfies
\begin{equation}
\label{eq:state_endpoint_surjective}
    (d\Psi_p)_{\boldsymbol 0}(\mathbb R^m) 
        = [\mathfrak g, \rho_p] 
        = T_{\rho_p} \mathcal O_\rho.
\end{equation}
Thus, $\Psi_p$ is a local submersion onto the unitary orbit $\mathcal O_\rho$ at $\boldsymbol x = \boldsymbol 0$. By  \cite[Proposition~9.6]{Boumal2023}, first- and second-order criticality of the parameterized loss $\ell_{(\boldsymbol x, \boldsymbol\theta_p)}$ at $\boldsymbol x = \boldsymbol 0$ is strictly equivalent to that of the orbit loss at $\rho_p$.

Since Proposition~\ref{prop:no_suboptimal_orbit_minima} asserts that all non global minima orbit critical points are saddle points or local maxima (excluding suboptimal second-order local minima), $\rho_p$ cannot be second-order optimal unless $\ell_{\boldsymbol\theta_p}(\rho, H_P) = \lambda_{\min}(H_P)$.  Thus either the parameter gradient is nonzero or the parameter Hessian has a negative direction. In either case, arbitrarily small parameter changes strictly decrease the loss.
\end{proof}

\end{document}